%% file: Brandt-Nbody.tex
\documentclass[conference, onecolumn, oneside, 12pt]{IEEEtran}
\usepackage{geometry}
\usepackage{fancyhdr}
\usepackage[backend=biber,giveninits=true,bibencoding=utf-8,maxbibnames=5]{biblatex}
\bibliography{references}

\input{settings}
\input{macros}

\begin{document}

\renewcommand{\thelstlisting}{\arabic{lstlisting}}

\setlength{\parskip}{1em}

\title{\huge{On Distributed Gravitational $N$-Body Simulations}}
\author{
\IEEEauthorblockN{Alexander Brandt}
\IEEEauthorblockA{\small{Department of Computer Science}\\
\small{University of Western Ontario}\\
\small{London, Canada}\\
\small{Email: abrandt5@uwo.ca}}
}

\maketitle
\input{abstract}
\input{intro}
\input{gravitational}
\input{background}

\input{parallel}

\input{experimentation}

\renewcommand{\bibname}{References}
\printbibliography

\end{document}

%% file: settings.tex
\usepackage{lhelp}
\usepackage{amsfonts,latexsym}
\usepackage{amsmath}
\usepackage{amssymb}
\usepackage{bm}
\usepackage{nicefrac}
\usepackage[font=footnotesize]{caption}
\usepackage{subcaption}
\usepackage{paralist}
\usepackage{booktabs}
\usepackage{varwidth}

\usepackage{algorithm}
\usepackage{algorithmicx}
\usepackage{algpseudocode}

\usepackage{verbatim}
\usepackage{enumerate}
\usepackage{hhline}

\usepackage{listings}

\usepackage[hidelinks]{hyperref}
\usepackage{cleveref}

\usepackage{tikz}
\usepackage{pgfplots}
\usetikzlibrary{positioning,calc}
\usetikzlibrary{tikzmark}
\usetikzlibrary{arrows.meta,chains,decorations.pathreplacing,scopes,positioning}
\usetikzlibrary{calc,shapes.multipart,chains,arrows}
\graphicspath{ {figures/} }
\DeclareGraphicsExtensions{.eps,.png}

\usepackage{scrextend} 

\usepackage[usetransparent=false]{svg}
\usepackage{etoolbox}
\usepackage{xcolor}
\definecolor{commentgreen}{RGB}{2,112,10}
\definecolor{eminence}{RGB}{158,58,180}
\definecolor{weborange}{RGB}{230,145,0}
\definecolor{frenchplum}{rgb}{0.55, 0.0, 0.55}

\newtoggle{InString}{}
\togglefalse{InString}

\usepackage{diagbox}
\usepackage{listings}

\newcommand\digitstyle{\color{eminence}}
\makeatletter
\newcommand{\ProcessDigit}[1]
{%
	\ifnum\lst@mode=\lst@Pmode\relax%
	{\digitstyle #1}%
	\else
	#1%
	\fi
}
\makeatother
\lstdefinestyle{mystyle} {
	language=C++,
	frame=tb,
	tabsize=4,
	showstringspaces=false,
	numbers=left,
	backgroundcolor=\color{black!3}, 
	belowcaptionskip=1\baselineskip,
	breaklines=true,
	basicstyle=\footnotesize\ttfamily,
	commentstyle=\color{commentgreen},
	keywordstyle=\bfseries\color{eminence},
    numberstyle=\tiny\color{gray},
	stringstyle=\color{weborange},
	basicstyle=\small\ttfamily, 
	emph={int,char,double,float,unsigned,void,bool,AltArrZ\_t, mpz\_t, degrees\_t, SMZP},
	emphstyle={\color{blue!95!black}},
	escapechar=\&,
	otherkeywords={>,<,.,;,-,!,=,~},
	morekeywords = [2]{>,<,.,;,-,!,=,~},
	keywordstyle = [2]\color{black},
}

%% file: macros.tex
\def\p {\ensuremath{\mathbf{p}}}

\def\x {\ensuremath{\mathbf{x}}}

\def\y {\ensuremath{\mathbf{y}}}

\def\0 {\ensuremath{\mathbf{0}}}

\algnewcommand\algorithmicswitch{\textbf{switch}}
\algnewcommand\algorithmiccase{\textbf{case}}
\algnewcommand\algorithmicassert{\texttt{assert}}
\algnewcommand\Assert[1]{\State \algorithmicassert(#1)}%
\algdef{SE}[SWITCH]{Switch}{EndSwitch}[1]{\algorithmicswitch\ #1\ \algorithmicdo}{\algorithmicend\ \algorithmicswitch}%
\algdef{SE}[CASE]{Case}{EndCase}[1]{\algorithmiccase\ #1}{\algorithmicend\ \algorithmiccase}%
\algtext*{EndSwitch}%
\algtext*{EndCase}%

\algdef{SE}[SUBALG]{Indent}{EndIndent}{}{\algorithmicend\ }%
\algtext*{Indent}
\algtext*{EndIndent}

\errorcontextlines\maxdimen

\newcommand{\ALGtikzmarkcolor}{black}
\newcommand{\ALGtikzmarkextraindent}{4pt}
\newcommand{\ALGtikzmarkverticaloffsetstart}{-1.0ex}
\newcommand{\ALGtikzmarkverticaloffsetend}{-0.3ex}
\makeatletter
\newcounter{ALG@tikzmark@tempcnta}

\newcommand\ALG@tikzmark@start{%
	\global\let\ALG@tikzmark@last\ALG@tikzmark@starttext%
	\expandafter\edef\csname ALG@tikzmark@\theALG@nested\endcsname{\theALG@tikzmark@tempcnta}%
	\tikzmark{ALG@tikzmark@start@\csname ALG@tikzmark@\theALG@nested\endcsname}%
	\addtocounter{ALG@tikzmark@tempcnta}{1}%
}

\def\ALG@tikzmark@starttext{start}
\newcommand\ALG@tikzmark@end{%
	\ifx\ALG@tikzmark@last\ALG@tikzmark@starttext
	\else
	\tikzmark{ALG@tikzmark@end@\csname ALG@tikzmark@\theALG@nested\endcsname}%
	\tikz[overlay,remember picture] \draw[\ALGtikzmarkcolor] let \p{S}=($(pic cs:ALG@tikzmark@start@\csname ALG@tikzmark@\theALG@nested\endcsname)+(\ALGtikzmarkextraindent,\ALGtikzmarkverticaloffsetstart)$), \p{E}=($(pic cs:ALG@tikzmark@end@\csname ALG@tikzmark@\theALG@nested\endcsname)+(\ALGtikzmarkextraindent,\ALGtikzmarkverticaloffsetend)$) in (\x{S},\y{S})--(\x{S},\y{E});%
	\fi
	\gdef\ALG@tikzmark@last{end}%
}

\apptocmd{\ALG@beginblock}{\ALG@tikzmark@start}{}{\errmessage{failed to patch}}
\pretocmd{\ALG@endblock}{\ALG@tikzmark@end}{}{\errmessage{failed to patch}}
\makeatother

\newcommand{\hidetext}[1]{\mbox{ \ }}

\definecolor{nicedarkgreen}{rgb}{0.05, 0.35, 0.12}

\newcommand{\textblockcomment}[1]{}
\renewcommand{\gets}{:=}

%% file: abstract.tex
\begin{abstract}
The $N$-body problem is a classic problem involving
a system of $N$ discrete bodies mutually interacting 
in a dynamical system.  At any moment in time there
are ${N(N-1)}/{2}$ such interactions occurring. 
This scaling as $N^2$ leads to computational difficulties
where simulations range from tens of thousands of bodies
to many millions.
Approximation algorithms, such as the famous Barnes-Hut algorithm,
simplify the number of interactions to scale as $N\log{N}$.
Even still, this improvement in complexity is insufficient to achieve
the desired performance for very large simulations on 
computing clusters with many nodes and many cores. 
In this work we explore a variety of algorithmic techniques 
for distributed and parallel variations on the Barnes-Hut algorithm 
to improve parallelism and reduce inter-process communication requirements.
Our MPI implementation of distributed gravitational $N$-body
simulation is evaluated on a cluster of 10 nodes, 
each with two 6-core CPUs, to test the effectiveness and 
scalability of the aforementioned techniques. 
\end{abstract}


%% file: intro.tex
\begin{figure}[h]
\centering
\includegraphics[width=0.9\linewidth, trim={600px, 400px, 600px, 390px}, clip]{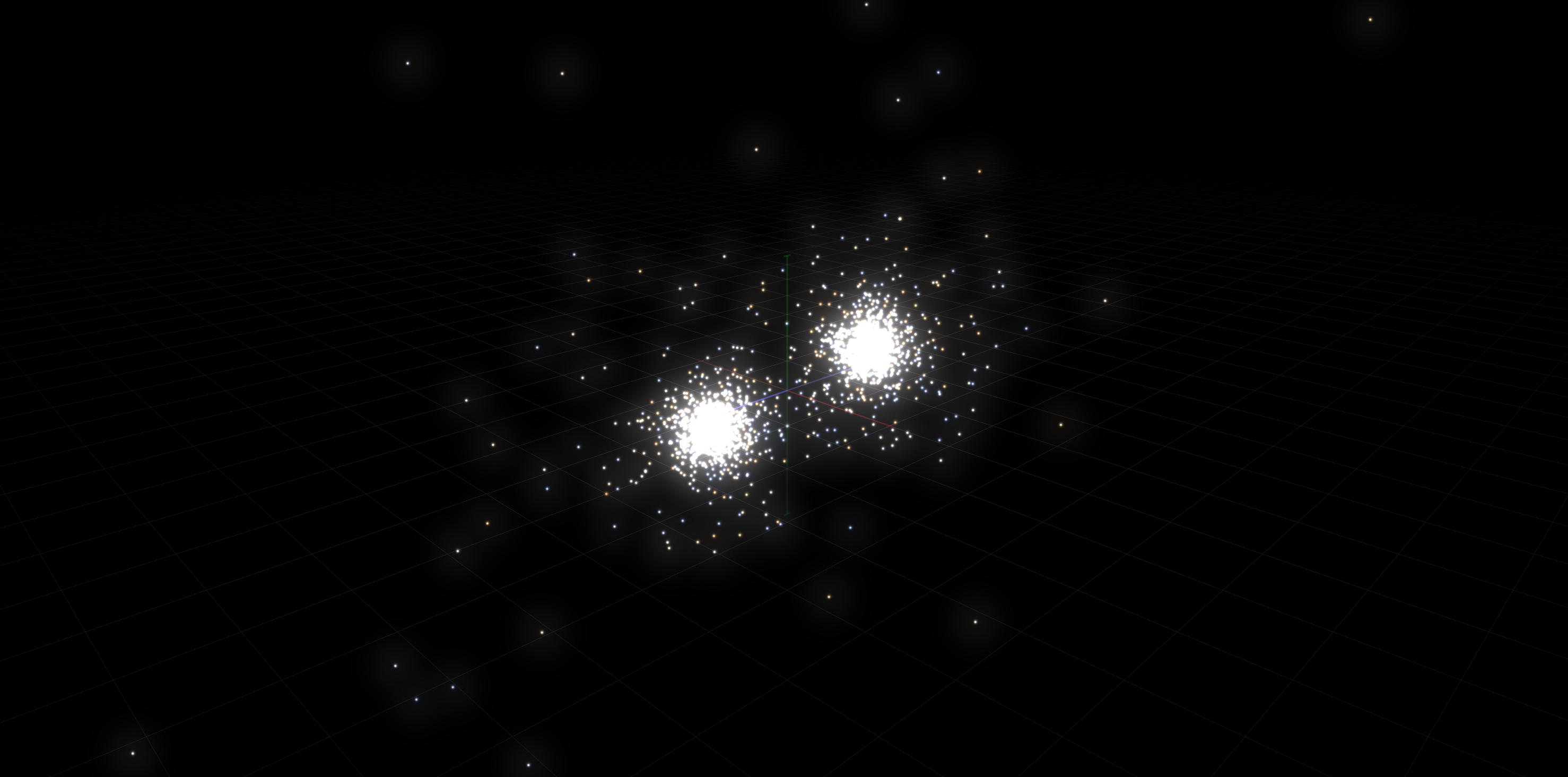}
\caption{Simulating the collision of two globular clusters, each consisting of 1000 stars.}\label{fig:ogltwoclusters}
\end{figure}

\section{Introduction}
\label{sec:intro}

$N$-body problems involve a large collection of problems 
wherein a set of $N$ bodies, particles, or points, interact under
mutually influential forces together in a dynamical system. 
Often, the dynamical system is governed by a potential field,
for example, electrical, magnetic, or gravitational.
The result is a system of second order differential equations 
where the acceleration of each body is computed from forces or potential 
created by the other $N-1$ bodies. Then, velocities and positions are updated
from this acceleration.
Targeted applications include fluid dynamics, 
molecular dynamics, quantum mechanics, astrophysics, and cosmology.
Historically, the $N$-body problem refers specifically to the problem where bodies are
governed by gravity and, moreover, will be the focus of this work.
Whether the physical scale of each body is very small or very large, these $N$-body problems 
share two common issues: the lack of analytic solutions, and scalability.

As the number of bodies in a system increases, the number
of differential equations describing the dynamical system grows proportionally.
Due to the mutual interaction of those bodies,
the number of interactions, and the number of terms
in the differential equations, grows as $N^2$. 
It is therefore prohibitively challenging to determine exact solutions
to these systems of equations as the value of $N$ increases. 
For example, in quantum mechanics, and when restricting the number of
dimensions to only 2, exact solutions are only known up to the order of $N=10$ \cite{sutherland2004beautiful}.
For bodies moving under the force of gravity, general solutions are known only for $N=2$, 
while for $N=3$ solutions exist for only very specific initial conditions \cite[Ch. 5]{gould1996computer}.
Simulations are therefore required to study systems with any practically meaningful number of bodies.
The efficiency of simulations with such a practical number of bodies is thus highly important.

In astronomy and gravitational $N$-body systems there are several different 
simulation scales of practical importance: 
\begin{enumerate}[$(i)$]
	\item celestial mechanics, where a few dozen bodies---planets, stars---typically orbit a single body
	of much higher mass than the others, e.g., a solar system;
	\item stellar dynamics, where a collection of roughly equal mass stars interact under their mutual gravity, 
	e.g., $10^4$--$10^6$ stars bound in a globular cluster;
	\item galaxy cluster evolution, where each body represents an entire galaxy and $10^2$--$10^3$ such galaxies are gravitationally-bound in a so-called galaxy cluster; and
	\item cosmological simulation, where $10^6$--$10^{11}$ bodies simulate the evolving structure of the universe.
\end{enumerate}
These latter two, however, must also account for dark matter and relativistic behaviour
(see~\cite{warren20142hot} and references therein).
The former two categories are dominated rather by Newtonian dynamics, 
posing less of a mathematical challenge and more of a computational one.
Cosmological simulation also requires extreme-scale computing (petascale computing and beyond).
For these reasons, we focus on stellar dynamics throughout this report.

Where the number of bodies in a simulation increases from hundreds to thousands, or millions, or even billions, 
the ability to simulate that number quickly becomes infeasible. 
While, technically, simulating a system with $N^2$ interactions 
is a polynomial-time algorithm, these $N^2$ interactions must
be computed at every time-step of the simulation. 
The result is a $\mathcal{O}(TN^2)$ running time, for $T$ 
simulation steps. $T$ is relatively unbounded as scientists 
wish to examine the long-term evolution of these systems.
As either $T$ or $N$ grows in orders of magnitude, 
\textit{direct simulation} of these $N^2$ interactions 
becomes impossible.

Research into approximation algorithms 
to reduce the running time complexity of such simulations,
while maintaining accurate results, has been ongoing for decades. 
Several different methods appeared in the late 1980s
which brought the complexity of computing interactions down to $\mathcal{O}(NlogN)$.
The \textit{Particle-Mesh} method \cite[Ch. 1]{hockney1988computer} represents
particles\footnote{Particles, points, bodies are all terms used interchangeably to mean
the same thing: discrete positions in space. The differences only come from different application domains 
and bear little difference in interpretation or understanding. Throughout this text we interchange the terms to keep the writing varied and interesting.} as a grid or mesh and then solves the potential on this density-mesh. 
This method is well-suited where separations between particles are large, but 
is insufficient where particle density is high.
Hierarchical methods, or \textit{treecodes}, on the other hand, look to balance 
approximating long-range interactions with using the direct method for close-range interactions.
The \textit{Fast Multipole Method} \cite{carrier1988fast} and 
the \textit{Barnes-Hut} algorithm \cite{barnes1986hierachical}
are two examples of hierarchical methods. We explore both in more detail in
Section~\ref{sec:bgoctree}.

As computer hardware began to evolve in the 1990s,
so did support for parallel execution of $N$-body simulations 
on multiprocessors and distributed systems.
Two different pioneering works appeared
simultaneously at \textit{Supercomputing '93}. 
The \textit{costzones} approach by Singh et al. \cite{singh1993parallel}, and the
\textit{hashed octree} method of Warren and Salmon \cite{warren1993parallel}.
Both works attempt to parallelize hierarchical methods through 
adaptive domain decomposition, dynamic load-balancing across processors, 
and parallel interaction computation. 
The costzones approach was initially developed for a (distributed) shared memory system, while
the hashed octree method was developed explicitly for distributed systems.
On modern multi-core multiprocessor systems, the shared memory costzones approach
is simple and efficient on a single compute node. 
Yet, as simulation sizes and durations grow, explicit distributed 
computing is required to achieve higher performance without limiting simulations
to a single node. Nonetheless, the costzones approach can be adapted to 
distributed computing, as we will explore.

Throughout this report we present a step-by-step guide to the implementation
of gravitational $N$-body simulation and, in particular, algorithmic
techniques which bolster a distributed and parallel 
implementation. We collect and coherently present the work of many papers: \cite{barnes1986hierachical}, \cite{warren1993parallel},
\cite{DBLP:journals/ijhpca/SalmonW94},  \cite{warren20142hot}, \cite{singh1993parallel}, \cite{DBLP:journals/jpdc/SinghHTGH95}, and \cite{DBLP:journals/tocs/SingHG95};
with plenty of otherwise missing details coming from \cite{aarseth2003gravitational}.
The parallelizations and optimization techniques of the costzones and hashed octree methods
are presented incrementally to build up to a robust distributed algorithm (see Section~\ref{sec:paralleltrees}).
The required details are made explicit through ample figures and algorithmic listings. 
Further, our implementation, written in C/C++ and making using of the OpenMPI~\cite{gabriel04:_open_mpi}
implementation of Message Passing Interface (MPI), is freely and openly available at
\textcolor{blue}{\url{https://github.com/alexgbrandt/Parallel-NBody/}}.
While no algorithm or technique here is cutting edge, nor at the level of modern super-computing,
it is nonetheless foundational and can be applied to small local area network clusters.

The remainder of this report is organized as follows.
We begin in Section~\ref{sec:math} by describing the mathematical model
underpinning gravitational simulations. 
Section~\ref{sec:bgoctree} is further background describing the basis
of hierarchical methods through their use of spatial tree representations
(quadtrees, octrees) and force approximations.
Section~\ref{sec:paralleltrees} explores the algorithmic techniques of 
the costzones and hashed octree methods to parallelize treecodes.
These techniques are applied incrementally to produce continually more complex code
but with continually increasing scalability, resulting in 
8 total algorithm configurations.
Finally, Section~\ref{sec:experimentation} 
evaluates the implementation of these algorithms 
and presents our experimental data., while Section~\ref{sec:conclusion}
concludes.

%% file: gravitational.tex
\section{Gravitational $N$-Body Simulation}
\label{sec:math}

\setlength{\abovedisplayskip}{1em}
\setlength{\belowdisplayskip}{1em}

This section reviews the mathematical background necessary to perform
an $N$-body simulation governed by the force of gravity. 
While this information is not strictly necessary to understand the 
distributed computing and algorithmic techniques discussed later in Section~\ref{sec:paralleltrees}, 
it does concretely define how ``interactions'' between bodies are computed.
Moreover, this section presents the mathematical reasoning, particularly gravitational potential, 
which allows for the development of the approximation algorithms described in Section~\ref{sec:bgoctree}.

We begin in Section~\ref{sec:gravforce} describing the direct interactions between
particles using Newton's law of universal gravitation. Then, Section~\ref{sec:multipole}
describes gravitational potential and the gravitational field. Through superposition of 
gravitational potentials, and a multipole expansion of that superposition, the 
acceleration due to the collective gravity of many bodies can be accurately approximated.
Section~\ref{sec:leapfrog} explains how this acceleration can be used to 
update the dynamical system over time (i.e. perform integration).
Finally, Section~\ref{sec:energy} explains how $N$-body simulations 
can be scaled and standardized for consistent comparison across simulation scales.

This section is largely based on a comprehensive text by Sverre Aarseth;
see \cite{aarseth2003gravitational} and references therein for further details.
Throughout this section, scalars are given by Latin characters, and vectors are three-dimensional
and given by bold Latin characters.
The vector norm $||\cdot||$ is the Euclidean norm.

\subsection{Gravitational Force}
\label{sec:gravforce}

Newton's law of universal gravitation describes the attractive forces
between any two particles. 
For two point masses with mass $m_i$ and $m_j$, respectively, separated
by a distance $r$, we know the magnitude of their mutual attractive gravitational force is given by
\begin{align}
F = \frac{G{m_i}{m_j}}{r^2}   \label{eqn:gforcesimple}
\end{align}
where $G$ is the gravitational constant.

For brevity, let us describe a particle whose mass is $m_i$ as the mass $m_i$.
Less mass $m_i$ be at position $\bm{r}_i$ and mass $m_j$ be at position $\bm{r}_j$.
Describing the force exerted on mass $m_i$ by mass $m_j$ can then be given by:
\begin{align}
\bm{F}_{ij} = \frac{G{m_i}{m_j}}{\lVert \bm{r}_i - \bm{r}_j \rVert^2} \cdot \frac{-(\bm{r}_i - \bm{r}_j)}{\lVert \bm{r}_i - \bm{r}_j \rVert}. \label{eqn:gforce}
\end{align}
Here, $\lVert \bm{r}_i - \bm{r}_j \rVert = r$ and $\nicefrac{-(\bm{r}_i - \bm{r}_j)}{\lVert \bm{r}_i - \bm{r}_j \rVert}$ is a unit vector pointing from mass $m_i$ to mass $m_j$. Using  Newton's second law,
the acceleration $\bm{a}_{ij}$ induced on mass $m_i$ by the force
$\bm{F}_{ij}$ is given by $\bm{F}_{ij} = m_i\bm{a}_{ij}$.

Using Newtonian dynamics
combined with this law of gravity yields the equations of motions for a system of bodies moving
under gravitational forces. 
For a system of $N$ bodies there exists $3N$ second-order differential equations 
describing the motion of the bodies. Using vector notation reduces the number of
equations to $N$ and improves readability. 
Then, the position of the body with index $i$ evolves as:
\begin{align}
\ddot{\bm{r}}_i = \bm{a}_i &= \sum_{\substack{j=1\\j\neq i}}^{N} {\bm{a}_{ij}} \label{eqn:motion1} \\ &= \frac{1}{m_i}\sum_{\substack{j=1\\j\neq i}}^{N} {\bm{F}_{ij}} \\ &= -G \sum_{\substack{j=1\\j\neq i}}^{N} \frac{m_j(\bm{r}_i - \bm{r}_j)}{\lVert\bm{r}_i - \bm{r}_j\rVert^3}.  \label{eqn:motion}
\end{align}
Therefore, determining the interaction between two bodies $i$ and $j$ 
in an $N$-body simulation is simply the computation of 
this $\bm{F}_{ij}$ or, equivalently, the acceleration $\bm{a}_{ij}$.
This defines a so-called \textit{particle-particle} interaction.
Summing (\ref{eqn:gforce}) over all other bodies, as given by (\ref{eqn:motion}), and dividing through by $m_i$, 
yields the acceleration of body $i$, $\bm{a}_i$, as in (\ref{eqn:motion1}).

Due to the finite and numerical nature of the simulation,
it is useful to introduce a so-called 
\textit{softening factor} to a particle-particle interaction.
Including a small value $\epsilon$
in computing particle separations helps reduce the effects of close-encounters.
One simply replaces  $\lVert \bm{r}_i - \bm{r}_j \rVert$ with 
 $(\lVert \bm{r}_i - \bm{r}_j \rVert^2 + \epsilon^2)^{1/2}$.
This softening avoids numerical errors associated with singularities
and prevents odd behaviour arising from discretizing time steps during close encounters \cite[Ch. 2]{aarseth2003gravitational}.

However, we know that to compute $\bm{F}_{ij}$ for every pair of points 
quickly becomes computationally infeasible.
One key observation, which is employed by hierarchical $N$-body methods, 
is the approximation of the gravitational force exerted by many discrete bodies at once.
The fundamental observation is that for some subset of bodies $J$, 
whose centre of mass is $\bm{r}_J$ and total mass is $m_J$, the force acting on particle $i$
from this ensemble of points is: 
\begin{align}
\bm{F}_{iJ} = \sum_{j \in J} \frac{-G{m_i}{m_j}(\bm{r}_i - \bm{r}_j)}{\lVert \bm{r}_i - \bm{r}_j \rVert^3} \approx 
\frac{-G{m_i}{m_J}(\bm{r}_i - \bm{r}_J)}{\lVert \bm{r}_i - \bm{r}_J \rVert^3} \label{eqn:approxforce}.
\end{align}
Recall computing the centre of mass is simply $\bm{r}_J = \sum_{j \in J}m_j\bm{r}_j$.
This approximation approaches equality as the separation distance $\lVert\bm{r}_i - \bm{r}_J\rVert$ grows larger.
However, this approximation can be further refined by using gravitational potential.

\subsection{Gravitational Field, Gravitational Potential, and Multipole Expansion}
\label{sec:multipole}

Expanding from Newton's original idea of an attractive force between point masses, 
a better model is that of the gravitational field. 
The gravitational field of a point mass $M$ is a vector field describing the 
force of gravity per unit mass felt by a body at any point in space
in the field. The gravitational field is typically denoted by $\bm{g}$ and,
for body with mass $m$ at position $\bm{r}$, is defined as:
\begin{align}
\bm{g}(\bm{r}) = \frac{\bm{F}(\bm{r})}{m} = -\frac{GM}{r^3}\bm{r}.
\end{align}
Notice that force per unit mass is precisely acceleration as from Newton's second law.
Hence, the acceleration due to gravity of mass $m$ by mass $M$ is precisely $\bm{g}$. 

The gravitational potential $\Phi$ relates
energy, work, and force. It can be defined 
as the gravitational potential energy per unit mass at a particular location.
Equivalently, it can be defined as the amount of
work per unit mass done against gravity 
to move a mass from infinity to that location.
In the frame of reference of the point mass $M$ 
inducing a gravitational field (i.e. putting the point mass at the origin), 
the gravitational potential at position $\bm{r}$ 
at a distance $r$ from the point mass is:
\begin{align}
\Phi(\bm{r}) = \frac{1}{m}\int_{\infty}^{r} \bm{F}\cdot d\bm{r} = \int_{\infty}^{r} \frac{GM}{r^2} dr = -\frac{GM}{r}.
\label{eqn:workpotential}
\end{align}

Gravitational potential is related to the gravitational field by the gradient.
Since gravity is an attractive force, gravitational potential
is negative while the gravitational field is positive.
The acceleration due to gravity of an object at position 
$\bm{r}$ is then given as the negative gradient of the potential:
\begin{align}
\bm{a} = \bm{g}(\bm{r}) = -\nabla\Phi(\bm{r}) = -\frac{GM}{r^3}\bm{r}. \label{eqn:accelpot}
\end{align}

This implies that we can use gravitational potential 
for the force calculations and interactions in our $N$-body system. 
The useful consequence of this is that gravitational potentials
can be combined via superposition. Given a subset of points $J$, as before, 
and a point $\bm{r}$ external to the sphere enclosing the points in $J$, the resulting potential is: 
\begin{align}
\Phi(\bm{r}) &= \sum_{j \in J} -\frac{Gm_j}{\lVert\bm{r} - \bm{r}_j\rVert} \label{eqn:exactpotentialsum} \\
&\approx -\frac{Gm_J}{\lVert\bm{r} - \bm{r}_J\rVert}. \label{eqn:approxpotentialsum}
\end{align}
Much like forces, we can approximate the superposition of
potentials using their total mass $m_J$ and their centre of mass $\bm{r}_J$ as seen in (\ref{eqn:approxpotentialsum}).
However, an even better approximation can be obtained using a \textit{multipole expansion}.

For simplicity of notation, let us take $G=1$ henceforth.
A multipole expansion is a series expansion of a function 
depending on angles. Often, the set of basis functions 
for this series expansion are the spherical harmonics, 
a set of orthogonal functions defined on the surface of a sphere. 
In Cartesian coordinates these functions are the 
Legendre polynomials. Following \cite[Ch. 3]{aarseth2003gravitational}, 
let us denote the 	Legendre  polynomial of order (degree) $k$ as $P_k$. 
The multipole expansion of (\ref{eqn:exactpotentialsum})
is then given by:
\begin{align}
\Phi(\bm{r}) = -\frac{1}{r}\sum_{j\in J}m_j \sum_{k=0}^n \left(\frac{r_j}{r}\right)^k P_k(\cos\theta_j), \label{eqn:multipole}
\end{align}
with $r_j = \lVert\bm{r}_j\rVert$ and $\cos\theta = {\bm{r}\cdot\bm{r}_j}/{rr_j}$.
The direct summation of (\ref{eqn:exactpotentialsum}) is approached with
higher order $n$ in the series (\ref{eqn:multipole}).

In this series expansion the first few terms are called the monopole, dipole, and quadrupole terms.
Including even higher-ordered terms in the approximation yields diminishing
returns since their contribution to the approximation decreases with higher order
meanwhile their computational cost increases.
Often only up to quadrupole terms are included, yet more
recent works have included up to octopole terms \cite{hubber2011seren}.

Using the compact notation of \cite{aarseth2003gravitational},
the multipole expansion of $\Phi$ to order 2 is:
\begin{align}
\Phi_2(\bm{r}) = -\frac{M_0}{r} - \frac{D_ax_a}{r^3} - \frac{Q_{ab}x_ax_b}{2r^5}, \label{eqn:truncmultipole}
\end{align}
where $\bm{r} = (x_1, x_2, x_3)$ and the summation over $\{1,2,3\}$ for indices $a$ and $b$
is implied. There are therefore 3 dipole coefficients $D_a$ and 9 quadrupole coefficients $Q_{ab}$.
The multipole coefficients hide the summation over the
bodies in $J$ as: 
\begin{equation*}
\begin{aligned}
M_0 &= \sum_{j \in J}m_j, \\
Q_{aa} &= \sum_{j \in J}m_j\left(3x_{a,j}^2 - r_j^2 \right), 
\end{aligned}
\qquad\quad
\begin{aligned}
 D_a &= \sum_{j \in J}m_jx_{a,j}, \\
Q_{ab} &= \sum_{j \in J}m_j\left(3x_{a,j}x_{b,j}\right), \ \ \text{if } (a \neq b), 
\end{aligned}
\end{equation*}
where $x_{a,j}$ is the $a$th coordinate of $\bm{r}_j$.

Notice that the monopole term  is precisely the sum of the masses in $J$
as in our initial approximation.
Next, notice that taking the frame of reference to be the centre of mass of 
the points in $J$ causes the dipole term to vanish ($D_a = 0$) by definition of $\bm{r}_J$.
Via the relation $\bm{g}(\bm{r}) = -\nabla\Phi(\bm{r})$,
the multipole expansion (\ref{eqn:truncmultipole}) can be used in place of 
the approximation (\ref{eqn:approxforce}) for a set of masses $J$.
Therefore, we need only to add the quadrupole terms 
to obtain an approximation which is better by two orders.

\input{leapfrog}

\subsection{Energy of the System and Standard Units}
\label{sec:energy}

In order to understand the accuracy of our simulation, 
and in particular our approximation method,
a quantitative measure of accuracy or stability is very useful.
In a gravitational $N$-body simulation, this value is the total system's energy.
This is natural following the law of conservation of energy.
Since our simulation, by construction, simulates an isolated system, the total
energy should be conserved. The total energy of an $N$-body
system is given by the kinetic energy and the potential energy
of the $N$ particles. For masses $m_i$, positions $\bm{r}_i$, and velocities $\bm{v}_i$, 
we have:
\begin{align}
E_{kin} = \sum_{i=1}^N \frac{1}{2}m_i\lVert \bm{v}_i\rVert^2, \qquad
E_{pot} = \sum_{i=1}^N \sum_{\substack{j=1\\j\neq i}}^N -\frac{Gm_im_j}{\lVert\bm{r}_i - \bm{r}_j\rVert}, \qquad E_{tot} = E_{kin} + E_{pot}.
\end{align}

By computing $E_{tot}$ at the beginning and the end of a simulation (or even throughout), 
it is possible to obtain a measure of how well the simulation performed. A perfect simulation 
would perfectly conserve energy. Of course this is not possible due to floating point errors,
truncation errors in the numerical integrator (see Section~\ref{sec:leapfrog}), 
and the approximations introduced in the approximation algorithm itself.

While looking at conservation of energy for any one system is a good measure of its 
stability, it is not completely adequate. 
Different systems, with different initial conditions,
can have vastly different values for their total energy.
We would like a way to measure accuracy agnostic to particular initial conditions or
simulation parameters like $N$ or total mass.
The few practitioners of $N$-body simulation in the 1980s then decided on the so-called 
\textit{standard units} for gravitational $N$-body simulations,
see \cite{heggie1986standardised} and \cite[Ch. 7]{aarseth2003gravitational}.

The standard units define a scaling whereby all simulations 
have an initial total energy $E_0 = -0.25$.
In the standard units, the total mass of the system is scaled to 1, the gravitational 
constant is 1, and the system's coordinates are shifted to be centred at its centre of mass.
This latter fact makes the system's centre of mass the origin and the
net velocity $\vec{0}$.
To scale a system to standard units it is first shifted to its centre of mass frame
and then its kinetic energy and potential energy calculated. 
From the virial theorem, a system in equilibrium 
has kinetic energy equal to a negative half of its potential energy.
One can scale velocities and positions to independently and respectively 
scale $E_{kin}$ and $E_{pot}$ to obtain $E_{kin} = \nicefrac{-1}{2}E_{pot}$.
Finally, the positions and velocities are scaled together
so that $E_0 = -0.25$;
see Algorithm~7.2 in \cite{aarseth2003gravitational}.

%% file: leapfrog.tex
\subsection{Simulating Dynamical Systems with Leapfrog Integration}
\label{sec:leapfrog}

Up to this point we have focused on how to calculate the interactions between bodies
in our simulation, and thus each body's acceleration.
Now we must use this acceleration to update the positions and velocities
of the bodies in our system so that it may evolve over time.

The simplest method is the basic Euler method. 
Velocities are updated from accelerations, and positions are updated from velocities. 
Let $\bm{r}_i^{(k)}$ and $\bm{v}_i^{(k)}$
be the position and velocity, respectively, of the $i$th particle at the $k$th time-step.
The forward Euler method uses values of derivatives at the current step
to approximate values at the next step:
\begin{align}
\bm{r}^{(k+1)} = \bm{r}^{(k)} + \bm{v}^{(k)}\Delta t, \\
\bm{v}^{(k+1)} = \bm{v}^{(k)} + \bm{a}^{(k)}\Delta t.
\end{align}

While simple, the forward Euler method is only a first-order method, 
meaning that the (local) error introduced at each time-step
is proportional to $\mathcal{O}(\Delta t^2)$ \cite{corless2013graduate}.
A higher-order method would see that the error reduces more quickly 
with smaller time-steps. 
For example, a common fourth-order method is the \textit{Hermite scheme} \cite[Ch. 2]{aarseth2003gravitational}.
This scheme is a more complex generalization 
of the classic and simpler scheme known as \textit{leapfrog}, which is sufficient
in many cases.

The leapfrog method is as simple and as computationally expensive as the Euler method, 
but is a second-order method, thus providing better accuracy.
Leapfrog specifically solves second-order differential equations of the form
$\ddot{\bm{r}} = \bm{a(\bm{r})}$ or, equivalently, $\dot{\bm{v}} =  \bm{a}(\bm{r}), \dot{\bm{r}} = \bm{v}$.
The method gets its name from the way that the two differential equations 
are solved at interleaved time-steps, so that each equation jumps over the other in time.
The classic formulation of leapfrog shows this jumping structure: 
\begin{align}
\bm{v}^{(k + \frac{1}{2})} &= \bm{v}^{(k - \frac{1}{2})} + \bm{a}^{(k)}\Delta t, \label{eqn:classicleapv} \\
\bm{r}^{(k + 1)} &= \bm{r}^{(k)} + \bm{v}^{(k + \frac{1}{2})}\Delta t. \label{eqn:classicleapr}
\end{align}
One of the key advantages to this scheme is its simplicity and time-symmetry.
That is, it provides the same results when run forward in time as 
when run backward in time. In the context of gravitational simulations, 
this time-symmetry prevents systematic build-up in error in the total energy of
the system over time \cite{hut1995building}. Why this is a favourable condition
is discussed next in Section~\ref{sec:energy}.

With a simple rearrangement and translation of time-steps, these equations
can be written at integer time-steps.
In particular, we can arrive at the 
``Kick-Drift-Kick'' (KDK) version\footnote{The KDK leapfrog is also known as the velocity Verlet method.} of leapfrog:
\begin{align}
\bm{v}^{(k+\frac{1}{2})} &= \bm{v}^{(k)} + \bm{a}^{(k)}\frac{\Delta t}{2}, \label{eqn:leapkick1}\\
\bm{r}^{(k+1)} &= \bm{r}^{(k)} + \bm{v}^{(k+\frac{1}{2})}\Delta t, \label{eqn:leapdrift}\\
\bm{v}^{(k+1)} &= \bm{v}^{(k + \frac{1}{2})} + \bm{a}^{(k+1)}\frac{\Delta t}{2}. \label{eqn:leapkick2}
\end{align}
The idea behind KDK leapfrog is to use acceleration from the previous step ($k$) to update velocity
a little bit, to time-step $k + \nicefrac{1}{2}$; this is the first kick.
Then, position is allowed to update by drifting along the trajectory induced by the updated velocity. 
Then, a second kick occurs to update the velocity based on the acceleration at the current time step ($k+1$).

This variation of leapfrog allows for variable time-steps. That is, 
changing the value of $\Delta t$ over the course of a simulation to give 
more precision during volatile points of the simulation (i.e. close encounters).
While the KDK leapfrog is employed in our implementation, 
we currently only use a fixed time-step.


%% file: background.tex

\clearpage

\section{Hierarchical $N$-body Methods}
\label{sec:bgoctree}

In $N$-body simulations, the dynamical system describing the 
bodies includes $N$ differential equations.
By the mutual interaction of these $N$ bodies, 
the differential equation describing each body includes $N-1$ terms 
for the $N-1$ interactions of that body with all others.
The details of this dynamical system were described in Section~\ref{sec:math}.

From the tools established in the previous section---force or potential calculation, 
acceleration calculation, and discretized updates to velocity and position 
via leapfrog integration---we have everything required to perform an $N$-body 
simulation.
Directly computing the $N(N-1)$ interaction terms (or $\nicefrac{1}{2}N(N-1)$,
if taking advantage of symmetries), of the dynamical is infeasible for large-scale simulations.
Nonetheless, Algorithm~\ref{alg:nbodysim} depicts this simple scheme.
For each discrete time step in the simulation, the acceleration
is computed for each body (Lines 2--5) and then each body's position and velocity are updated (Lines 6--7).

\begin{center}
	\vspace{-0.5em}
	\begin{minipage}{0.65\textwidth}
		\begin{algorithm}[H]
			\centering
			\footnotesize
			\caption{\scriptsize\textsc{Direct $N$-Body Simulation}($P$, $\Delta t$, $t_{end}$)\newline The general scheme loops over small time intervals, first computing the forces on and the acceleration of each body for that interval, then updating each body's position. $P$ is the list of bodies and $t_{end}$ and $\Delta t$ are simulation parameters for the total simulation time and time-step, respectively.}\label{alg:nbodysim}
			\algnotext{EndIf}
			\algnotext{EndFor}
			\algnotext{EndWhile}
			\begin{algorithmic}[1]
				\For {$t$ \textbf{from} 0 \textbf{to} $t_{end}$ \textbf{by} $\Delta t$}
				\For {\textbf{each} particle $p$ in $P$} \Comment{Force Calculation}
				\State Set the acceleration of $p$ to $\vec{0}$.
				\For {\textbf{each} particle $q$ in $P \setminus \{p\}$}
				\State Add the force contribution of $q$ acting on $p$ to the acceleration of $p$.
				\EndFor
				\EndFor
				\For {\textbf{each} particle $p$ in $P$} \Comment{Position Update}
				\State Update position and velocity of $p$ using its current acceleration.
				\EndFor
				\EndFor
			\end{algorithmic}
		\end{algorithm}
	\end{minipage}
\end{center}
\vspace{0.5em}

Computing these direct particle-particle interactions scales as $N^2$, 
and is computationally infeasible in practice.
Much research has been dedicated to reducing the number
of interactions computed while maintaining accuracy in the simulation.
As we have seen in Sections~\ref{sec:gravforce} and \ref{sec:multipole}, 
a collection of bodies, or a \textit{mass distribution}, can be approximated
by its total mass and centre of mass. The multipole expansion 
can also be used for even better approximation.
Moreover, recall that the approximation (\ref{eqn:approxforce}) in Section~\ref{sec:gravforce}
approaches equality with larger distances from this centre of mass.
Hierarchical methods (treecodes) take advantage of this approximation 
to reduce the number of interaction terms computed in the force calculations of each time step
in the simulation.
These methods only modify the force calculation loop of the simulation,
leaving the position update, i.e. numerical integration, unchanged.

In the late 1980s two such hierarchical methods arose to reduce the computational efforts
required in the force calculation step of $N$-body simulations. 
The first method to appear was by Barnes and Hut \cite{barnes1986hierachical}
and eventually named the Barnes-Hut algorithm.
The second method of Greengard and Rokhlin is the Fast Multipole Method (FMM) \cite{greengard1987fast}.
Both of these approximation algorithms---approximate in the way force acting on a body is
calculated, not by numerical or truncation errors---are based on a hierarchical 
tree representation of the geometric space containing the bodies; hence, treecodes.
In two-dimensions these methods employ quadtrees, while in three-dimensions octrees are used.

The key idea in treecodes is to describe an ensemble of bodies, a mass distribution, 
by a single representative pseudo-body in place of its
constituents. Then, a far-enough away test body can approximate its interactions 
with each of the constituent bodies by a single interaction with the pseudo-body.
These are so-called long-range interactions.
Applying this technique hierarchically, the further away a target point is from 
a group of points, the more of those points can be included in an ensemble
to further reduce the number of interactions to be computed for the target point.
Whenever two points are considered to be too close, and thus the approximation too inaccurate, 
computations fall back to the direct method of computing particle-particle interactions.
Hierarchical methods thus drastically reduce the total number of interactions needed to be computed, 
down to $\mathcal{O}(NlogN)$.

\begin{figure}[htb]
\centering
\begin{subfigure}{0.45\textwidth}
	\centering
	\begin{tikzpicture}[scale=0.75]
	\draw
	[step=2cm] (4,4) grid +(4,4)
	[step=2cm] (0,4) grid +(4,4)
	[step=1cm] (2,4) grid +(2,2)
	[step=1cm] (4,4) grid +(2,2)
	[step=2cm] (4,0) grid + (4,4)
	[step=0.5cm] (4, 2) grid +(1,1)
	[step=0.5cm] (4, 3) grid +(1,1)
	[step=1cm] (5, 2) grid +(1,1)
	[step=0.5cm] (2, 4) grid +(1,1);
	\draw
	[step=4cm, semithick, line cap=round] (0,0) grid (8,8);
	
\draw[fill] (0.88, 1.77) circle (0.06cm);
\draw[fill] (0.43, 4.65) circle (0.06cm);
\draw[fill] (0.77, 7.23) circle (0.06cm);
\draw[fill] (2.11, 4.13) circle (0.06cm);
\draw[fill] (2.41, 4.81) circle (0.06cm);
\draw[fill] (2.75, 4.25) circle (0.06cm);
\draw[fill] (2.70, 4.60) circle (0.06cm);

\draw[fill] (3.60, 4.44) circle (0.06cm);
\draw[fill] (3.44, 5.77) circle (0.06cm);

\draw[fill] (4.20, 2.78) circle (0.06cm);
\draw[fill] (4.65, 2.42) circle (0.06cm);
\draw[fill] (4.33, 3.27) circle (0.06cm);
\draw[fill] (4.26, 3.88) circle (0.06cm);
\draw[fill] (5.18, 2.77) circle (0.06cm);
\draw[fill] (5.80, 3.13) circle (0.06cm);

\draw[fill] (7.22, 0.8) circle (0.06cm);
\draw[fill] (6.77, 3.2) circle (0.06cm);

\draw[fill] (4.64, 4.76) circle (0.06cm);
\draw[fill] (5.72, 5.31) circle (0.06cm);
\draw[fill] (6.22, 4.81) circle (0.06cm);
\draw[fill] (6.4, 6.81) circle (0.06cm);

	\end{tikzpicture}
	\subcaption{The particle distribution of 21 points in two dimensions contained in a quadtree.}\label{fig:quadtree}
\end{subfigure}
\hspace{0.5em}
\begin{subfigure}{0.4\textwidth}
\centering
\includegraphics[width=0.975\textwidth, 
]{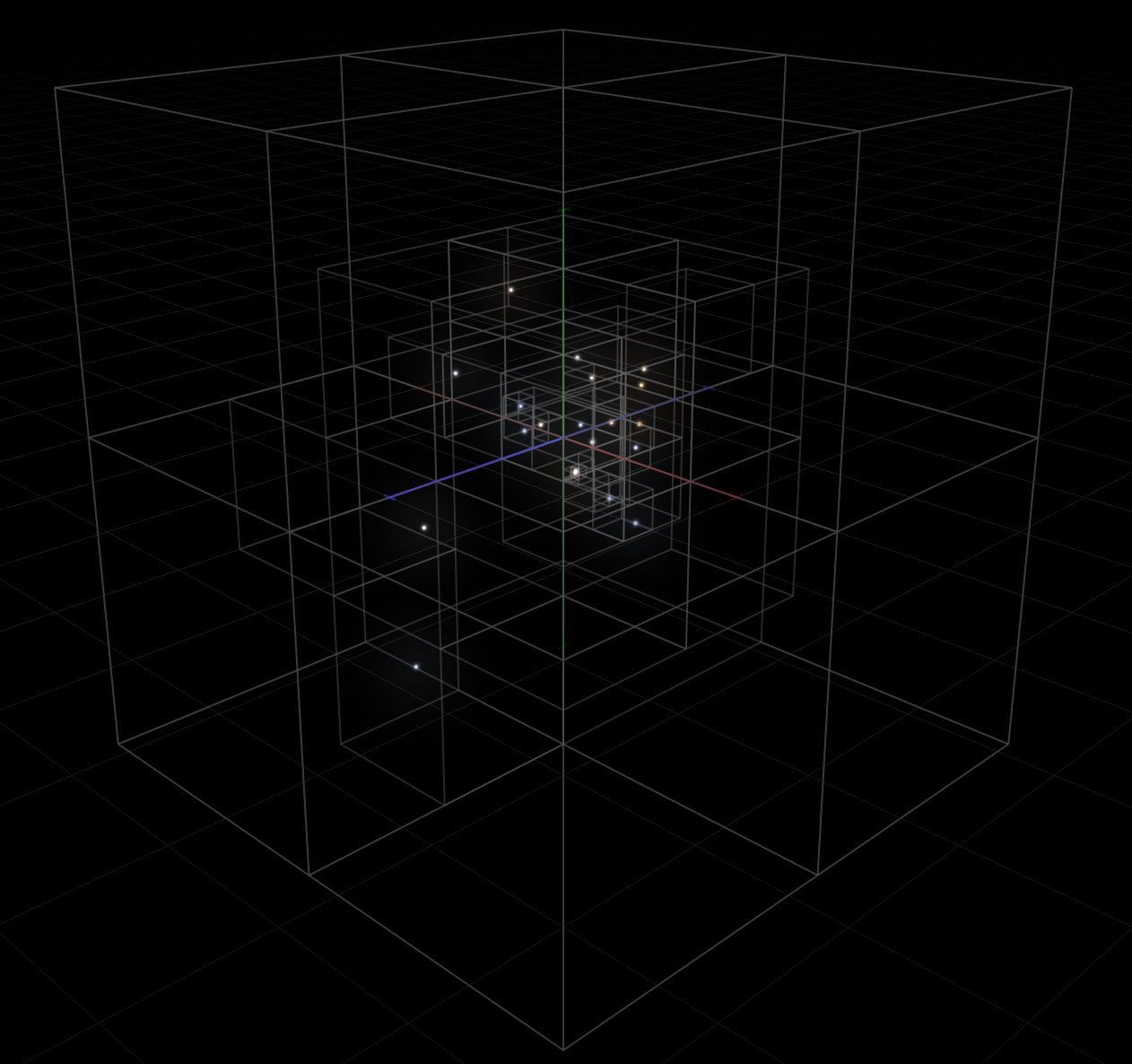}
\subcaption{The particle distribtuion of 20 points in three dimensions contained in an octree.}\label{fig:octree}
\end{subfigure}
\end{figure}

The Barnes-Hut and FMM methods are both hierarchical methods. 
They differ in two places: ($i$) how an interaction is determined to be ``long-range'', 
and ($ii$) how long-range interactions are specifically computed. 
However, let us begin with the commonalities. 
Both methods begin by constructing a hierarchical tree representing the entire geometric space
which bounds the particles being simulated. 
The root of this tree is precisely the bounding box of all the particles. 
The root is then split in half in each of its dimensions. In two dimensions, 
a \textit{quadtree} has each node split into 4 children, the area of each being $\nicefrac{1}{4}$
of its parent. In three dimensions, an \textit{octree} splits each of its nodes into 8 children, 
the volume of each being $\nicefrac{1}{8}$ of its parent.
This splitting continues recursively until each leaf node contains exactly one particle.
A quadtree is shown in Fig.~\ref{fig:quadtree} while an octree is shown in Fig.~\ref{fig:octree}.

With each particle assigned to a leaf node, each method now beings parsing the tree in a bottom-up fashion.
Starting from the leaves, for each level in the tree, a single representative particle 
is computed for the ensemble of particles in the child (and grandchild, etc.) nodes below it.
To compute this representative particle, 
one can simply place it at the the centre of mass of the child particles and
set its mass equal to the sum of masses of its children.
This is the original formulation of the Barnes-Hut method \cite{barnes1986hierachical}.
However, in the FMM method and later variations of the Barnes-Hut method \cite{DBLP:journals/jpdc/SinghHTGH95,warren20142hot,DBLP:journals/ijhpca/SalmonW94},
the representative particle is instead described by a \textit{multipole expansion}
of the gravitational potential of the ensemble of points. Recall
that the centre of mass formulation is equivalent to only the monopole term
in the multipole expansion, and including more terms is only required for increased
accuracy in approximation; see Section~\ref{sec:multipole}.

%
When including terms in the multipole expansion beyond the centre of mass, this representative particle, 
or pseudo-body, at each level is no longer really a single particle
but rather a formula expressing the mass distribution of its children.
For a better description we then refer to such a pseudo-body as simply a cell, or node, 
since there is a one-to-one correspondence between these
and the nodes or cells of a tree structure.\footnote{By both
\textit{node} and \textit{cell} we refer to the discrete data units of a tree structure.
Cell is more common in the context of treecodes, while node is more common for generic data structures.}
In treecodes, the tree structure thus fulfills two aspects: 
\begin{inparaenum}[$(i)$]
	\item a spatial decomposition of the bodies being simulated; and 
	\item succinctly encoding hierarchical groupings of bodies and the approximations of those mass distributions. 
\end{inparaenum}

\begin{figure}[htb]
\centering
\usetikzlibrary{angles, quotes, decorations.pathreplacing}
\begin{tikzpicture}[scale=0.8]
\draw
[step=2cm] (4,4) grid +(4,4)
[step=2cm] (0,4) grid +(4,4)
[step=1cm] (2,4) grid +(2,2)
[step=1cm] (4,4) grid +(2,2)
[step=2cm] (4,0) grid + (4,4)
[step=0.5cm] (4, 2) grid +(1,1)
[step=0.5cm] (4, 3) grid +(1,1)
[step=1cm] (5, 2) grid +(1,1)
[step=0.5cm] (2, 4) grid +(1,1);
\draw
[step=4cm, semithick, line cap=round] (0,0) grid (8,8);

	\draw[fill] (0.88, 1.77) circle (0.06cm);
\draw[fill] (0.43, 4.65) circle (0.06cm);
\draw[fill] (0.77, 7.23) circle (0.06cm);
\draw[fill] (2.11, 4.13) circle (0.06cm);
\draw[fill] (2.41, 4.81) circle (0.06cm);
\draw[fill] (2.75, 4.25) circle (0.06cm);
\draw[fill] (2.70, 4.60) circle (0.06cm);

\draw[fill] (3.60, 4.44) circle (0.06cm);
\draw[fill] (3.44, 5.77) circle (0.06cm);

\draw[fill] (4.20, 2.78) circle (0.06cm);
\draw[fill] (4.65, 2.42) circle (0.06cm);
\draw[fill] (4.33, 3.27) circle (0.06cm);
\draw[fill] (4.26, 3.88) circle (0.06cm);
\draw[fill] (5.18, 2.77) circle (0.06cm);
\draw[fill] (5.80, 3.13) circle (0.06cm);

\draw[fill] (7.22, 0.8) circle (0.06cm);
\draw[fill] (6.77, 3.2) circle (0.06cm);

\draw[fill] (4.64, 4.76) circle (0.06cm);
\draw[fill] (5.72, 5.31) circle (0.06cm);
\draw[fill] (6.22, 4.81) circle (0.06cm);
\draw[fill] (6.4, 6.81) circle (0.06cm);

\coordinate (particle) at (0.77, 7.23);
\coordinate (leftcorner) at (3,4);
\coordinate (rightcorner) at (2,5);
\draw[draw=red, thick] (leftcorner) rectangle (rightcorner);


\draw[blue, |-|, thick] (2.48, 4.4425) -- node[above=0.09, xshift=0.9, black] {$\,d_1$} (particle);
\draw[decorate,decoration={brace, amplitude=3pt, raise=2pt}] (2,4) to node[black,midway,left= 0.12cm] {$\ell_1$} ++(0, 1);
\draw[fill=yellow] (0.77, 7.23) circle (0.06cm);

\draw[red, thick] (4,0) rectangle (8,4);
\draw[decorate, decoration={brace, amplitude=5pt, raise=2pt, aspect=0.75}] (4,0) to node[black, pos=0.75, left=0.15cm] {$\ell_2$} (4,4);
\draw[|-|, thick, blue] (0.88, 1.77) -- node[above, black, pos=0.4] {$d_2$} (5.322, 2.661);
\draw[fill=purple] (0.88, 1.77) circle (0.06cm);

\draw[blue, thick, |-|] (5.72, 5.31) -- (6.22, 4.81);
\draw[fill=green] (5.72, 5.31) circle (0.06cm);
\draw[fill=green] (6.22, 4.81) circle (0.06cm);
\end{tikzpicture}
		\caption{Determining particle-particle and particle-cell interactions in the Barnes-Hut method. The yellow point is well-separated from the cell with side length $\ell_1$ if $\nicefrac{\ell_1}{d_1} < \theta$. The purple point is well-separated from the large cell with side length $\ell_2$ if $\nicefrac{\ell_2}{d_2} < \theta$. The non-particle endpoints of the two line segments are at the centre of mass of the respective ensemble of points. The two green points are not well-separated and their close-range interaction must be computed directly.}\label{fig:separatedness}
\end{figure}
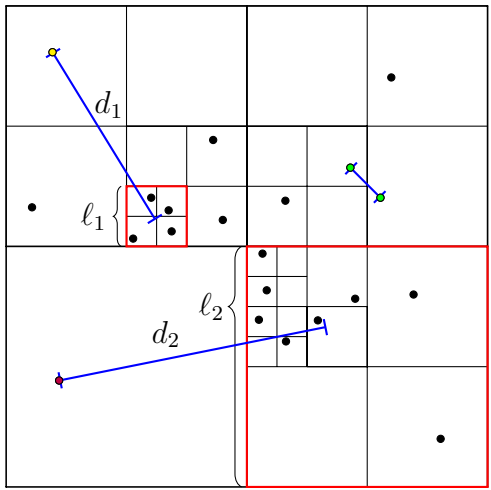

With the tree structure fully constructed, 
it is now used for force calculations for each body.  
Recall that the key behaviour of treecodes is to approximate the interaction between
a body and an ensemble of bodies.
Hierarchical ensembles are precisely what are encoded as the internal cells of the tree.
To accurately approximate the interaction between a body and an internal cell, 
they must be ``far enough away'' or \textit{well-separated}. 
This notion depends on the particular method.
\begin{enumerate}[($i$)]
	\item In the FMM method, a cell $d$ is statically determined to be well-separated from another cell $c$ if the separation is greater
	than the side length of $c$. Note that this definition is not symmetric.
	\item In the Barnes-Hut method, a parameter $\theta$, called the \textit{opening angle} or \textit{multipole acceptance criterion} (MAC), controls whether a cell should be considered well-separated. If the ratio between a cell's side length $\ell$ and the separation $d$ between a point and the cell's centre of mass is less than $\theta$ (i.e. $\nicefrac{\ell}{d} < \theta$), then they are well-separated. This is depicted in Fig.~\ref{fig:separatedness}.
\end{enumerate}

These two definitions also imply the second major difference between the two methods. 
While both methods directly compute the so-called particle-particle interactions when two particles are close together, 
the approximation of long-range interactions differ. In the Barnes-Hut method, all long-range interactions
are computed as particle-cell interactions. That is, a cell contributes to the net force acting on a particle. 
In contrast, in FMM, cell-cell interactions are possible, where all bodies in an ensemble experience the same exerted force by 
another ensemble. This additional type of interaction allows the FMM method to reach an asymptotic complexity of
$\mathcal{O}(N)$. Despite this attractive computational complexity, this method is more mathematically difficult
and more challenging to implement \cite{DBLP:journals/jpdc/SinghHTGH95}.
Further, particularly in three-dimensions, the constants discarded by the big-$\mathcal{O}$ notation in fact 
dominate running time for practical values of $N$ \cite{DBLP:journals/ijhpca/SalmonW94}.
For these reasons, we focus on the Barnes-Hut algorithm. 
We refer the reader to \cite{greengard1987fast, singh1993parallel, aarseth2003gravitational}
for further information on the FMM method.

In the Barnes-Hut method, the forces acting on each body are 
computed via a top-down tree traversal which can be implemented recursively, starting at the root node.
Let $\ell$ and $d$ be defined as before for the current node.
If the multipole approximation is acceptable, $\nicefrac{\ell}{d} < \theta$ 
(i.e. the particle is far enough away from the cell),
then a particle-cell interaction is computed and the traversal along 
this branch of the tree is complete.
Otherwise, the cell is ``opened'' and the traversal continues down
each of the current node's children.
If a leaf node is encountered, a typical particle-particle interaction is computed. 

The entire Barnes-Hut method is detailed in Algorithm~\ref{alg:BHloop}.
It beings by building the tree which describes the current positions of bodies (Lines 2--7).
Notice that the tree construction here proceeds differently as previously mentioned.
Rather than loading all particles into the tree and then recursively dividing cells until the leaves
contain exactly one particle, the tree is constructed one particle at a time, 
splitting cells and pushing particles down the tree until each particle is alone in its leaf node.
Once the tree is constructed, it is parsed bottom-up to compute the internal cell multipole moments (Lines 8--9).
Next, forces acting on each particle are computed via a tree traversal (Lines 13--22), where 
particle-particle interactions or particle-cell interactions are computed as appropriate. This is repeated
for every body in the simulation.
Finally, just as in the direct method, velocities and positions are updated (Lines 23--24).

We conclude the discussion of the classic Barnes-Hut method with a final remark.
The essential benefit of the Barnes-Hut method is
that to traverse the tree and compute the forces on a single particle in this hierarchical manner 
takes time $\mathcal{O}(\log N)$. Thus, to compute forces acting on all $N$ particles is $\mathcal{O}(N\log N)$.
First, notice that constructing the tree also takes time $\mathcal{O}(N\log N)$ since $N$ bodies
are inserted, on average, to a depth of $\log N$ in the tree. 
This depth arises from the fact that the average size of a leaf node is on the order of
the average inter-particle spacing. The best case is a uniform distribution of particles
which creates a balanced tree with $\log_8 N$ levels. 
Then, traversing the tree to compute the forces acting on a particle is also
on the order of $\mathcal{O}(\log N)$. In the worse case, the target particle is ``close''
to every other particle and no approximations can be made, thus 
computing $N$ interactions for that target particle. However, it is not possible
for every particle to be close to every particle. Indeed, 
for branches of the tree not including the target particle, 
it is much more common for those branches to be approximated with a single 
particle-cell interaction. Since well-separateness is controlled
by the parameter $\theta$, the amount of work spent in 
the force calculation phase is dependent on $\theta$. 
It is found that the force calculation for a single particle
typically scales as $\nicefrac{1}{\theta^2}\log N$ \cite{hernquist1998hierarchical};
a smaller value of $\theta$ requires more cells to be opened and thus more interactions to be computed.
Typical values of $\theta$ are on the order of $\nicefrac{1}{2}$ \cite{aarseth2003gravitational,barnes1986hierachical}.

\begin{center}
	\begin{minipage}{0.75\linewidth}
		\begin{algorithm}[H]
			\scriptsize
			\centering
			\caption{\scriptsize \textsc{Barnes-Hut Simultation Loop Body}($P$, $\theta$)\newline Given a list of particles $P$ and a MAC criterion $\theta$, compute forces and update positions for the current time-step. This is repeated for each time-step in the simulation.}\label{alg:BHloop}
			\algnotext{EndIf}
			\algnotext{EndFor}
			\algnotext{EndWhile}
			\begin{algorithmic}[1]
				\Statex Build Tree: 
				\Indent
				\State Create a tree $T$ with a single root node that is the bounding box of all particles. 
				\For {\textbf{each} particle $p$ in $P$} \Comment{Spatial decomposition of bodies}
				\State Insert $p$ into the leaf node $L$ of $T$ which encloses the point's coordinates. 
				\If {$L$ now holds two particles}
				\State Split $L$ in half in each of its dimensions to create children, $L$ is now an internal node.
				\State Insert each particle of $L$ into the appropriate child.
				\State Repeat until each particle is in a separate leaf node.
				\EndIf
				\EndFor
				\For {\textbf{each} node in $T$ from leaves to root} \Comment{Compute mass distribution approximations}
				\State Compute sum of masses of contained points, centre of mass, and multipole values.
				\EndFor
				\vspace{0.3em}
				\EndIndent
				\Statex Compute Forces:
				\Indent
				\For {\textbf{each} particle $p$ in $P$}
				\State Set the acceleration of p to $\vec{0}$.
				\State $Node$ $\gets$ the root node of $T$ \vspace{0.3em}
				\EndFor
				\EndIndent
				\Statex Compute Cell Force:
				\Indent
				\Indent
				\For {\textbf{each} child cell $C$ of $Node$}
				\If {$C$ contains 1 particle} \Comment{$C$ is a leaf}
				\State Add to $p$'s acceleration the contribution from $C$'s particle. \Comment{particle-particle}
				\Else
				\State $\ell$ $\gets$ $C$'s side length
				\State $d$ $\gets$ distance between $C$'s centre of mass (c.o.m.) and $p$'s position.
				\If {${\ell} / {d} < \theta$}
				\State Add to $p$'s acceleration the contribution from $C$. \Comment{particle-cell}
				\Else 
				\State Compute Cell Force with $Node \gets C$ \Comment{recursive call}
				\EndIf
				\EndIf
				\EndFor
				\EndIndent
				\EndIndent
				\vspace{0.3em}
				\Statex Integration step (update velocities and positions):
				\Indent
				\For {\textbf{each} particle $p$ in $P$} 
				\State Update $p$'s position and velocity using its acceleration.
				\EndFor
				\EndIndent
			\end{algorithmic}
		\end{algorithm}
	\end{minipage}
\end{center}
\vspace{0.5em}

%% file: parallel.tex
\section{Parallel Implementation of $N$-Body Simulations and Treecodes}
\label{sec:paralleltrees}

To gain further performance in $N$-body simulations,
and hierarchical methods for such simulations, 
parallel execution is a natural choice. 
Toward that goal we will examine two pioneering works:
the costzones approach of Singh et al. \cite{singh1993parallel,DBLP:journals/jpdc/SinghHTGH95}, and the hashed octree method 
of Warren and Salmon \cite{warren1993parallel,DBLP:journals/ijhpca/SalmonW94}.
The costzones approach was first developed for a shared-memory architecture, meanwhile
the hashed octree method was explicitly developed for a distributed system. 
Nonetheless, all the techniques we will discuss can be applied to either a shared-memory system
or a distributed system thanks to the message passing model being agnostic to the underlying
communication protocol.

We begin this section with a high-level overview of the 
designs and parallelization techniques of the costzones and hashed octree methods.
The details follow in the below subsections.

Just as with any parallel algorithm, both methods are concerned with:
\begin{enumerate}[($i$)]
	\item minimizing \textit{synchronization}---where processes must wait for all others before proceeding;
	\item \textit{load-balancing}---ensuring each process has an equal amount of work, thus reducing wait times at synchronization points; 
	\item reducing \textit{communication overheads}---the amount data and number of messages to be shared between processes; and
	\item reducing \textit{span}---the maximum amount of work done by any one process.
\end{enumerate}

Reducing span means parallelizing as many parts of the algorithm as possible. 
Both the costzones and hashed octree algorithms attempt to parallelize the Barnes-Hut algorithm by: 
\begin{enumerate}[($a$)]
	\item parallelizing the building of the octree;
	\item parallelizing the force calculation on each body, i.e. parallelizing the tree traversal; and
	\item parallelizing the integration step.
\end{enumerate}

An obvious and simple approach to parallelizing an $N$-body simulation is to simply
assign $\nicefrac{N}{s}$ bodies to each of the $s$ parallel processes.
This would be sufficient for the integration step, where each body's
position and velocity is easily updated from its own acceleration. 
However, due to the mutual interaction between all bodies, 
the computation of that acceleration requires a global view of the entire simulation 
domain, and thus a global view of the octree.
Both methods have a solution to this global view requirement.

In the costzones method, each process, in parallel, builds a so-called local octree 
for its assigned bodies. Then, these local octrees are merged into a global octree which is then
shared with all processes. Finally, with each process having a global view of the simulation domain, 
each process can proceed in parallel to compute the forces on each of its assigned bodies.

In the hashed octree method, again, each process builds its local octree in parallel
for its assigned bodies. This method, however, explicitly avoids creating a single global tree.
Instead, during the tree traversal and force calculations, when a local particle
needs information about the position or multipole values of a non-local particle or cell, 
that information is requested dynamically from the owning process. An optional improvement
is to use asynchronous communication and a latency-hiding traversal to avoid
stalling computations while waiting for communications (see Section~\ref{sec:parallel7}). 
To facilitate consistent addressing of octree cells stored on different processes, 
the octree is implemented as a \textit{hashed octree} or \textit{linear octree} \cite{DBLP:journals/cacm/Gargantini82};
hence the naming of this method.

Both methods are in fact very similar in design. They differ only in how a process 
receives octree information for non-local bodies.
The costzones approach uses upfront communication to construct and share the global octree, 
and then proceeds with tree traversals in parallel, without any communication.
On the other hand, the global tree is not strictly needed to calculate 
the force on any one particle due the particle-cell interactions of the Barnes-Hut method.  
The hashed octree method thus avoids computing the entire global tree at the beginning
and instead uses extra communication during the tree traversal to obtain only
the essential data for its assigned bodies.
In principle, this latter approach is the most scalable, where simulation sizes may
grow so large that the entire domain cannot be stored in the memory of a single process. 

Nonetheless, both techniques share many similarities to reduce span and improve load-balancing.
Firstly, a \textit{spatial decomposition} of the 
simulation domain is performed. Each process is then assigned a partition of the spatial domain,
and the bodies it contains, rather than statically assigning $\nicefrac{N}{s}$ bodies to each process. 
Having the bodies assigned to one process be spatially local greatly improves the 
tree merging of the costzones method 
and reduces the number of communications required during the 	tree traversal 
step of the hashed octree method. This latter fact is obvious 
considering when particle-particle interactions are computed in a Barnes-Hut algorithm.
Secondly, dynamic load-balancing is achieved by re-partitioning the spatial domain at every time step.
Particles in regions dense with others require more work during force computation 
due to the higher number of particle-particle interactions. Those spatial partitions
should be smaller and contain fewer particles so that each process
computes roughly the same number of interactions. 
Thirdly, the spatial partitioning of bodies is implemented by linearizing three-dimensional 
space via a space-filling curve, sorting the list of bodies based on this linearization,
and then partitioning the sorted list. 
This sort can be parallelized to further reduce span.

Clearly, there are many different opportunities for parallelization in 
the Barnes-Hut algorithm and many different algorithmic techniques which can be implemented 
to improve this parallelization through improved load-balancing, etc.
We now detail each parallelization and technique, one at a time, in the following subsections.
These techniques are added incrementally to our implementation 
to step-by-step improve the parallelization and scalability of the algorithm;
see Section~\ref{sec:experimentation} for that experimentation.

We begin in Section~\ref{sec:parallel0} by describing a simple parallelization
of the \textit{direct method} and the parallelization of the integration step.
This serves as a point of comparison for the remaining methods to show
that the additional work of creating and using the octree is worthwhile.
Section~\ref{sec:parallel1} presents a first parallelization
of the Barnes-Hut method where parallel force calculation via
tree traversal is added to the previously described parallel integration step.
Next, building the octree in parallel and the octree merging process is detailed in Section~\ref{sec:parallel2}.
The method of spatial decomposition is then detailed in Section~\ref{sec:parallel3}.
Spatial decomposition is further enhanced by dynamic load-balancing in Section~\ref{sec:parallel4}
and by distributed sorting in Section~\ref{sec:parallel5}.
Section~\ref{sec:parallel6} introduces the hashed octree and the techniques required to 
build local octrees which are globally consistent. 
Finally, we add latency-hiding asynchronous communication to the hashed octree tree traversal 
in Section~\ref{sec:parallel7}.
Therefore, the algorithm described in Section~\ref{sec:parallel5} describes the ``final''
version of our costzones-based method, meanwhile the algorithm described in Section~\ref{sec:parallel7}
describes the ``final'' version of our hashed octree method. 

In the follow sections we use common terminology from parallel processing
and distributed computing, e.g., gather, scatter, and broadcast. For definitions
and further details see \cite{mccool2012structured}.

\subsection{Naive Parallelization of Integration and Particle-Particle Interactions}
\label{sec:parallel0}

We have already seen the serial direct method as
Algorithm~\ref{alg:nbodysim} in Section~\ref{sec:bgoctree}.
We now formulate this algorithm as a distributed algorithm, 
as shown in Algorithm~\ref{alg:directdistrib}.
We begin by partitioning the bodies evenly across the $s$ processes.
No care is taken for spatial decomposition because all $\mathcal{O}(N^2)$ interactions
will be computed.
Next, we compute the mutual forces on the local bodies by the other local bodies (Lines 6--8).
Each process sets a temporary list of particles to its own list of local particles.
Then, $s - 1$ rounds of message passing occur (Lines 9--14). Assuming a ring topology, 
each process sends its current temporary list of particles to the process on the right
and receives a new list from the process on the left. The force
from each particle in this temporary list is then added to each particle 
in the local list. Then, the next round begins. After $s-1$ rounds process $id-1$ has
received process $id$'s particles and all other particles. This concludes 
the force calculation in this time-step.
Each process then updates the positions and velocities of each local particle (Lines 15--16).
In this algorithm and all algorithms which follow, we assume the processes are ranked $0,\ldots,s-1$.

\begin{center}
\vspace{-0.5em}
\begin{minipage}{0.85\textwidth}
\begin{algorithm}[H]
	\centering
	\footnotesize
	\caption{\scriptsize\textsc{Parallel Direct $N$-Body Simulation}($id$, $s$, $\Delta t$, $t_{end}$, $N$, $P_{all}$)\newline $P_{all}$ is the list of size $N$ of bodies to simulate from 0 to $t_{end}$ going by $\Delta t$ time-steps. $id$ is the executing process's id, $s$ is the number of processes. Assume $s \mid N$.}\label{alg:directdistrib}
	\algnotext{EndIf}
	\algnotext{EndFor}
	\algnotext{EndWhile}
	\begin{algorithmic}[1]
		\State $N' \gets N / s$
		\State $P \gets$ \textbf{Scatter} $P_{all}$, sending to process $id$ $P_{all}[id\cdot N', \ldots, id\cdot N' + N' - 1]$ \Comment{Get Local Bodies}
		\For {$t$ \textbf{from} 0 \textbf{to} $t_{end}$ \textbf{by} $\Delta t$}
		\State $P' \gets P$
		\State \textbf{for each} particle $p$ in $P$ set acceleration to $\vec{0}$.
		\For {\textbf{each} particle $p$ in $P$} \Comment{Force Calculation with Locals}
		\For {\textbf{each} particle $q$ in $P \setminus \{p\}$}
			\State Add the force contribution of $q$ acting on $p$ to the acceleration of $p$.
		\EndFor
		\EndFor
		\For {$i$ \textbf{from} 0 \textbf{to} $s-1$}
			\State Send $P'$ to process $id + 1 \mod s$
			\State $P' \gets $ particles received from process $id - 1 \mod s$
			\For {\textbf{each} particle $p$ in $P$} \Comment{Force Calculation with Remotes}
			\For {\textbf{each} particle $q$ in $P'$}
			\State Add the force contribution of $q$ acting on $p$ to the acceleration of $p$.
			\EndFor
		\EndFor
		\EndFor
		\For {\textbf{each} particle $p$ in $P$} \Comment{Position Update}
		\State Update position and velocity of $p$ using its current acceleration.
		\EndFor
		\EndFor
	\end{algorithmic}
\end{algorithm}
\end{minipage}
\end{center}
\vspace{0.5em}

In this algorithm, each process computes only $N'(N'-1) + N'(N')(s-1) = \nicefrac{1}{s}N(N-1)$ interactions. 
Further, each process integrates for only $\nicefrac{N}{s}$ particles. 
The span and load-balance of this algorithm are thus optimal, since the total work is $N(N-1)$ interactions
plus $N$ integrations. Nonetheless, the number of interactions still scales as $\mathcal{O}(N^2)$.
We thus look to parallelize the Barnes-Hut method with its $\mathcal{O}(N\log N)$ number of interactions.

\subsection{Parallel Tree Traversal}
\label{sec:parallel1}

Having seen our first distributed algorithm, we are now ready to modify 
the Barnes-Hut algorithm to be executed in parallel. 
As seen at the beginning of Section~\ref{sec:paralleltrees}, 
there are three aspects to parallelize: $(a)$ the building of the octree, 
$(b)$ the force calculation, and $(c)$ the integration step. 
Parallelizing the integration is \textit{embarrassingly parallel} and
is as simple as assigning a subset of the bodies to each process as 
was just seen in Algorithm~\ref{alg:directdistrib}.
The next simplest procedure is to parallelize the force calculation.
In the Barnes-Hut method, if the octree encompassing all bodies in the simulation has already been constructed, 
then parallel force calculation is also embarrassingly parallel as 
each process simply traverses the octree for each of its local bodies.

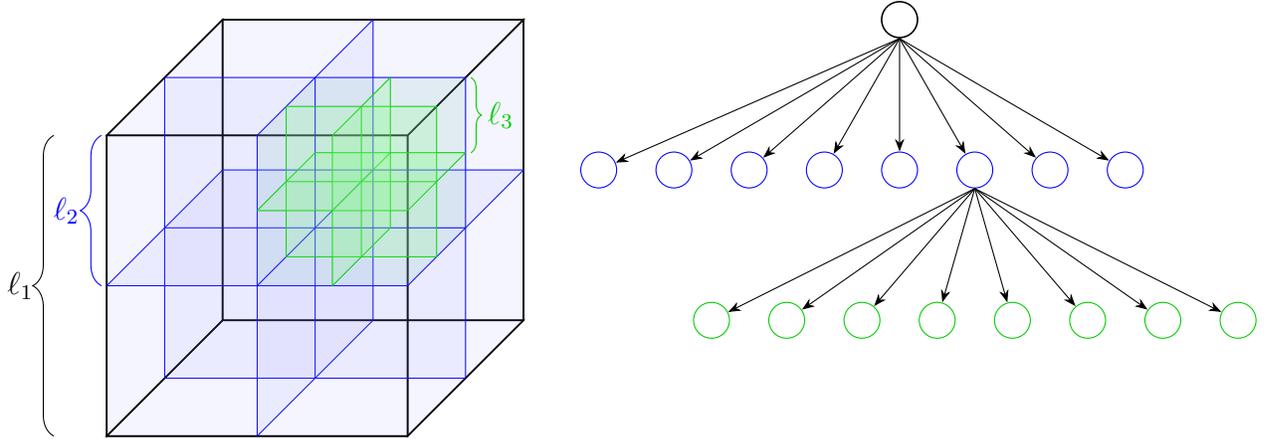
\begin{figure}[htb]
\begin{tikzpicture}

\node[circle, semithick, draw, scale=1.2] (root) at (9,4) {};
\foreach \x in {5,...,12} 
\node[circle, draw, scale=1.2, blue] (\x-1) at (\x,2) {};
\foreach \x in {6,...,13} 
\node[circle, draw,scale=1.2, green!80!black] (\x-2) at (\x.5,0) {};
\foreach \x in {5,...,12}
\draw[->, >={Stealth}] (root.south) -- (\x-1);
\foreach \x in {6,...,13}
\draw[->, >={Stealth}] (10-1.south) -- (\x-2);

\pgfmathsetmacro{\Depth}{4}
\pgfmathsetmacro{\Height}{4}
\pgfmathsetmacro{\Width}{4}

\coordinate (O) at (0,0,0);
\coordinate (A) at (0,\Width,0);
\coordinate (B) at (0,\Width,\Height);
\coordinate (C) at (0,0,\Height);
\coordinate (D) at (\Depth,0,0);
\coordinate (E) at (\Depth,\Width,0);
\coordinate (F) at (\Depth,\Width,\Height);
\coordinate (G) at (\Depth,0,\Height);

\draw[black, semithick, line cap=round] (O) -- (C) -- (G) -- (D) -- cycle;
\draw[black, semithick, line cap=round] (O) -- (A) -- (E) -- (D) -- cycle;
\draw[black, semithick, line cap=round] (O) -- (A) -- (B) -- (C) -- cycle;
\draw[black, semithick, line cap=round] (D) -- (E) -- (F) -- (G) -- cycle;
\draw[black, semithick, line cap=round] (C) -- (B) -- (F) -- (G) -- cycle;
\draw[black, semithick, line cap=round] (A) -- (B) -- (F) -- (E) -- cycle;

\draw[blue] (2, 0, 0) -- (2, 4, 0);
\draw[blue] (0, 2, 0) -- (4, 2, 0);
\draw[blue] (4, 2, 0) -- (4, 2, 4);
\draw[blue] (4, 0, 2) -- (4, 4, 2);
\draw[blue] (2, 0, 0) -- (2, 0, 4);
\draw[blue] (0, 0, 2) -- (4, 0, 2);
\draw[blue] (0, 2, 0) -- (0, 2, 4);
\draw[blue] (0, 0, 2) -- (0, 4, 2);
\draw[blue] (2, 4, 0) -- (2, 4, 4);
\draw[blue] (0, 4, 2) -- (4, 4, 2);
\draw[blue] (2, 0, 4) -- (2, 4, 4);
\draw[blue] (0, 2, 4) -- (4, 2, 4);
\draw[blue] (0, 2, 2) -- (4, 2, 2);
\draw[blue] (2, 2, 0) -- (2, 2, 4);
\draw[blue] (2, 0, 2) -- (2, 4, 2);

\draw[fill=blue!50,fill opacity=0.08, draw opacity=0] (0,4,0) -- (4,4,0) -- (4,0,0) -- (4,0,4) -- (0,0,4) -- (0,4,4) -- (0,4,0);
\draw[fill=blue!50,fill opacity=0.08,draw opacity=0] (0,2,0) -- (4,2,0) -- (4,2,4) -- (0,2,4) -- cycle;
\draw[fill=blue!50,fill opacity=0.08,draw opacity=0] (0,0,2) -- (4,0,2) -- (4,4,2) -- (0,4,2) -- cycle;
\draw[fill=blue!50,fill opacity=0.08,draw opacity=0] (2,0,0) -- (2,4,0) -- (2,4,4) -- (2,0,4) -- cycle;

\draw[blue,fill=green!50, opacity=0.1] (2,2,2) -- (2,4,2) -- (2,4,4) -- (2,2,4) -- cycle;
\draw[blue,fill=green!50,opacity=0.1] (2,2,2) -- (4,2,2) -- (4,4,2) -- (2,4,2) -- cycle;
\draw[blue,fill=green!50,opacity=0.1] (2,2,2) -- (4,2,2) -- (4,2,4) -- (2,2,4) -- cycle;

\draw[green!80!black] (3, 2, 2) -- (3, 4, 2);
\draw[green!80!black] (2, 3, 2) -- (4, 3, 2);
\draw[green!80!black] (4, 3, 2) -- (4, 3, 4);
\draw[green!80!black] (4, 2, 3) -- (4, 4, 3);
\draw[green!80!black] (3, 2, 2) -- (3, 2, 4);
\draw[green!80!black] (2, 2, 3) -- (4, 2, 3);
\draw[green!80!black] (2, 3, 2) -- (2, 3, 4);
\draw[green!80!black] (2, 2, 3) -- (2, 4, 3);
\draw[green!80!black] (3, 4, 2) -- (3, 4, 4);
\draw[green!80!black] (2, 4, 3) -- (4, 4, 3);
\draw[green!80!black] (3, 2, 4) -- (3, 4, 4);
\draw[green!80!black] (2, 3, 4) -- (4, 3, 4);
\draw[green!80!black] (2, 3, 3) -- (4, 3, 3);
\draw[green!80!black] (3, 3, 2) -- (3, 3, 4);
\draw[green!80!black] (3, 2, 3) -- (3, 4, 3);

\draw[fill=green!50,fill opacity=0.2,draw opacity=0] (2,3,2) -- (4,3,2) -- (4,3,4) -- (2,3,4) -- cycle;
\draw[fill=green!50,fill opacity=0.2,draw opacity=0] (2,2,3) -- (4,2,3) -- (4,4,3) -- (2,4,3) -- cycle;
\draw[fill=green!50,fill opacity=0.2,draw opacity=0] (3,2,2) -- (3,4,2) -- (3,4,4) -- (3,2,4) -- cycle;

\draw[decorate,decoration={brace, amplitude=8pt, raise=20pt}] (0,0,4) to node[black,midway,left=0.83] {$\ell_1$} (0, 4, 4);
\draw[decorate,decoration={brace, amplitude=8pt, raise=2pt}, blue] (0,2,4) to node[black,midway,left=0.22,blue] {$\ell_2$} (0, 4, 4);
\draw[decorate,decoration={brace, amplitude=4pt, raise=2pt, mirror},green!80!black] (4,3,2) to node[black,midway,right=0.15,green!80!black] {$\ell_3$} (4, 4, 2);

\end{tikzpicture}
\caption{An octree with three levels of spatial division, and its corresponding encoding as
a rooted tree. The root is black, with side length $\ell_1$; this side length
is the extents of the entire simulation domain. The next division is blue, with side length $\ell_2$, 
and the next green, with side length $\ell_3$.}\label{fig:octree-tree}
\end{figure}

Let us begin with a short review of the implementation of an octree.
A simple recursive implementation based on pointers is a direct generalization
of a rooted binary tree based on linked lists; 
see, e.g., \cite[Ch. 11]{DBLP:books/daglib/0023376}.
Each cell in an octree is represented by its data or \textit{payload} and 8 pointers
to its children. Since the octree is deeply connected with the spatial domain, 
each node is also affixed with its \textit{spatial centre} and its
\textit{size}---the side length of the cell. 
The payload of each node, for the purpose of $N$-body simulation, 
also includes the position and mass of the body
it contains, if a leaf node, or the center of mass and multipole coefficients,
if an internal node; see Section~\ref{sec:multipole}.
Fig.~\ref{fig:octree-tree} shows the encoding
of an octree with three spatial divisions as a rooted tree of 
nodes. 

Algorithm~\ref{alg:BH1} shows the Barnes-Hut method with simple parallel tree traversal. 
It begins by assigning $\nicefrac{N}{s}$ bodies to each process. Then, the root process (id 0),
gathers all bodies (since their positions were updated in the previous simulation step).
The root process simply builds the global tree from all bodies at once. This global tree
is then broadcast to all other processes (Lines 5--7). 
With each process having the global octree, they can proceed in parallel to 
compute the forces on their local bodies with respect to the entire octree (Lines 8-20).
Integration steps then proceed in parallel as before (Lines 21-22).

\begin{figure}[htb]
\begin{center}
\begin{minipage}{0.85\textwidth}
	\begin{algorithm}[H]
		\centering
		\footnotesize
		\caption{\scriptsize\textsc{Parallel Barnes-Hut V1}($id$, $s$, $\Delta t$, $t_{end}$, $N$, $P_{all}$)\newline $P_{all}$ is the list of size $N$ of bodies to simulate from 0 to $t_{end}$ going by $\Delta t$ time-steps. $id$ is the executing process's id, $s$ is the number of processes. Assume $s \mid N$.}\label{alg:BH1}
		\algnotext{EndIf}
		\algnotext{EndFor}
		\algnotext{EndWhile}
		\begin{algorithmic}[1]
			\State $N' \gets N / s$
			\State $P \gets$ \textbf{Scatter} $P_{all}$, sending to process $id$ $P_{all}[id\cdot N', \ldots, id\cdot N' + N' - 1]$ \Comment{Get Local Bodies}
			\For {$t$ \textbf{from} 0 \textbf{to} $t_{end}$ \textbf{by} $\Delta t$}
			\State $P_{all} \gets$ \textbf{Gather} $P$ to process 0.
			\If {$id = 0$}
				\State $T \gets$ build octree from $P_{all}$
			\EndIf
			\State \textbf{Broadcast} $T$ from process 0 to all others.
			\For {\textbf{each} particle $p$ in $P$} \Comment{Compute Forces}
			\State Set the acceleration of $p$ to $\vec{0}$.
			\State $Node$ $\gets$ the root node of $T$ \vspace{0.3em}
			\EndFor
			\Statex $\quad\ $Compute Cell Force:
			\Indent
			\For {\textbf{each} child cell $C$ of $Node$}
			\If {$C$ contains 1 particle} \Comment{$C$ is a leaf}
			\State Add to $p$'s acceleration the contribution from $C$'s particle. \Comment{particle-particle}
			\Else
			\State $\ell$ $\gets$ $C$'s side length
			\State $d$ $\gets$ distance between $C$'s c.o.m. and $p$'s position.
			\If {${\ell} / {d} < \theta$}
			\State Add to $p$'s acceleration the contribution from $C$. \Comment{particle-cell}
			\Else 
				\State Compute Cell Force with $Node \gets C$ \Comment{recursive call}
			\EndIf
			\EndIf
			\EndFor
			\EndIndent
			\For {\textbf{each} particle $p$ in $P$} \Comment{Position Update}
			\State Update position and velocity of $p$ using its current acceleration.
			\EndFor
			\EndFor
		\end{algorithmic}
	\end{algorithm}
\end{minipage}
\end{center}
\end{figure}
\clearpage

Notice that for this algorithm to fit within the message passing model for 
distributed algorithms, we must have a way to \textit{serialize}
the octree into a message. Since its encoding is based on pointers and linked lists, 
this is not immediately obvious. 
One solution, which we implement, 
is to add to each node an 8-bit mask whose bits encode 
whether any of the 8 children actually exist. 
Then, serializing the octree is as simple as serializing each node 
and concatenating them in the order of pre-order tree traversal.
Deseralizing is also made simple by following a pre-order traversal 
in combination with the deserialized bitmask.

In this simple parallel Barnes-Hut algorithm, the span 
is not very good since the root process must construct the entire octree
in time $\mathcal{O}(N\log N)$. The force calculation is still done
in parallel, but now taking time $\mathcal{O}(\nicefrac{N}{s}\log N)$ thanks to the 
octree and particle-cell interactions. Updating the particle positions
and velocities remains at $\nicefrac{N}{s}$ integrations.

\subsection{Parallel Octree Construction and Merging}
\label{sec:parallel2}

Parallelizing the octree construction is a non-trivial task. 
Naive attempts, such as all processes simultaneously inserting their 
assigned bodies into a single global octree, requires an excessive amount of 
synchronization. This technique was explored, and abandoned, in \cite{singh1993parallel}.
Their improved variation involves each process creating
its own local octree, from its local bodies, and then merging
trees together.
One fundamental aspect of this merge operation is that each local tree
must cover the same spatial domain. That is, their root nodes
must have the same size. Then, all cells in all of the local trees
have the same sizes. This allows entire subtrees to be inserted into
a merged tree where there is an otherwise empty child node.
Algorithm~\ref{alg:octreemerge} presents this merge algorithm;
Lines 5--6 and Lines 11-12 show this subtree replacement.

\begin{center}
\vspace{-0.5em}
\begin{minipage}{0.75\textwidth}
\begin{algorithm}[H]
	\centering
	\footnotesize
	\caption{\scriptsize\textsc{Octree Merge}($T_1$, $T_2$)\newline Merges the two octrees rooted at nodes $T_1$ and $T_2$ by moving data from $T_2$ into $T_1$. This assumes both $T_1$ and $T_2$ have at least one level of children.}\label{alg:octreemerge}
	\algnotext{EndIf}
	\algnotext{EndFor}
	\algnotext{EndWhile}
	\begin{algorithmic}[1]
		\For {$i$ \textbf{from} 0 \textbf{to} 7}
			\If {$T_1$ and $T_2$ both have an $i$th child}
				\State $(C_1,C_2) \gets$  (the $i$th child of $T_1$, the $i$th child of $T_2$)
				\If {$C_1$ is a leaf node}
					\State Insert the particle of $C_1$ into $C_2$ \Comment{see Build Tree of Algorithm~\ref{alg:BHloop}}
					\State Replace $C_1$ in $T_1$ with $C_2$
				\ElsIf {$C_2$ is a leaf node}
					\State Insert the particle of $C_2$ into $C_1$ 
				\Else 
					\State \textsc{Octree Merge}($C_1$, $C_2$) \Comment{Recursive Call}
				\EndIf
			\ElsIf {$T_2$ has an $i$th child} \Comment{$T_1$ does not have an $i$th child}
				\State Set the $i$th child of $T_1$ to $C_2$
			\EndIf
		\EndFor
	\end{algorithmic}
\end{algorithm}
\end{minipage}
\end{center}
\vspace{0.5em}

In the original octree merge of \cite{singh1993parallel}, 
each process's local tree was merged one at a time into the global tree. 
This was feasible in a shared-memory model.
In a message passing model, or in any more generic parallel scheme, 
this would require excessive communication and synchronization.
Instead, we perform merges of local trees pairwise, in a \textit{reduction} (see \cite[Ch. 5]{mccool2012structured}), 
until a single tree contains all $N$ bodies. This single tree is then broadcast to all other processes.
Note that while this is phrased in terms of a Map-Reduce pattern, the reduction in fact
does not ``reduce'' the size of data, but moreso performs a gather of the individual local trees.
Algorithm~\ref{alg:BH2} shows this scheme, where the parallel tree building occurs in Lines 4--6, 
and the remaining lines are the same as Algorithm~\ref{alg:BH1}.
Notice that the reduce step hides many serializations, inter-process messages, and deserializations.

\begin{center}
\vspace{-0.5em}
\begin{minipage}{0.85\textwidth}
\begin{algorithm}[H]
	\centering
	\footnotesize
	\caption{\scriptsize\textsc{Parallel Barnes-Hut V2}($id$, $s$, $\Delta t$, $t_{end}$, $N$, $P_{all}$)\newline $P_{all}$ is the list of size $N$ of bodies to simulate from 0 to $t_{end}$ going by $\Delta t$ time-steps. $id$ is the executing process's id, $s$ is the number of processes. Assume $s \mid N$.}\label{alg:BH2}
	\algnotext{EndIf}
	\algnotext{EndFor}
	\algnotext{EndWhile}
	\begin{algorithmic}[1]
		\State $N' \gets N / s$
		\State $P \gets$ \textbf{Scatter} $P_{all}$, sending to process $id$ $P_{all}[id\cdot N', \ldots, id\cdot N' + N' - 1]$ \Comment{Get Local Bodies}		\For {$t$ \textbf{from} 0 \textbf{to} $t_{end}$ \textbf{by} $\Delta t$}
		\State $T' \gets$ build octree from $P$ \Comment{Parallel Tree Build}
		\State $T \gets $ \textbf{Reduce}, via \textsc{Octree Merge}, $T'$ to process 0.
		\State \textbf{Broadcast} $T$ from process 0 to all others.
		\For {\textbf{each} particle $p$ in $P$} \Comment{Compute Forces; unchanged from Algorithm~\ref{alg:BH1}}
		\State ... 
		\EndFor
		\For {\textbf{each} particle $p$ in $P$} \Comment{Position Update}
		\State Update position and velocity of $p$ using its current acceleration.
		\EndFor
		\EndFor
	\end{algorithmic}
\end{algorithm}
\end{minipage}
\end{center}
\vspace{0.5em}

While the above algorithm parallelizes the tree building step,
and building each local octree now only incurs time $\mathcal{O}(\nicefrac{N}{s}\log(\nicefrac{N}{s}))$,
the span of the reduce still requires $\log(s)$ merges to occur. In the worst case, 
where each process's local bodies are uniformly distributed across
the entire spatial domain, merging octrees is a laborious process
where both trees must be traversed all the way down to their leaves
(see the recursive call in Algorithm~\ref{alg:octreemerge}), 
and then leaf nodes inserted into the other subtree
(see Build Tree step of Algorithm~\ref{alg:BHloop}).
A single merge of two trees with $n$ bodies each thus requires time
$\mathcal{O}(n\log n)$, resulting in a 
span of the parallel octree build of $\mathcal{O}(\log s (N\log N))$.
This does not exhibit \textit{strong scaling}---where the span 
would reduce with an increased number of processes---since the 
dominant term is still $N\log N$.
A better parallel solution would result in a span 
with a dominant term of $\mathcal{O}(\nicefrac{N}{s}\log(\nicefrac{N}{s}))$

\subsection{Spatial Decomposition via Morton's Ordering}
\label{sec:parallel3}

A key consideration in both the
costzones and hashed octree methods is the spatial decomposition of bodies 
rather than static assignment of bodies to processes. 
Both methods are based on the idea that each process
is responsible for a spatial subdomain of the simulation, and the bodies 
contained therein, rather than a set of bodies
spread across the entire domain.
In the costzones method this allows for dynamic load-balancing into
the namesake \textit{costzones} (see Section~\ref{sec:parallel4})
and improves the octree merge, as we will soon see.
In the hashed octree method, the spatial decomposition minimizes 
inter-process communication (see Section~\ref{sec:parallel6}).

However, one key challenge for $N$-body simulation 
is its dynamicity and irregularity. It would be wholly
insufficient to simply divide the spatial domain
into regular pieces, and assign one piece to each process.
As bodies move throughout the spatial domain during the simulation, 
some areas would have more bodies, some would have less, and 
some may not contain any bodies at all.
This load-balancing problem is further exacerbated by the formation of
highly clustered areas as the gravitational system evolves \cite{warren1993parallel}.

A first attempt then at spatial decomposition involves partitioning
the spatial domain so that each process's domain contains
the same number of bodies. This will naturally create different
shapes and sizes for each spatial partition. 
First, a linear ordering is given to the leaf nodes of the octree
or, equivalently, the bodies themselves.
This is done through the use of a space-filling curve---Peano-Hilbert
curve in the case of costzones, and Morton ordering in the case of hashed octree.
This linear ordering on the bodies is then partitioned evenly and each partition assigned
to a process. 

The more simple Morton ordering, or Z-ordering, involves
transforming a multi-dimensional coordinate tuple into a single integer, 
which can then be sorted in a typical integer order.
This single integer \textit{spatial key} is created by interleaving
the bits of each coordinate's value.
For example, a 2-dimension ($x,y$) coordinate $(3,5) = (0b\textcolor{blue}{0011},0b\textcolor{green!60!black}{0101})$
can be interleaved into the 8-bit \textit{spatial key} $0b(\textcolor{blue}{0}\textcolor{green!60!black}{0})(\textcolor{blue}{0}\textcolor{green!60!black}{1})(\textcolor{blue}{1}\textcolor{green!60!black}{0})(\textcolor{blue}{1}\textcolor{green!60!black}{1}) = 27$.
Notice this interleave orders the bits such that those from the x-coordinate 
are more significant, thus we say this ordering has $x > y$. A different orientation would be
produced with $y > x$, but it would still create a space-filling curve.

With such a bit interleave, in $d$ dimensions, the spatial key
requires $d\cdot b$ bits to encode spatial coordinates
which range over the integers $[0, 2^b)$. 
Moreover, spatial coordinates are often encoded using floating point
values, not integers. To remedy this, we perform several mappings
to obtain integer spatial coordinates in the range $[0, 2^{21}-1]$;
$2^{21}$ is chosen so that the three-dimensional spatial key of $3\cdot21=63$ bits can be
stored in a single 64-bit machine word.
First, we must know the extents of the spatial domain. Let us
assume the position of each body falls in the range $(-R,R)$, for some integer $R$, 
in each dimension. 
We first translate each coordinate of each body to the range $(0,2R)$.
Then, we map the range $(0,2R)$ to $(0,2^{21})$. Notice
this mapping causes a loss of precision, with at most $2^{21}$
discrete values in each dimension. This is sufficient for simulations
with up to several million bodies. In larger simulations,
a 128-bit key is more than sufficient. 

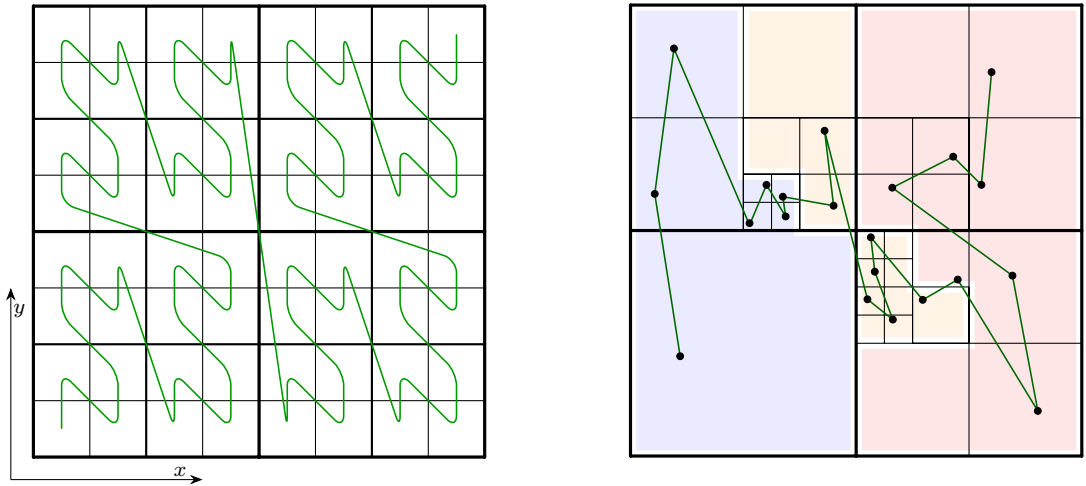
\begin{figure}[h]
	\centering
\begin{subfigure}{0.45\textwidth}
	\centering
\begin{tikzpicture}[scale=0.75]
\draw [step=1, thin] (0,0) grid (8,8);
\draw [step=2,thick] (0,0) grid (8,8);
\draw [step=4,very thick, line cap=round] (0,0) grid (8,8);

\draw[green!60!black, semithick, rounded corners=5] (0.5,0.5) -- (0.5,1.5) -- (1.5,0.5) -- (1.5,1.5) -- (0.5,2.5) -- (0.5,3.5) -- (1.5,2.5) -- (1.5,3.5) -- (2.5,0.5) -- (2.5,1.5) -- (3.5,0.5) -- (3.5,1.5) -- (2.5,2.5) -- (2.5,3.5) -- (3.5,2.5) -- (3.5,3.5)
-- (0.5,4.5) -- (0.5,5.5) -- (1.5,4.5) -- (1.5,5.5) -- (0.5,6.5) -- (0.5,7.5) -- (1.5,6.5) -- (1.5,7.5) -- (2.5,4.5) -- (2.5,5.5) -- (3.5,4.5) -- (3.5,5.5) -- (2.5,6.5) -- (2.5,7.5) -- (3.5,6.5) -- (3.5,7.5) --
(4.5,0.5) -- (4.5,1.5) -- (5.5,0.5) -- (5.5,1.5) -- (4.5,2.5) -- (4.5,3.5) -- (5.5,2.5) -- (5.5,3.5) -- (6.5,0.5) -- (6.5,1.5) -- (7.5,0.5) -- (7.5,1.5) -- (6.5,2.5) -- (6.5,3.5) -- (7.5,2.5) -- (7.5,3.5) --
(4.5,4.5) -- (4.5,5.5) -- (5.5,4.5) -- (5.5,5.5) -- (4.5,6.5) -- (4.5,7.5) -- (5.5,6.5) -- (5.5,7.5) -- (6.5,4.5) -- (6.5,5.5) -- (7.5,4.5) -- (7.5,5.5) -- (6.5,6.5) -- (6.5,7.5) -- (7.5,6.5) -- (7.5,7.5);

\draw[->, >=Stealth] (-0.4, -0.4) -- (-0.4, 3);
\node[font={\scriptsize}] at (-0.25, 2.6) {$y$};
\draw[->, >=Stealth] (-0.4, -0.4) -- (3, -0.4);
\node[font={\scriptsize}] at (2.6, -0.25) {$x$};
%
%
%
%
\end{tikzpicture}
\caption{The Morton ordering of a quadtree with three levels of spatial division. This ordering treats the $x$ dimension as more significant than the $y$ dimension in the bit interleave. $y > x$ would instead create a mirrored ``Z'' shape.}\label{fig:morton}
\end{subfigure}
\hspace{0.5em}
\begin{subfigure}{0.45\textwidth}
	\centering
	\begin{tikzpicture}[scale=0.75]

	\draw[fill=blue!80, fill opacity=0.1, draw opacity=0] (0.1,0.1) -- (0.1, 7.9) -- (1.9, 7.9) -- (1.9, 4.9) -- (2.9, 4.9) -- (2.9,3.9) -- (3.9,3.9) -- (3.9,0.1) -- cycle;
	\draw[fill=orange!90!yellow, fill opacity=0.1, draw opacity=0] (3.1,5.1) -- (2.1,5.1) -- (2.1, 7.9) -- (3.1,7.9) -- (3.9,7.9) -- (3.9,4.1) -- (3.1,4.1) -- (3.1, 5.1) -- cycle;
	\draw[fill=orange!90!yellow, fill opacity=0.1, draw opacity=0] (4.1, 3.9) -- (4.9, 3.9) -- (4.9, 2.9) -- (5.9, 2.9) -- (5.9, 2.1) -- (4.1,2.1) -- cycle;
	
	\draw[fill=red, fill opacity=0.1, draw opacity=0] (4.1,1.9) -- (4.1, 0.1) -- (7.9,0.1) -- (7.9, 7.9) -- (4.1, 7.9) -- (4.1, 4.1) -- (5.1, 4.1) -- (5.1, 3.1) -- (6.1, 3.1) -- (6.1, 1.9) -- cycle;

	\draw
	[step=2cm, thin] (4,4) grid +(4,4)
	[step=2cm, thin] (0,4) grid +(4,4)
	[step=1cm, thin] (2,4) grid +(2,2)
	[step=1cm, thin] (4,4) grid +(2,2)
	[step=2cm, thin] (4,0) grid + (4,4)
	[step=0.5cm, thin] (4, 2) grid +(1,1)
	[step=0.5cm, thin] (4, 3) grid +(1,1)
	[step=1cm, thin] (5, 2) grid +(1,1)
	[step=0.5cm, thin] (2, 4) grid +(1,1);
	\draw [step=4cm, very thick, line cap=round] (0,0) grid (8,8);
	
		\draw[green!40!black,semithick, rounded corners=2](0.88, 1.77) -- (0.43, 4.65) -- (0.77, 7.23) -- (2.11, 4.1) -- (2.41, 4.81)-- (2.75, 4.25)-- (2.70, 4.60) -- (3.60, 4.44) -- (3.44, 5.77) 
	-- (4.20, 2.78) -- (4.65, 2.42) -- (4.33, 3.27) -- (4.26, 3.88)-- (5.18, 2.77) -- (5.80, 3.13) 	-- (7.22, 0.8) -- (6.77, 3.2) -- (4.64, 4.76) -- (5.72, 5.31)  -- (6.22, 4.81) -- (6.4, 6.81);

	\draw[fill] (0.88, 1.77) circle (0.06cm);
\draw[fill] (0.43, 4.65) circle (0.06cm);
\draw[fill] (0.77, 7.23) circle (0.06cm);
\draw[fill] (2.11, 4.13) circle (0.06cm);
\draw[fill] (2.41, 4.81) circle (0.06cm);
\draw[fill] (2.75, 4.25) circle (0.06cm);
\draw[fill] (2.70, 4.60) circle (0.06cm);

\draw[fill] (3.60, 4.44) circle (0.06cm);
\draw[fill] (3.44, 5.77) circle (0.06cm);

\draw[fill] (4.20, 2.78) circle (0.06cm);
\draw[fill] (4.65, 2.42) circle (0.06cm);
\draw[fill] (4.33, 3.27) circle (0.06cm);
\draw[fill] (4.26, 3.88) circle (0.06cm);
\draw[fill] (5.18, 2.77) circle (0.06cm);
\draw[fill] (5.80, 3.13) circle (0.06cm);

\draw[fill] (7.22, 0.8) circle (0.06cm);
\draw[fill] (6.77, 3.2) circle (0.06cm);

\draw[fill] (4.64, 4.76) circle (0.06cm);
\draw[fill] (5.72, 5.31) circle (0.06cm);
\draw[fill] (6.22, 4.81) circle (0.06cm);
\draw[fill] (6.4, 6.81) circle (0.06cm);

\draw[->, >=Stealth,opacity=0] (-0.4, -0.4) -- (-0.4, 3);
\node[font={\scriptsize},opacity=0] at (-0.25, 2.6) {$x$};
\draw[->, >=Stealth,opacity=0] (-0.4, -0.4) -- (3, -0.4);
\node[font={\scriptsize},opacity=0] at (2.6, -0.25) {$y$};	
	\end{tikzpicture}
	\caption{The Morton ordering of 21 points in two dimensions. The spatial domain is partitioned into three pieces according to the cell colors. Notice that the second domain contains a spatial discontinuity.}\label{fig:spatialdecomp}
\end{subfigure}
\caption{The Morton ordering on the cells of a quad tree and on a collection of bodies in 2 dimensions.}
\end{figure}
\clearpage

Fig.~\ref{fig:morton} presents the Morton ordering, for $x > y$, for 
a quadtree with three levels of spatial division. Fig.~\ref{fig:spatialdecomp}
then shows this Morton ordering applied to a collection of bodies in two-dimensional space;
the line connects the bodies in order. In this same figure, a spatial decomposition 
into 3 partitions is shown where the list of bodies is separated into 3 sections, each
containing 7 bodies. Notice that
a partition may span a spatial discontinuity. This is a result of the
Morton ordering; a solution would be to use the more complicated Peano-Hilbert ordering.

Algorithm~\ref{alg:BH3} presents the parallel Barnes-Hut algorithm which makes use of 
spatial decomposition. At each simulation step, all bodies
must be \textit{gathered} to the root process so that
their spatial keys can be computed, the bodies sorted, 
and then the bodies \textit{scattered} again to each process (Lines 4--8).
The sort followed by the scatter implies the spatial decomposition, where
the scatter evenly distributes the sorted bodies to each process.
The algorithm proceeds then as Algorithm~\ref{alg:BH2}, constructing
and sharing a global octree by a reduce and broadcast.

\begin{center}
	\vspace{-1.0em}
	\begin{minipage}{0.85\textwidth}
		\begin{algorithm}[H]
			\centering
			\footnotesize
			\caption{\scriptsize\textsc{Parallel Barnes-Hut V3}($id$, $s$, $\Delta t$, $t_{end}$, $N$, $P_{all}$)\newline $P_{all}$ is the list of size $N$ of bodies to simulate from 0 to $t_{end}$ going by $\Delta t$ time-steps. $id$ is the executing process's id, $s$ is the number of processes. Assume $s \mid N$.}\label{alg:BH3}
			\algnotext{EndIf}
			\algnotext{EndFor}
			\algnotext{EndWhile}
			\begin{algorithmic}[1]
				\State $N' \gets N / s$
				\State $P \gets$ \textbf{Scatter} $P_{all}$, sending to process $id$ $P_{all}[id\cdot N', \ldots, id\cdot N' + N' - 1]$ \Comment{Get Local Bodies}
				\For {$t$ \textbf{from} 0 \textbf{to} $t_{end}$ \textbf{by} $\Delta t$}
				\State $P_{all} \gets$ \textbf{Gather} $P$ to process 0. \Comment{Spatial Decomposition}
				\If {id = 0} 
					\State $keys \gets $ \textsc{ConstructSpatialKeys}($P_{all}$).
					\State Sort $P_{all}$ by $keys$.
				\EndIf
				\State $P \gets$ \textbf{Scatter} $P_{all}$, sending to process $id$ $P_{all}[id\cdot N', \ldots, id\cdot N' + N' - 1]$
				\State $T' \gets$ build octree from $P$ \Comment{Parallel Tree Build}
				\State $T \gets $ \textbf{Reduce}, via \textsc{Octree Merge}, $T'$ to process 0.
				\State \textbf{Broadcast} $T$ from process 0 to all others.
				\For {\textbf{each} particle $p$ in $P$} \Comment{Compute Forces; unchanged from Algorithm~\ref{alg:BH2}}
				\State ... 
				\EndFor
				\For {\textbf{each} particle $p$ in $P$} \Comment{Position Update}
				\State Update position and velocity of $p$ using its current acceleration.
				\EndFor
				\EndFor
			\end{algorithmic}
		\end{algorithm}
	\end{minipage}
\end{center}
\vspace{0.25em}

One crucial side effect of the spatial decomposition is its effect 
on the octree merge. Recall that, in the worst case, each
merge requires $\mathcal{O}(n\log n)$ steps to merge two
trees which each have $n$ uniformly distributed bodies.
As noted in \cite[Section 5.3]{singh1993parallel}, since
each process now contains a distinct spatial partition, 
the leaf nodes in each local octree do not overlap. 
Each local octree thus only shares a few high-level nodes
with other local octrees. A merge
between two octrees then occurs in time $\mathcal{O}(\log n)$, 
since only one traversal is needed to find where to place
entire subtrees.
The entire reduce then has a span of $\mathcal{O}(\log N)$, meanwhile 
the span of the local tree construction remains as $\mathcal{O}(\nicefrac{N}{s}\log(\nicefrac{N}{s}))$
The dominant term is the latter, and thus the span
of the entire parallel tree construction becomes $\mathcal{O}(\nicefrac{N}{s}\log(\nicefrac{N}{s}))$.

Notice further that, for a sufficiently small time-step $\Delta t$,
the bodies will only move very slightly between time-steps.
Thus, a sorted list of bodies will, after the integration step,
still be mostly sorted. 
We thus implement an insertion sort to sort the particles and keys.
Sorting therefore only requires a few swaps and time $\mathcal{O}(N)$.
Yet, it still requires high amounts of communication, and sub-optimal scalability,
where the particles are gathered to the root process for sorting.
We tackle this problem in Section~\ref{sec:parallel5}. 
However, we first explore a more crucial part of the algorithm: load-balancing
the force calculations.

\subsection{Dynamic Load-Balancing}
\label{sec:parallel4}

The dominant operation in an $N$-body simulation is the force calculation. 
In the direct method it takes time $\mathcal{O}(N^2)$ and in a
hierarchical method $\mathcal{O}(N \log N)$. However, 
not every interaction is created equal when computing forces acting on a particle.
A particle in a dense region will have more particle-particle interactions, 
thus approaching an $N$ number of interactions. 
A very distant particle in a sparse region may have only particle-cell interactions,
thus requiring only $\mathcal{O}(1)$
interactions, depending on the value of the MAC; see Section~\ref{sec:bgoctree}.
Moreover, the big-$\mathcal{O}$ notation can be misleading.
The force calculation step is by far the most dominant part of the $N$-body
simulation---despite it and tree building both taking $\mathcal{O}(N\log N)$
time in serial---taking between 70\% and 90\% of the time of each simulation time-step \cite{warren1993parallel,singh1993parallel}.
Therefore, our spatial decomposition scheme, which assigns 
an equal $\nicefrac{N}{s}$ number of bodies to each of the
$s$ processes, is likely insufficient. 

The resolution is that of the aptly named costzones, 
to determine zones---spatial partitions---of equal cost.
The exact same operation is performed in the hashed octree method, 
but is not named nor emphasized.
The method of determining costzones is simple. 
Begin with a work estimate $w_i$ for each body as the number of 
interactions computed for that body in the previous time-step
(in the very first time-step, assume $w_i=N$).
Then, each partition (costzone) should be responsible for
$w = \left(\sum_{i=0}^N w_i\right) / s$ amount of work.
Determining costzones proceeds by iterating over
the spatially-linear list of bodies 
(the list of bodies after sorting by the Morton ordering)
and partitioning it into sets whose work estimate is at least $w$.
Algorithm~\ref{alg:costzones} shows this iteration.

\begin{center}
\vspace{-0.5em}
\begin{minipage}{0.85\textwidth}
	\begin{algorithm}[H]
		\centering
		\footnotesize
		\caption{\scriptsize\textsc{Costzones}($W$, $s$)\newline $W$ is a list of size $N$ of work estimates
			such that $W[i]$ is the work estimate of the $i$th body in the Morton (or other) ordering; $s$ is the number of processes.\newline Returns a list $Z$ of size $s+1$, returning the index of the first body in process $k$'s domain partition as $Z[k]$.}\label{alg:costzones}
		\algnotext{EndIf}
		\algnotext{EndFor}
		\algnotext{EndWhile}
		\begin{algorithmic}[1]
			\State $w_{tot} \gets 0$
			\For {$i$ \textbf{from} 0 \textbf{to} $N-1$}
				\State $w_{tot} \gets w_{tot} + W[i]$
			\EndFor
			\State $w_{targ} \gets w_{tot} / s$
			\State $k \gets 0$; $w \gets 0$
			\State $j \gets 0$; $Z \gets [\ ]$
			\For {$i$ \textbf{from} 0 \textbf{to} $N-1$}
				\State $w \gets w + W[i]$
				\If {$w \geq w_{targ}$}
					\State $Z[k] \gets j$; $j \gets i+1$
					\State $k \gets k + 1$; $w \gets 0$
				\EndIf
			\EndFor
			\State $Z[s-1] \gets j$ \Comment{Commit final partition}
			\State $Z[s] \gets N$ \Comment{Used to compute end of partition for $id$ $s-1$}
			\State \Return $Z$
		\end{algorithmic}
	\end{algorithm}
\end{minipage}
\end{center}
\vspace{0.5em}

Algorithm~\ref{alg:BH4} adds
the \textsc{Costzones} load balancing
to Algorithm~\ref{alg:BH3}.
It begins by gathering the work estimates of each particle 
to the root process (Line 6).
This list of work estimates $W_{all}$ is then sorted
along with $P_{all}$ as parallel arrays (Lines 8--9).
Costzones are computed using $W_{all}$ and then 
$P_{all}$ is scattered to all processes
based on the costzones partitioning (Lines 10--11).
Lastly, the tree traversal is modified 
to accumulate the number of interactions of
each particle in the array $W$ (Lines 16, 22, 28).
A subtle optimization for the work estimate is to 
weight the work of a particle-cell interaction
twice as much as a particle-particle interaction.
This is because of the multipole approximation incurred by a 
particle-cell interaction; computing acceleration 
from a particle-cell interaction involves twice as many floating point
operations as a simple particle-particle interaction (see Section~\ref{sec:multipole}).

\begin{center}
\vspace{-0.5em}
\begin{minipage}{0.85\textwidth}
\begin{algorithm}[H]
	\centering
	\footnotesize
	\caption{\scriptsize\textsc{Parallel Barnes-Hut V4}($id$, $s$, $\Delta t$, $t_{end}$, $N$, $P_{all}$)\newline $P_{all}$ is the list of size $N$ of bodies to simulate from 0 to $t_{end}$ going by $\Delta t$ time-steps. $id$ is the executing process's id, $s$ is the number of processes. Assume $s \mid N$.}\label{alg:BH4}
	\algnotext{EndIf}
	\algnotext{EndFor}
	\algnotext{EndWhile}
	\begin{algorithmic}[1]
		\State $N' \gets N / s$
		\State $P \gets$ \textbf{Scatter} $P_{all}$, sending to process $id$ $P_{all}[id\cdot N', \ldots, id\cdot N' + N' - 1]$ \Comment{Get Local Bodies}
		\State $W \gets [N,\dots,N]$ \Comment{$W$ of size $N'$}
		\For {$t$ \textbf{from} 0 \textbf{to} $t_{end}$ \textbf{by} $\Delta t$}
		\State $P_{all} \gets$ \textbf{Gather} $P$ to process 0. \Comment{Spatial Decomposition}
		\State $W_{all} \gets$ \textbf{Gather} $W$ to process 0.
		\If {id = 0} 
		\State $keys \gets $ \textsc{ConstructSpatialKeys}($P_{all}$).
		\State Sort $P_{all}$ and $W_{all}$ by $keys$.
		\State $Z \gets$ \textsc{Costzones}($W_{all}$, $s$)
		\EndIf
		\State $P \gets$ \textbf{Scatter} $P_{all}$ sending to process $id$ $P_{all}[Z[id], \ldots, Z[id+1]-1]$
		\State $T' \gets$ build octree from $P$ \Comment{Parallel Tree Build}
		\State $T \gets $ \textbf{Reduce}, via \textsc{Octree Merge}, $T'$ to process 0.
		\State \textbf{Broadcast} $T$ from process 0 to all others.
		\For {$i$ \textbf{from} 0 \textbf{to} $|P|-1$} \Comment{Compute Forces}
		\State $p \gets P[i]$; $W[i] \gets 0$
		\State Set the acceleration of $p$ to $\vec{0}$.
		\State $Node$ $\gets$ the root node of $T$ \vspace{0.3em}
		\EndFor
		\Statex $\quad\ $Compute Cell Force:
		\Indent
		\For {\textbf{each} child cell $C$ of $Node$}
		\If {$C$ contains 1 particle} \Comment{$C$ is a leaf}
		\State Add to $p$'s acceleration the contribution from $C$'s particle. \Comment{particle-particle}
		\State $W[i] \gets W[i] + 1$
		\Else
		\State $\ell$ $\gets$ $C$'s side length
		\State $d$ $\gets$ distance between $C$'s c.o.m. and $p$'s position.
		\If {${\ell} / {d} < \theta$}
		\State Add to $p$'s acceleration the contribution from $C$. \Comment{particle-cell}
		\State $W[i] \gets W[i] + 2$
		\Else 
		\State Compute Cell Force with $Node \gets C$ \Comment{recursive call}
		\EndIf
		\EndIf
		\EndFor
\EndIndent
		\For {\textbf{each} particle $p$ in $P$} \Comment{Position Update}
		\State Update position and velocity of $p$ using its current acceleration.
		\EndFor
		\EndFor
	\end{algorithmic}
\end{algorithm}
\end{minipage}
\end{center}
\vspace{0.5em}

The load balancing of costzones more evenly distributes the amount of work
performed by each process during the force calculation step. 
This comes at the expense of an uneven number of bodies
assigned to each process for the octree construction 
and integration steps. Experiments suggest the number of 
bodies assigned to any one process is roughly twice as many as another.
In long-term evolution, clustering may lead to the maximum number of bodies 
on one process being 10 times as many as the one with the fewest \cite[Section 4]{warren1993parallel}.
Nonetheless, since the force calculation is such a dominant part of the algorithm,
load-balancing the force calculation results in better overall performance
despite the loss of load-balancing in the other steps.

\subsection{Parallel and Distributed Sorting}
\label{sec:parallel5}

Our next step towards improved scalability is to avoid
the gather of bodies and work estimates to the root process.
This gather occurs so that the spatial decomposition via sorting,
and load-balancing via costzones, can occur.
This gather also incurs high amounts of communication cost, 
particularly as the number of processes increases.
We propose here a solution which provides a distributed algorithm for both 
the sorting and load-balancing steps so as to avoid excess communication.
We note that methods detailed in this section did not appear
in the original costzones method \cite{singh1993parallel}, but are
a result of applying this method to a distributed system.

We begin with the sorting algorithm. One could implement 
a generalized and highly scalable distributed sort, e.g. 
methods based on sample sort; see \cite{DBLP:conf/ipps/SolomonikK10,warren20142hot}.
However, we have already made the observation that
bodies remain mostly sorted between time-steps. 
Hence we implement a modification of bubble sort to a distributed system.
Simply stated, elements out of place 
bubble up and are exchanged with the process's right neighbour, or sink down and
are exchanged with the process's left neighbour.

Assume a linear topology of the processes. 
Neighbour processes exchange fixed amounts of data, 
both neighbours sort this exchanged data, 
and then each keeps their respective halves. 
For example, process with $id$ 0 would send its
$k$ largest elements to $id$ 1, and $id$ 1 would send
its $k$ smallest elements. These $2k$ elements are locally sorted 
by both processes and $id$ 0 keeps the smallest $k$, while $id$ 1 keeps the largest $k$.
Finally, the new data is sorted into the local data collection on each process
to update which elements are now the $k$ largest or $k$ smallest.
These exchanges repeat between all sets of neighbours until an iteration
occurs with no changes on any process.

This distributed sort is shown in Algorithm~\ref{alg:parallelsort}. 
Local sorting is implemented
by insertion sort.
Again, since bodies are already mostly sorted,
this bubble sort generalization should
behave as in the best case, resulting
in only $\mathcal{O}(1)$ rounds of 
exchanges and $\mathcal{O}(\nicefrac{N}{s})$ total work.

With the addition of a distributed sort, we would like to continue to avoid gathering the bodies and work estimates
to the root process for the computation of the costzones. 
We thus propose a modified costzones approach which load-balances
without a gather, requiring only exchanges between neighbouring processes.
This is a simple greedy algorithm, shown in Algorithm~\ref{alg:parallelcostzones}.
Each process begins by determining the work target via a reduce and 
broadcast (Lines 1--6). Then, iterating from process 0 to process $s-2$, each process
``pushes'' (Lines 14--16) or ``pulls''  (Lines 17--19) data to or from its right neighbour
to achieve it's own load balance.
Process $s-1$ is left with whatever remains.
This redistribution and data transfer must be careful to
not change the ordering. Notice the ordering of the concatenations 
upon receiving data (Lines 19 and 24).

Having modified the spatial decomposition and load-balancing
to be in a distributed manner, 
Algorithm~\ref{alg:BH5} presents
this finalized costzones method, with reduced communications
and still parallelized tree construction, tree traversal, and integration steps.
This presents the ``final'' version of a costzones-based Barnes-Hut simulation.
The next improvement to make on top of Algorithm~\ref{alg:BH5}
is to avoid the octree merge and broadcast. Indeed, for the highest scalability
on the largest simulations, we should avoid having the entire global tree
present on one process. This is precisely the intent of the hashed-octree method.

\begin{center}
	\vspace*{-3em}
	\begin{minipage}{0.85\textwidth}
		\begin{algorithm}[H]
			\algdef{SE}[DOWHILE]{Do}{doWhile}{\algorithmicdo}[1]{\algorithmicwhile\ #1}%

			\centering
			\footnotesize
			\caption{\scriptsize\textsc{Parallel Sort}($V$, $id$, $s$, $k=64$)\newline $V$ is a list of size $n$ of values to sort, in place. $id$ is the executing process's id, $s$ is the number of processes. $k$ controls the number of values to exchange between neighbours at each step. The result is any element of $V$ on process $id$ is less than any element of $V$ on process $id+1$, and $V$ is locally sorted. This assumes $n$ > $2k$. An experimentally determined default is $k=64$.}\label{alg:parallelsort}
			\algnotext{EndIf}
			\algnotext{EndFor}
			\algnotext{EndWhile}
			\begin{algorithmic}[1]
				\Do
				\State Sort $V$.
				\State $Left \gets V[0,...,k-1] $ concat $V[0,...,k-1]$
				\State $Right \gets V[n-k,...,n-1] $ concat $V[n-k,...,n-1]$
				\If {$id - 1 > 0$} 
				\State Send $Left[0,...,k-1]$ to $id - 1$.
				\State $Left[0,...,k-1] \gets k$ values from $id - 1$.
				\State Sort $Left$.
				\State $V[0,...,k-1] \gets Left[k,...,2k-1]$ \Comment{Keep upper half}
				\EndIf
				\If {$id + 1 < s$}
				\State Send $Right[k,...,2k-1]$ to $id + 1$.
				\State $Right[k,...,2k-1] \gets k$ values from $id + 1$.
				\State Sort $Right$.
				\State $V[n-k,...,n-1] \gets Right[0,...,k-1]$ \Comment{Keep lower half}
				\EndIf
				\State Sort $V$.
				\State $c' \gets V$ modified during sort?
				\State $c \gets $\textbf{Reduce}, via logical OR, $c'$ to process 0.
				\State $c \gets $\textbf{Broadcast} $c$ from process 0 to all others.
				\doWhile{$c \neq 0$}
			\end{algorithmic}
		\end{algorithm}
	\vspace{-1em}
	\begin{algorithm}[H]
		\centering
		\footnotesize
		\caption{\scriptsize\textsc{ParallelCostzones}($P$, $W$, $id$, $s$)\newline $P$ and $W$ are lists of size $n$ of particles and work estimates such that $W[i]$ is the work estimate for $P[i]$. $id$ is the executing process's id, $s$ is the number of processes. Distributes work between neighbours to achieve costzones-like load-balancing.}\label{alg:parallelcostzones}
		\algnotext{EndIf}
		\algnotext{EndFor}
		\algnotext{EndWhile}
		\begin{algorithmic}[1]
			\State $w_{tot} \gets 0$
			\For {$i$ \textbf{from} 0 \textbf{to} $N-1$}
			\State $w_{tot} \gets w_{tot} + W[i]$
			\EndFor
			\State $w_{tot} \gets $\textbf{Reduce}, via addition, $w_{tot}$ to process 0.
			\State $w_{tot} \gets $\textbf{Broadcast} $w_{tot}$ from process 0 to all others.
			\State $w_{targ} \gets w_{tot} / s$
			\For {$k$ \textbf{from} 0 \textbf{to} $s-2$}
				\If {$id = k$}
					\State $w \gets 0$; $j \gets 0$
					\For {$i$ \textbf{from} 0 \textbf{to} $n-1$}
						\State $w \gets w + W[i]$
						\If {$w \geq w_{targ}$ \textbf{and} $j = 0$} $j \gets i$ \EndIf
					\EndFor
					\State Send to process $id + 1$ $(w - w_{targ})$
					\If {$w > w_{targ}$} \Comment{Push excess work to right neighbour}
						\State $P,P' \gets P[0,...,j], P[j+1,...,n]$; $W,W' \gets W[0,...,j], W[j+1,...,n]$
						\State Send to process $id + 1$ $(P',W')$
					\EndIf 
					\If {$w < w_{targ}$} \Comment{Pull work from right neighbour}
						\State $P',W' \gets $ Receive data from process $id + 1$
						\State $P \gets P $ \textbf{concat} $P'$; $W \gets W $ \textbf{concat} $W'$
					\EndIf
				\ElsIf {$id = k + 1$}
					\State $w' \gets $ Receive work difference from $id - 1$
					\If {$w' > 0$} \Comment{Pull excess work from left neighbour}
						\State $P',W' \gets $ Receive data from process $id - 1$
						\State $P \gets P' $ \textbf{concat} $P$; $W \gets W' $ \textbf{concat} $W$
					\ElsIf {$w' < 0$} \Comment{Push work to left neighbour}
						\State $w \gets 0$; $j \gets n$
						\For {$i$ \textbf{from} 0 \textbf{to} $n-1$}
							\State $w \gets w + W[i]$
							\If {$w \geq |w'|$} $j \gets i$; break; \EndIf
						\EndFor
						\State $P',P \gets P[0,...,j], P[j+1,...,n]$; $W',W \gets W[0,...,j], W[j+1,...,n]$
						\State Send to process $id - 1$ $(P',W')$
					\EndIf
					\State $n \gets |P|$
				\EndIf
				\State \textbf{Synchronize} all processes \Comment{Ensure process $k$ has finished before continuing}
			\EndFor
		\end{algorithmic}
	\end{algorithm}
\end{minipage}

	\begin{minipage}{0.9\textwidth}
		\begin{algorithm}[H]
			\centering
			\footnotesize
			\caption{\scriptsize\textsc{Parallel Barnes-Hut V5}($id$, $s$, $\Delta t$, $t_{end}$, $N$, $P_{all}$)\newline $P_{all}$ is the list of size $N$ of bodies to simulate from 0 to $t_{end}$ going by $\Delta t$ time-steps. \newline $id$ is the executing process's id, $s$ is the number of processes. Assume $s \mid N$.}\label{alg:BH5}
			\algnotext{EndIf}
			\algnotext{EndFor}
			\algnotext{EndWhile}
			\begin{algorithmic}[1]
				\State $N' \gets N / s$
				\State $P \gets$ \textbf{Scatter} $P_{all}$, sending to process $id$ $P_{all}[id\cdot N', \ldots, id\cdot N' + N' - 1]$ \Comment{Get Local Bodies}
				\State $W \gets [N,\dots,N]$ \Comment{$W$ of size $N'$}
				\For {$t$ \textbf{from} 0 \textbf{to} $t_{end}$ \textbf{by} $\Delta t$}
				\State $keys \gets $ \textsc{ConstructSpatialKeys}($P$)
				\State \textsc{ParallelSort}($\{P,W,keys\}, id, s$)\Comment{Spatial Decomp.; sorting particle-work-key tuples by key}
				\State \textsc{ParallelCostzones}($P, W, id, s$) \Comment{Load balance}
				\State $T' \gets$ build octree from $P$ \Comment{Parallel Tree Build}
				\State $T \gets $ \textbf{Reduce}, via \textsc{Octree Merge}, $T'$ to process 0.
				\State \textbf{Broadcast} $T$ from process 0 to all others.
				\For {$i$ \textbf{from} 0 \textbf{to} $|P|-1$} \Comment{Compute Forces; unchanged from Algorithm~\ref{alg:BH4}}
				\State ...
				\EndFor
				\For {\textbf{each} particle $p$ in $P$} \Comment{Position Update}
				\State Update position and velocity of $p$ using its current acceleration.
				\EndFor
				\EndFor
			\end{algorithmic}
		\end{algorithm}
	\end{minipage}
\end{center}

\subsection{A Distributed Octree Implementation}
\label{sec:parallel6}

The hashed octree method is concerned with keeping the
scalability of the simulation high and avoiding
excess communication caused by broadcasting the global octree.
Rather, this method sees that
each process dynamically requests the data it needs
from other processes during the tree traversal of its local tree.

Such a distributed data structure requires a consistent addressing
scheme such that an octree node on one process
can be translated into a node on a different process. 
Moreover, a data structure which allows direct access to a particular node
would be beneficial. Recall the basic implementation of
an octree based on pointers and linked lists from Section~\ref{sec:parallel1}.
Accessing this structure involves traversing pointers and moving from root to leaf---
accessing a leaf thus requires time $\mathcal{O}(\log N)$.
Both aspects do not fulfill the needs of a distributed data structure.
However, the \textit{linear octree} or \textit{hashed octree} (originally presented in \cite{DBLP:journals/cacm/Gargantini82}) does serve this purpose.

In a hashed octree, each cell is given a unique key. This key
directly serves as the index---modulo the table size---into a hash table.
The hash table is a standard implementation using chaining and linked lists
for conflict resolution; see, e.g., \cite[Ch. 12]{DBLP:books/daglib/0023376} 
or \cite[Ch. 3]{sedgewick2011algorithms}.
A child cell's key is derived from its parent's key plus
its linear order in the parent's list of children. 
These keys will be represented in octal notation for simplicity.
The root has key 1, its 8 children have keys: 10, 11, 12, 13, 14, 15, 16, 17.
This procedure continues recursively; for example the children of node 12 are:
120, 121, 122, 123, 124, 125, 126, 127.
Notice that this scheme allows for determining the key of a parent 
or child of a particular node through simple bit operations.
The parent's key is obtained by a 3-bit right shift;
the child's key is obtained by a 3-bit left shift followed
by a bit-wise or with the child's index (0--7).

Any ordering may be used when assigning the indices 0--7 to the children
of a particular node, as long as the same ordering is used at each level. 
A natural ordering in our context is of course the
Morton ordering of Section~\ref{sec:parallel3}.
This Morton ordering of a cell's position to a spatial key
serves the purpose of a traditional hash function, 
distributing cells of the octree evenly across the hash table. 
The hash function is then just a mod function over the hash table's size.
Moreover, the spatial keys computed for each body using 
the Morton ordering directly correspond
to the key of the octree node---or the node's child, grandchild, etc.---in which they are contained.
That is to say, the body's key corresponds to a leaf node's key
if all levels of spatial division are used. 
It is often not needed to actually divide
the octree into that many levels.

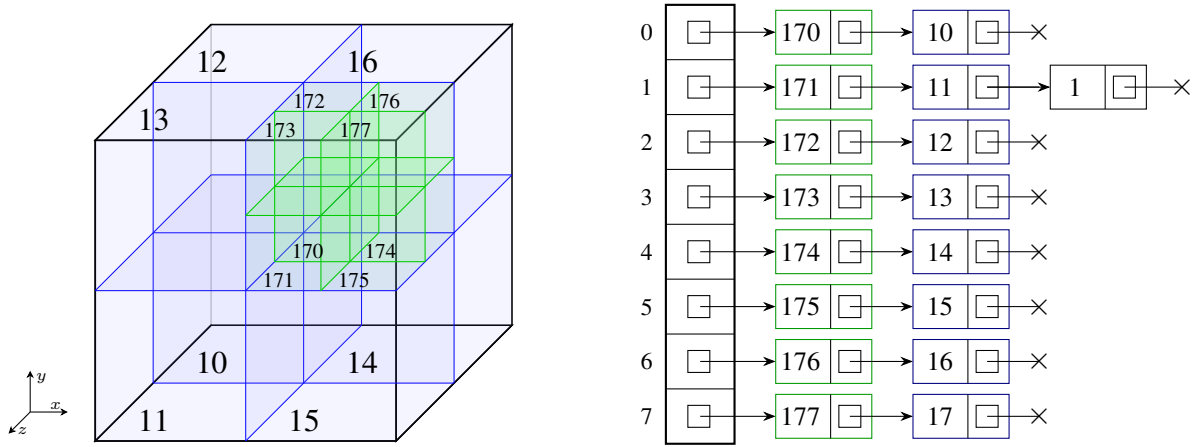
\begin{figure}[htb]
	\centering
\begin{tikzpicture}
\draw[->,>={Stealth[width=1mm,length=1mm]}] (-1.25, 0, 3) -- (-1.25, 0, 3.75);
\draw[->,>={Stealth[width=1mm,length=1mm]}] (-1.25, 0, 3) -- (-1.25, 0.55, 3);
\draw[->,>={Stealth[width=1mm,length=1mm]}] (-1.25, 0, 3) -- (-0.75, 0, 3);
\node[font=\tiny] at (-1.1, 0.45, 3) {$y$};
\node[font=\tiny] at (-1.1, 0, 3.65) {$z$};
\node[font=\tiny] at (-0.91, 0.09, 3) {$x$};
\pgfmathsetmacro{\Depth}{4}
\pgfmathsetmacro{\Height}{4}
\pgfmathsetmacro{\Width}{4}

\coordinate (O) at (0,0,0);
\coordinate (A) at (0,\Width,0);
\coordinate (B) at (0,\Width,\Height);
\coordinate (C) at (0,0,\Height);
\coordinate (D) at (\Depth,0,0);
\coordinate (E) at (\Depth,\Width,0);
\coordinate (F) at (\Depth,\Width,\Height);
\coordinate (G) at (\Depth,0,\Height);

\draw[black, semithick, line cap=round] (O) -- (C) -- (G) -- (D) -- cycle;
\draw[black, semithick, line cap=round] (A) -- (B) -- (C) -- (O);
\draw[black, semithick, opacity=0.2] (A) -- (O);
\draw[black, semithick, line cap=round] (D) -- (E) -- (F) -- (G) -- cycle;
\draw[black, semithick, line cap=round] (C) -- (B) -- (F) -- (G) -- cycle;
\draw[black, semithick, line cap=round] (A) -- (B) -- (F) -- (E) -- cycle;

\draw[blue] (2, 0, 0) -- (2, 4, 0);
\draw[blue] (0, 2, 0) -- (4, 2, 0);
\draw[blue] (4, 2, 0) -- (4, 2, 4);
\draw[blue] (4, 0, 2) -- (4, 4, 2);
\draw[blue] (2, 0, 0) -- (2, 0, 4);
\draw[blue] (0, 0, 2) -- (4, 0, 2);
\draw[blue] (0, 2, 0) -- (0, 2, 4);
\draw[blue] (0, 0, 2) -- (0, 4, 2);
\draw[blue] (2, 4, 0) -- (2, 4, 4);
\draw[blue] (0, 4, 2) -- (4, 4, 2);
\draw[blue] (2, 0, 4) -- (2, 4, 4);
\draw[blue] (0, 2, 4) -- (4, 2, 4);
\draw[blue] (0, 2, 2) -- (4, 2, 2);
\draw[blue] (2, 2, 0) -- (2, 2, 4);
\draw[blue] (2, 0, 2) -- (2, 4, 2);

\draw[fill=blue!50,fill opacity=0.08, draw opacity=0] (0,4,0) -- (4,4,0) -- (4,0,0) -- (4,0,4) -- (0,0,4) -- (0,4,4) -- (0,4,0);
\draw[fill=blue!50,fill opacity=0.08,draw opacity=0] (0,2,0) -- (4,2,0) -- (4,2,4) -- (0,2,4) -- cycle;
\draw[fill=blue!50,fill opacity=0.08,draw opacity=0] (0,0,2) -- (4,0,2) -- (4,4,2) -- (0,4,2) -- cycle;
\draw[fill=blue!50,fill opacity=0.08,draw opacity=0] (2,0,0) -- (2,4,0) -- (2,4,4) -- (2,0,4) -- cycle;

\draw[blue,fill=green!50, opacity=0.1] (2,2,2) -- (2,4,2) -- (2,4,4) -- (2,2,4) -- cycle;
\draw[blue,fill=green!50,opacity=0.1] (2,2,2) -- (4,2,2) -- (4,4,2) -- (2,4,2) -- cycle;
\draw[blue,fill=green!50,opacity=0.1] (2,2,2) -- (4,2,2) -- (4,2,4) -- (2,2,4) -- cycle;

\draw[fill=green!50,fill opacity=0.2,draw opacity=0] (2,3,2) -- (4,3,2) -- (4,3,4) -- (2,3,4) -- cycle;
\draw[fill=green!50,fill opacity=0.2,draw opacity=0] (2,2,3) -- (4,2,3) -- (4,4,3) -- (2,4,3) -- cycle;
\draw[fill=green!50,fill opacity=0.2,draw opacity=0] (3,2,2) -- (3,4,2) -- (3,4,4) -- (3,2,4) -- cycle;

\draw[green!80!black] (3, 2, 2) -- (3, 4, 2);
\draw[green!80!black] (2, 3, 2) -- (4, 3, 2);
\draw[green!80!black] (4, 3, 2) -- (4, 3, 4);
\draw[green!80!black] (4, 2, 3) -- (4, 4, 3);
\draw[green!80!black] (3, 2, 2) -- (3, 2, 4);
\draw[green!80!black] (2, 2, 3) -- (4, 2, 3);
\draw[green!80!black] (2, 3, 2) -- (2, 3, 4);
\draw[green!80!black] (2, 2, 3) -- (2, 4, 3);
\draw[green!80!black] (3, 4, 2) -- (3, 4, 4);
\draw[green!80!black] (2, 4, 3) -- (4, 4, 3);
\draw[green!80!black] (3, 2, 4) -- (3, 4, 4);
\draw[green!80!black] (2, 3, 4) -- (4, 3, 4);
\draw[green!80!black] (2, 3, 3) -- (4, 3, 3);
\draw[green!80!black] (3, 3, 2) -- (3, 3, 4);
\draw[green!80!black] (3, 2, 3) -- (3, 4, 3);


\node[] at (0.7,0.2,1.8) {10};
\node[] at (0.7,0.2,3.8) {11};
\node[] at (0.7,4.2,1.8) {12};
\node[] at (0.7,4.2,3.8) {13};
\node[] at (2.7,0.2,1.8) {14};
\node[] at (2.7,0.2,3.8) {15};
\node[] at (2.7,4.2,1.8) {16};

\node[font=\scriptsize] at (2.4, 2.1, 2.9) {170};
\node[font=\scriptsize] at (2.4, 2.1, 3.9) {171};
\node[font=\scriptsize] at (2.42, 4.1, 2.9) {172};
\node[font=\scriptsize] at (2.42, 4.1, 3.9) {173};
\node[font=\scriptsize] at (3.37, 2.1, 2.9) {174};
\node[font=\scriptsize] at (3.4, 2.1, 3.9) {175};
\node[font=\scriptsize] at (3.42,4.1, 2.9) {176};
\node[font=\scriptsize] at (3.42, 4.1, 3.9) {177};
\end{tikzpicture}
\hspace{3em}
\begin{tikzpicture} [scale=0.73]
\foreach[count=\x from 0] \y in {0,...,-7} 
\node[font=\footnotesize] at (-0.35,\y-0.5) {\x};
\draw[thick] (0,0) -- (1.25,0) -- (1.25,-8) -- (0,-8) -- cycle;
\foreach \y in {0,...,-7} 
\draw (0,\y) -- (1.25,\y);
\foreach \y in {0,...,-7}
\draw (0.4,\y-0.3) rectangle (0.8,\y-0.7);
\foreach \y in {0,...,-7} \draw[->, >=Stealth] (0.625,\y-0.5)   -- (2.0,\y-0.5);
\foreach \y in {0,...,-7} \draw[green!60!black] (2.0,\y-0.1)  rectangle (3.75,\y-0.9);
\foreach[count=\x from 0] \y in {0,...,-7} \node[font=\small] at (2.45,\y-0.5) {17\x};
\foreach \y in {0,...,-7} \draw (3.0,\y-0.1) -- (3.0,\y-0.9);
\foreach \y in {0,...,-7}
\draw (3.15,\y-0.3) rectangle (3.55,\y-0.7);
\foreach \y in {0,...,-7} \draw[->, >=Stealth] (3.35,\y-0.5)   -- (4.5,\y-0.5);

\foreach \y in {0,...,-7} \draw[blue!50!black] (4.5,\y-0.1)  rectangle (6.25,\y-0.9);
\foreach[count=\x from 0] \y in {0,...,-7} \node[font=\small] at (4.99,\y-0.5) {1\x};
\foreach \y in {0,...,-7} \draw (5.5,\y-0.1) -- (5.5,\y-0.9);

\foreach \y in {0,...,-7}
\draw (5.65,\y-0.3) rectangle (6.05,\y-0.7);

\foreach \y in {0,...,-7} \draw (5.85,\y-0.5)   -- (6.8,\y-0.5);

\node at (6.8,-0.5) {$\times$};
\foreach \y in {-2,...,-7} \node at (6.8,\y-0.5) {$\times$};

\draw[->,>=Stealth] (5.85,-1.5) -- (7.0,-1.5);
\draw[black] (7.0,-1-0.1)  rectangle (8.75,-1-0.9);
\node[font=\small] at (7.45,-1-0.5) {1};
\draw (8.0,-1-0.1) -- (8.0,-1-0.9);
\draw (8.15,-1-0.3) rectangle (8.55,-1-0.7);
\draw (8.35,-1.5) -- (9.4,-1.5);
\node at (9.4,-1.5) {$\times$};

\end{tikzpicture}
\caption{An octree with cells labelled by their hashed octree key, written in octal. The root node's key of 1 is not shown. Keys are created
	from a Morton ordering with $x > y > z$. That octree's encoding as a hash table---an array of linked lists---with
	8 entries is shown to the right. Note that the payload of each cell is not shown.}\label{fig:hot}
\end{figure}

Fig.~\ref{fig:hot} shows the key assignment via Morton's ordering 
for up to three levels 
of an octree and an encoding of that octree as a hash table. 
The nodes of the pointer-based octree can be repurposed 
to be nodes of the hash table linked list with a simple transformation:
$(i)$ the list of pointers to children is replaced by a single pointer 
to the next cell for collision resolution via chaining; and $(ii)$ 
the cell's key is included in the payload. 

A hashed octree implementation allows for consistent indexing of nodes 
in the tree, rather than reliance on pointer traversal. Further, 
it allows access to any node in the tree in time $\mathcal{O}(1)$.
This scheme allows the $N$-body simulation 
to avoid the creation of a global tree
and for each process to dynamically and efficiently request remote data
from other processes as needed during the tree traversal. 

The last requirement is for each process to share the boundary
of its spatial domain so that each process knows from where
to request data during during the tree traversal.
This sharing proceeds in three steps.
\begin{enumerate}[($i$)]
	\item Assume a linear topology of processes, just as in distributed sorting and costzones. In the first step, each process shares the first body in its spatial domain with its left neighbour and the last body in its spatial domain with its right neighbour, if such a neighbour exists. The ``neighbour bodies'' (or single body for process 0 and $s-1$) are inserted into the local octree, splitting as needed so that each leaf node contains exactly one body. Due to the spatial ordering of bodies, this insertion will cause the required splitting of leaf nodes so that the union of all leaf nodes across the $s$ local octrees is exactly the leaf nodes that would be created if a global octree were to be constructed. That is to say, all leaf nodes (excluding the leaves containing neighbour bodes) are distinct across all processes and a consistent spatial partitioning has been created.
	\item In the second step, each process computes and broadcasts its so-called \textit{branch nodes}. These are the set of nodes of the octree which are closest to the root, but which do not contain a neighbour body in their domain.
	They represent the entire spatial domain of a process at the coarsest level possible. Branch nodes can be found easily by a depth-first traversal of the the local octree: if a cell's domain does not contain a neighbour body, but the cell's parent's domain does, then that cell is a branch node.
	\item In the third step, a process receives sets of branch nodes as they are broadcast. For each set, those branch nodes are inserted into the local octree and marked with their source process. This source id will allow for dynamically requesting data during tree traversal. If a branch node is received for which its parent is not in the local octree, the parent, grandparent, etc. nodes are ``filled'' in to the local octree as well.
\end{enumerate}

\begin{figure}[htb]
\hspace{1.5em}
\begin{subfigure}{0.45\textwidth}
	\centering
\begin{tikzpicture}[scale=0.75]
%

\draw
[step=2cm, thin] (0,4) grid +(4,4)
[step=1cm, thin] (2,4) grid +(2,2)
[step=0.5cm, thin] (2, 4) grid +(1,1);
\draw [step=4cm, very thick, line cap=round] (0,0) grid (8,8);


\draw[fill] (0.88, 1.77) circle (0.06cm);
\draw[fill] (0.43, 4.65) circle (0.06cm);
\draw[fill] (0.77, 7.23) circle (0.06cm);
\draw[fill] (2.11, 4.13) circle (0.06cm);
\draw[fill] (2.41, 4.81) circle (0.06cm);
\draw[fill] (2.75, 4.25) circle (0.06cm);
\draw[fill] (2.70, 4.60) circle (0.06cm);


%
%
%
\end{tikzpicture}
\subcaption{}\label{fig:branch1}
\end{subfigure}
\begin{subfigure}{0.45\textwidth}
		\centering
\begin{tikzpicture}[scale=0.75]
%

\draw[fill=blue!80, fill opacity=0.1] (0.1,0.1) -- (0.1,3.9) -- (3.9,3.9) -- (3.9, 0.1) -- cycle;
\draw[fill=blue!80, fill opacity=0.1] (0.1,4.1) -- (1.9,4.1) -- (1.9,5.9) -- (0.1,5.9) -- cycle;
\draw[fill=blue!80, fill opacity=0.1] (0.1,6.1) -- (1.9,6.1) -- (1.9,7.9) -- (0.1,7.9) -- cycle;
\draw[fill=blue!80, fill opacity=0.1] (2.1,4.1) -- (2.9,4.1) -- (2.9,4.9) -- (2.1,4.9) -- cycle;

\draw
[step=2cm, thin] (0,4) grid +(4,4)
[step=1cm, thin] (2,4) grid +(2,2)
[step=0.5cm, thin] (2, 4) grid +(1,1);
\draw [step=4cm, very thick, line cap=round] (0,0) grid (8,8);


\draw[fill] (0.88, 1.77) circle (0.06cm);
\draw[fill] (0.43, 4.65) circle (0.06cm);
\draw[fill] (0.77, 7.23) circle (0.06cm);
\draw[fill] (2.14, 4.14) circle (0.06cm);
\draw[fill] (2.41, 4.81) circle (0.06cm);
\draw[fill] (2.75, 4.25) circle (0.06cm);
\draw[fill] (2.70, 4.60) circle (0.06cm);

\draw[fill=orange!30!yellow,draw=orange!30!yellow!40!black] (3.60, 4.44) circle (0.07cm);

%
%
%
\end{tikzpicture}
\subcaption{}\label{fig:branch2}
\end{subfigure}
\vspace{1.5em}

\hspace{1.5em}
\begin{subfigure}{0.45\textwidth}
	\centering
\begin{tikzpicture}[scale=0.75]
%

\draw
[step=2cm, thin] (0,4) grid +(4,4)
[step=1cm, thin] (2,4) grid +(2,2)
[step=2cm, thin] (4,0) grid + (4,4)
[step=0.5cm, thin] (4, 2) grid +(1,1)
[step=0.5cm, thin] (4, 3) grid +(1,1)
[step=1cm, thin] (5, 2) grid +(1,1);
\draw [step=4cm, very thick, line cap=round] (0,0) grid (8,8);



\draw[fill] (3.60, 4.44) circle (0.06cm);
\draw[fill] (3.44, 5.77) circle (0.06cm);

\draw[fill] (4.20, 2.78) circle (0.06cm);
\draw[fill] (4.65, 2.42) circle (0.06cm);
\draw[fill] (4.33, 3.27) circle (0.06cm);
\draw[fill] (4.26, 3.88) circle (0.06cm);
\draw[fill] (5.18, 2.77) circle (0.06cm);
%
%
%
\end{tikzpicture}
\subcaption{}\label{fig:branch3}
\end{subfigure}
\begin{subfigure}{0.45\textwidth}
		\centering
\begin{tikzpicture}[scale=0.75]
%

\draw[fill=orange!90!yellow, fill opacity=0.1] (3.1,5.1) -- (3.9,5.1) -- (3.9,5.9) -- (3.1,5.9) -- cycle;
\draw[fill=orange!90!yellow, fill opacity=0.1] (3.1,4.1) -- (3.9,4.1) -- (3.9,4.9) -- (3.1,4.9) -- cycle;
\draw[fill=orange!90!yellow, fill opacity=0.1] (4.1,3.1) -- (4.9,3.1) -- (4.9,3.92) -- (4.1,3.92) -- cycle;
\draw[fill=orange!90!yellow, fill opacity=0.1] (4.1,2.1) -- (4.9,2.1) -- (4.9,2.9) -- (4.1,2.9) -- cycle;
\draw[fill=orange!90!yellow, fill opacity=0.1] (5.1,2.1) -- (5.9,2.1) -- (5.9,2.9) -- (5.1,2.9) -- cycle;
\draw
[step=2cm, thin] (0,4) grid +(4,4)
[step=1cm, thin] (2,4) grid +(2,2)
[step=2cm, thin] (4,0) grid + (4,4)
[step=0.5cm, thin] (4, 2) grid +(1,1)
[step=0.5cm, thin] (4, 3) grid +(1,1)
[step=1cm, thin] (5, 2) grid +(1,1);
\draw [step=4cm, very thick, line cap=round] (0,0) grid (8,8);


\draw[fill=blue!60,draw=blue!80] (2.70, 4.60) circle (0.07cm);

\draw[fill] (3.60, 4.44) circle (0.06cm);
\draw[fill] (3.44, 5.77) circle (0.06cm);

\draw[fill] (4.20, 2.78) circle (0.06cm);
\draw[fill] (4.65, 2.42) circle (0.06cm);
\draw[fill] (4.33, 3.27) circle (0.06cm);
\draw[fill] (4.26, 3.85) circle (0.06cm);
\draw[fill] (5.18, 2.77) circle (0.07cm);
\draw[fill=red!80!blue,fill opacity=0.5,draw=red!20!black,very thin] (5.80, 3.13) circle (0.07cm);
%
%
%
\end{tikzpicture}
\subcaption{}\label{fig:branch4}
\end{subfigure}

\vspace{1.5em}

\hspace{1.5em}
\begin{subfigure}{0.45\textwidth}
		\centering
\begin{tikzpicture}[scale=0.75]
%

\draw
[step=2cm, thin] (4,4) grid +(4,4)
[step=1cm, thin] (4,4) grid +(2,2)
[step=2cm, thin] (4,0) grid + (4,4);
\draw [step=4cm, very thick, line cap=round] (0,0) grid (8,8);



%
\draw[fill] (5.80, 3.13) circle (0.06cm);

\draw[fill] (7.22, 0.8) circle (0.06cm);
\draw[fill] (6.77, 3.2) circle (0.06cm);

\draw[fill] (4.64, 4.76) circle (0.06cm);
\draw[fill] (5.72, 5.31) circle (0.06cm);
\draw[fill] (6.22, 4.81) circle (0.06cm);
\draw[fill] (6.4, 6.81) circle (0.06cm);
\end{tikzpicture}
\subcaption{}\label{fig:branch5}
\end{subfigure}
\begin{subfigure}{0.45\textwidth}
\centering
\begin{tikzpicture}[scale=0.75]
%

\draw[fill=red, fill opacity=0.1] (4.1, 4.1) -- (4.1,7.9) -- (7.9, 7.9) -- (7.9,4.1) -- cycle;
\draw[fill=red, fill opacity=0.1] (6.1, 0.1) -- (7.9,0.1) -- (7.9, 1.9) -- (6.1,1.9) -- cycle;
\draw[fill=red, fill opacity=0.1] (6.1, 2.1) -- (7.9,2.1) -- (7.9, 3.9) -- (6.1,3.9) -- cycle;
\draw[fill=red, fill opacity=0.1] (5.07, 3.07) -- (5.93,3.07) -- (5.93, 3.9) -- (5.07,3.9) -- cycle;

\draw
[step=2cm, thin] (4,4) grid +(4,4)
[step=1cm, thin] (4,4) grid +(2,2)
[step=2cm, thin] (4,0) grid + (4,4)
[step=1cm, thin] (4, 2) grid +(2,2);
\draw [step=4cm, very thick, line cap=round] (0,0) grid (8,8);



%
\draw[fill=orange!30!yellow,draw=orange!30!yellow!40!black] (5.18, 2.77) circle (0.07cm);
\draw[fill] (5.80, 3.13) circle (0.06cm);

\draw[fill] (7.22, 0.8) circle (0.06cm);
\draw[fill] (6.77, 3.2) circle (0.06cm);

\draw[fill] (4.64, 4.76) circle (0.06cm);
\draw[fill] (5.72, 5.31) circle (0.06cm);
\draw[fill] (6.22, 4.81) circle (0.06cm);
\draw[fill] (6.4, 6.81) circle (0.06cm);
\end{tikzpicture}
\subcaption{}\label{fig:branch6}
\end{subfigure}
\caption{Three local quadtrees, (a), (b), (c), before sharing neighbour bodies, and their respective updated quadtrees (d), (e), (f), after sharing neighbour bodies. The neighbour bodies are colored corresponding to their owning process. Notice in (f) that the addition of the neighbour body causes a leaf node to split, creating consistent leaf nodes
between (d) and (f). The branch nodes of each process are colored in (b), (d), (f). The branch nodes collectively cover every body in the simulation domain.}\label{fig:branchnodes} 
\end{figure}
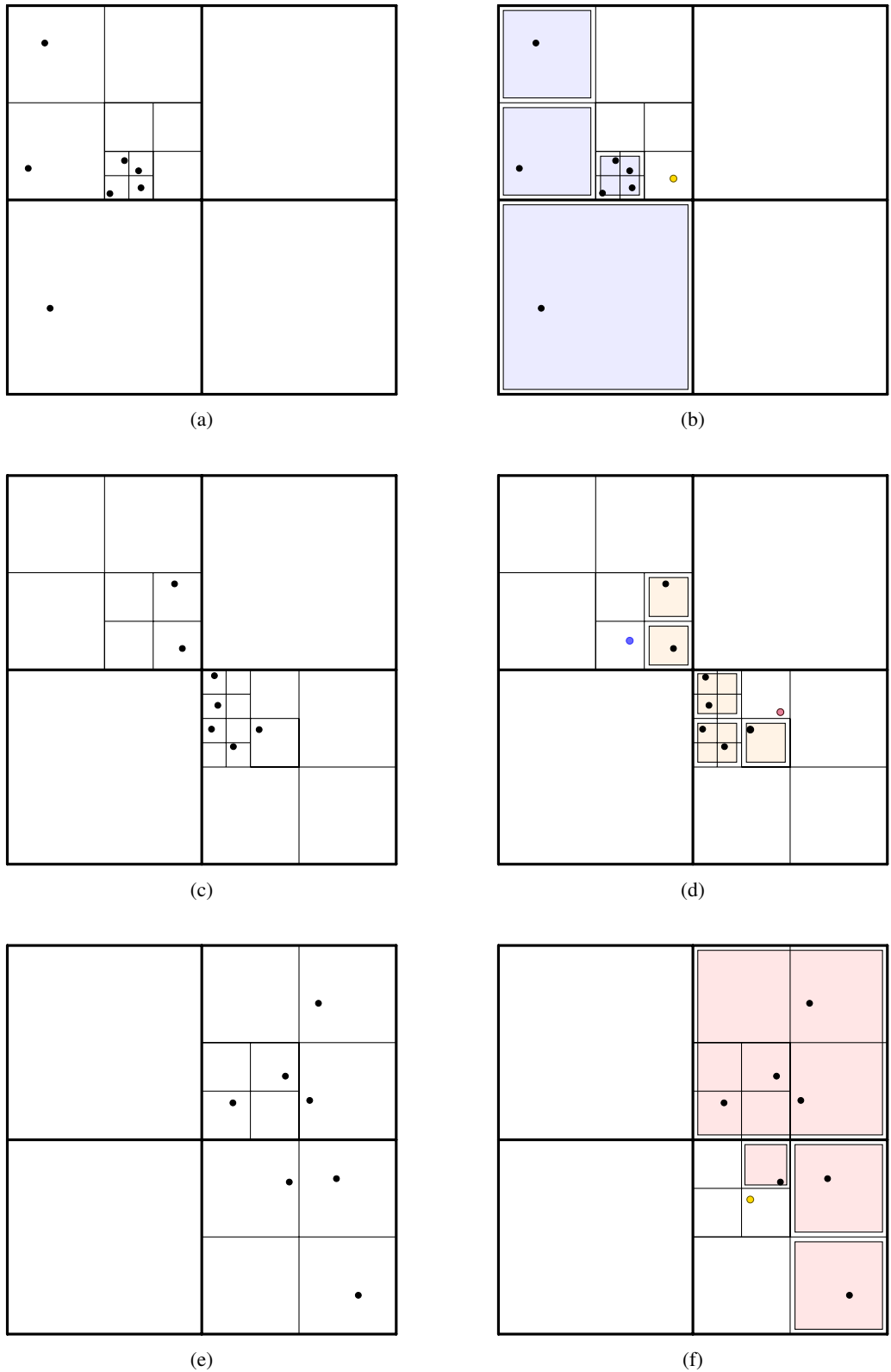

Upon inserting all the branch nodes, each process has a local octree which 
is globally and mutually consistent with the other local trees. 
Fig.~\ref{fig:branchnodes} shows this neighbour exchange
and resulting branch nodes for three processes. 
After sharing and inserting these branch nodes, 
each local tree contains the required information 
to request nodes below that of the received branch nodes. 
Such nodes must be requested during tree traversal when nodes 
fail the multiple acceptance criterion and must be ``opened'' (see Section~\ref{sec:bgoctree}).

\begin{center}
	\vspace{-0.5em}
	\begin{minipage}{0.7\textwidth}
		\begin{algorithm}[H]
			\centering
			\footnotesize
			\caption{\scriptsize\textsc{DistributeBranchNodes}($P$, $T$, $id$, $s$)\newline $P$ is the list of local bodies and $T$ is the octree containing those bodies. \newline $id$ is the executing process's id, $s$ is the number of processes.}\label{alg:branchnodes}
			\algnotext{EndIf}
			\algnotext{EndFor}
			\algnotext{EndWhile}
			\begin{algorithmic}[1]
				\If {$id - 1 \geq 0$} 
				\State Send $P[0]$ to process $id -1$ 
				\State $p \gets $ Receive neighbour body from $id -1$
				\State Insert $p$ into $T$, splitting as required.
				\EndIf
				\If {$id + 1 < s$} 
				\State Send $P[n]$ to process $id + 1$ 
				\State $p \gets $ Receive neighbour body from $id + 1$
				\State Insert $p$ into $T$, splitting as required.
				\EndIf
				\State $B \gets$ Branch nodes of $T$, the coarsest noes which do not contain a neighbour body.
				\For {$i$ \textbf{from} 0 \textbf{to} $s-1$}
				\State Broadcast $B$ from process $i$ to all others.
				\If {$id \neq i$}
					\State $B' \gets$ Receive branch nodes from $i$
					\For {\textbf{each} node $b$ in $B'$} insert $b$ into $T$ and fill in parents \EndFor
				\EndIf
				\EndFor
			\end{algorithmic}
		\end{algorithm}
	\end{minipage}
\end{center}
\vspace{0.5em}

\clearpage

With the hashed octree data structure and the required 
branch nodes fully specified, we can now take advantage of this
distributed data structure to remove the final reduce and broadcast
(apart from a few branch nodes) of our parallel Barnes-Hut algorithm.
The span of the simulation is now minimized.
Algorithm~\ref{alg:BH6} presents this modified Barnes-Hut method 
which--implicitly--makes use of a hashed octree encoding
of the octree. 
First, the reduce-broadcast of the global tree is replaced
with a call to \textsc{DistributeBranchNodes} (Line 9).
Second, the tree traversal is slightly modified to
check if a node's children are local or not. 
If not, they are requested from their source process, as determined
by their containing branch node (Lines 14--16).
Finally, all processes synchronize before moving to the integration step
to ensure all requests from all processes have been fulfilled (Lines 19--20).

In this configuration the requesting process stalls until its request has been fulfilled.
Therefore, before a request is actually sent, the process fulfills any 
and all requests made to it. This avoids the situation
where two processes are waiting on each other to fulfill a request.
In the next section we will improve this
to avoid stalling and instead perform asynchronous communication.

\begin{center}
	\vspace{-0.5em}
	\begin{minipage}{0.9\textwidth}
		\begin{algorithm}[H]
			\centering
			\footnotesize
			\caption{\scriptsize\textsc{Parallel Barnes-Hut V6}($id$, $s$, $\Delta t$, $t_{end}$, $N$, $P_{all}$)\newline $P_{all}$ is the list of size $N$ of bodies to simulate from 0 to $t_{end}$ going by $\Delta t$ time-steps. \newline $id$ is the executing process's id, $s$ is the number of processes. Assume $s \mid N$.}\label{alg:BH6}
			\algnotext{EndIf}
			\algnotext{EndFor}
			\algnotext{EndWhile}
			\begin{algorithmic}[1]
				\State $N' \gets N / s$
				\State $P \gets$ \textbf{Scatter} $P_{all}$, sending to process $id$ $P_{all}[id\cdot N', \ldots, id\cdot N' + N' - 1]$ \Comment{Get Local Bodies}
				\State $W \gets [N,\dots,N]$ \Comment{$W$ of size $N'$}
				\For {$t$ \textbf{from} 0 \textbf{to} $t_{end}$ \textbf{by} $\Delta t$}
				\State $keys \gets $ \textsc{ConstructSpatialKeys}($P$)
				\State \textsc{ParallelSort}($\{P,W,keys\}, id, s$)\Comment{Spatial Decomp.; sorting particle-work-key tuples by key}
				\State \textsc{ParallelCostzones}($P, W, id, s$) \Comment{Load balance}
				\State $T \gets$ build hashed octree from $P$ \Comment{Parallel Tree Build}
				\State \textsc{DistributeBranchNodes}($P$, $T$, $id$, $s$)
				\For {$i$ \textbf{from} 0 \textbf{to} $|P|-1$} \Comment{Compute Forces}
				\State $p \gets P[i]$; $W[i] \gets 0$
				\State Set the acceleration of $p$ to $\vec{0}$.
				\State $Node$ $\gets$ the root node of $T$ \vspace{0.3em}
				\EndFor
				\Statex $\quad\ $Compute Cell Force:
				\Indent
				\If {children of $Node$ are not stored locally}
					\State Fulfill any waiting requests for local children from other processes \Comment{avoid deadlock}
					\State Request and receive the children of $Node$ from its owning process
				\EndIf
				\For {\textbf{each} child cell $C$ of $Node$} \Comment{Tree traversal unchanged from Algorithm~\ref{alg:BH5}}
					\State ...
				\EndFor
				\EndIndent
				\While {not all processes done force computation}
					\State Fulfill any waiting requests for local children from other processes
				\EndWhile
				\For {\textbf{each} particle $p$ in $P$} \Comment{Position Update}
				\State Update position and velocity of $p$ using its current acceleration.
				\EndFor
				\EndFor
			\end{algorithmic}
		\end{algorithm}
	\end{minipage}
\end{center}

\subsection{Latency-Hiding Tree Traversal}
\label{sec:parallel7}

The hashed octree method 
avoids each process from needing
the global octree for force computation.
Instead, branch nodes are used as entry points
to remote trees, where any child node
can be dynamically requested from the source process
during the tree traversal.
In Algorithm~\ref{alg:BH6}
that dynamic request is made simple through synchronous
and blocking communication. 
We now suggest a \textit{latency-hiding} scheme
which uses asynchronous communication to avoid communication overheads. 

The concept is simple: 
during the tree traversal for a particle, 
maintain a \textit{walk list}---a queue of nodes which have yet to be traversed---and a
\textit{defer list}---the queue of nodes whose children have been requested from other processes.
This strategy also transforms the tree traversal 
from a recursive method to an iterative one.
Algorithm~\ref{alg:asynctraversal} shows this iterative tree traversal
to compute the acceleration of one particular body.
The walk list begins containing the root node. 
For each iteration of the traversal,
the children of the active node are iterated over, 
testing for multipole acceptance and thus a particle-cell
interaction. If the particle-cell interaction is not acceptable, 
the child is added to the walk list.
If, at any point, a node is encountered whose
child cells are not stored locally, 
then an asynchronous request is sent to the source process
for the children, and the node is enqueued to the defer list.
The traversal then skips to the next iteration.
If, at any point, the walk list becomes empty, 
then a node is dequeued from the defer list.
If the child data of that dequeued node has not yet arrived, the traversal waits synchronously for it.
The traversal then continues as normal.

\begin{center}
	\vspace{-0.5em}
	\begin{minipage}{0.80\textwidth}
		\begin{algorithm}[H]
			\centering
			\footnotesize
			\caption{\scriptsize\textsc{AsyncHashedOctreeTraversal}($p$, $T$)\newline $p$ is the body 
				for which to compute its acceleration, $T$ is the local octree. \newline Returns the work estimate for $p$ for the next time step.}\label{alg:asynctraversal}
			\algnotext{EndIf}
			\algnotext{EndFor}
			\algnotext{EndWhile}
			\begin{algorithmic}[1]
				\State $w \gets 0$; Set the acceleration of $p$ to $\vec{0}$.
				\State $walk \gets [\text{the root node of}\ T]$; $defer \gets [\ ]$
				\While {$walk$ is not empty \textbf{or} $defer$ is not empty}
				\If {$walk$ is not empty} 
					\State $Node \gets$ \textbf{dequeue} $walk$
				\Else 
					\State $Node \gets$ \textbf{dequeue} $defer$
					\State Fulfill any waiting requests for local children from other processes
					\State Wait for receipt of children of $Node$ from its owning process
				\EndIf
				\If {Children of $Node$ are not stored locally}
					\State Fulfill any waiting requests for local children from other processes
					\State Send synchronous request for children of $Node$ to its owning process
					\State \textbf{enqueue} $Node$ to $defer$
					\State \textbf{continue}
				\EndIf
				\For {\textbf{each} child cell $C$ of $Node$}
					\If {$C$ contains 1 particle} \Comment{$C$ is a leaf}
					\State Add to $p$'s acceleration the contribution from $C$'s particle. \Comment{particle-particle}
					\State $w \gets w + 1$
					\Else
					\State $\ell$ $\gets$ $C$'s side length
					\State $d$ $\gets$ distance between $C$'s c.o.m. and $p$'s position.
					\If {${\ell} / {d} < \theta$}
					\State Add to $p$'s acceleration the contribution from $C$. \Comment{particle-cell}
					\State $w \gets w + 2$
					\Else 
					\State \textbf{enqueue} $C$ to $walk$
					\EndIf
				\EndIf
				\EndFor
				\EndWhile
				\State \Return $w$
			\end{algorithmic}
		\end{algorithm}
	\end{minipage}
\end{center}
\vspace{1.5em}

The individual processes operating in this asynchronous 
request-and-reply scheme must act cooperatively to achieve best results. 
To ensure no request is left waiting for too long, other processes
should frequently check for and fulfill incoming requests for data. 
Such a fulfillment occurs every time a process wishes to make a request itself,
thus minimizing access to the communication subprograms and improving locality of operation.

The asynchronous tree traversal easily slots into 
the previous Barnes-Hut algorithm based on hashed octrees (Algorithm~\ref{alg:BH6}).
The modified algorithm is presented in Algorithm~\ref{alg:BH7}.
Notice that each force calculation step still must end with a synchronization 
so that all requests are fulfilled before any process continues
to the integration step (Lines 12--13).
This algorithm is the final version of our many parallel Barnes-Hut methods.

\begin{center}
	\vspace{-0.5em}
	\begin{minipage}{0.9\textwidth}
		\begin{algorithm}[H]
			\centering
			\footnotesize
			\caption{\scriptsize\textsc{Parallel Barnes-Hut V7}($id$, $s$, $\Delta t$, $t_{end}$, $N$, $P_{all}$)\newline $P_{all}$ is the list of size $N$ of bodies to simulate from 0 to $t_{end}$ going by $\Delta t$ time-steps. \newline $id$ is the executing process's id, $s$ is the number of processes. Assume $s \mid N$.}\label{alg:BH7}
			\algnotext{EndIf}
			\algnotext{EndFor}
			\algnotext{EndWhile}
			\begin{algorithmic}[1]
				\State $N' \gets N / s$
				\State $P \gets$ \textbf{Scatter} $P_{all}$, sending to process $id$ $P_{all}[id\cdot N', \ldots, id\cdot N' + N' - 1]$ \Comment{Get Local Bodies}
				\State $W \gets [N,\dots,N]$ \Comment{$W$ of size $N'$}
				\For {$t$ \textbf{from} 0 \textbf{to} $t_{end}$ \textbf{by} $\Delta t$}
				\State $keys \gets $ \textsc{ConstructSpatialKeys}($P$)
				\State \textsc{ParallelSort}($\{P,W,keys\}, id, s$)\Comment{Spatial Decomp.; sorting particle-work-key tuples by key}
				\State \textsc{ParallelCostzones}($P, W, id, s$) \Comment{Load balance}
				\State $T \gets$ build hashed octree from $P$ \Comment{Parallel Tree Build}
				\State \textsc{DistributeBranchNodes}($P$, $T$, $id$, $s$)
				\For {$i$ \textbf{from} 0 \textbf{to} $|P|-1$} \Comment{Compute Forces}
				\State $W[i] \gets$ \textsc{AsyncHashedOctreeTraversal}($P[i]$, $T$)
				\EndFor
				\While {not all processes done force computation}
				\State Fulfill any waiting requests for local children from other processes
				\EndWhile
				\For {\textbf{each} particle $p$ in $P$} \Comment{Position Update}
				\State Update position and velocity of $p$ using its current acceleration.
				\EndFor
				\EndFor
			\end{algorithmic}
		\end{algorithm}
	\end{minipage}
\end{center}

%% file: experimentation.tex
\section{Experimental Results and Discussion}
\label{sec:experimentation}

Throughout the last section, we 
presented and developed many algorithms 
for parallel $N$-body simulation. 
These algorithms saw various techniques applied incrementally 
to improve scalability and load-balancing. 
This lead to one parallel version of the direct method
and seven versions of the Barnes-Hut method. 
We summarize those methods here, and rename them as follows.
\begin{itemize}[align=parleft]
	\setlength{\itemsep}{0.5em}
	\item[\textit{Algorithm 0}:] The parallelization of the direct $N^2$ method.
	Each process is assigned an equal number of bodies, 
	and the force calculations and integration steps 
	are computed in parallel. This method passes each process's local set
	of bodies around in a ring formation to compute forces at each time-step.
	
	\item[\textit{Algorithm 1}:] Barnes-Hut parallelization version 1. Bodies are gathered to the root process, a global octree is constructed serially, and then broadcast to all processes. Force computation (tree traversals) and integration steps are performed in parallel.
	
	\item[\textit{Algorithm 2}:] Parallel octree construction is added to \textit{Algorithm 1}
	where each process builds a local octree from its local bodies, and then a global tree
	is constructing through pair-wise merges. The global octree is then broadcast. Force computation and integration steps continue to be computed in parallel.
	
	\item[\textit{Algorithm 3}:] Spatial decomposition is added to \textit{Algorithm 2}. Each process is now responsible for some partition of the spatial domain at each time-step rather than a fixed set of local bodies. 
	This domain decomposition follows a Morton ordering on the bodies to assign an equal number of bodies, 
	but an unequal amount of space, to each process.  The re-partitioning occurs every time step.
	
	\item[\textit{Algorithm 4}:] Load-balancing via \textit{costzones} is added to \textit{Algorithm 4}. Bodies are constantly re-distributed such that each process computes roughly the same number of interactions per time-step, rather than having the same number of bodies. The spatial partitioning is still applied. 
	
	\item[\textit{Algorithm 5}:] Collective gather and scatter operations are removed from \textit{Algorithm 4}
	with the addition of a distributed sort for the spatial decomposition and a greedy distributed re-balance 
	into costzones. This is the final variation of our costzones-based method.
	
	\item[\textit{Algorithm 6}:] The final collective operations are removed from \textit{Algorithm 5}: the octree merge and broadcast of the global octree. The basic octree implementation is reformed as a hashed octree implementation to allow for unique and consistent indexing into all local octrees. The global octree
	thus remains distributed. During tree traversal, 
	when a process requires non-local data, it is requested dynamically from the owning remote process.
	
	\item[\textit{Algorithm 7}:] The tree traversal of \textit{Algorithm 6} is improved by asynchronous communication and a latency-hiding tree traversal. Rather than waiting for requested remote data to be returned, those requested nodes are put on a so-called defer list and the algorithm proceeds in the meantime by traversing 
	other branches.
\end{itemize}

\begin{figure}[htb]
	\centering
	\begin{subfigure}{0.4\textwidth}
		\centering
		\includegraphics[width=\textwidth]{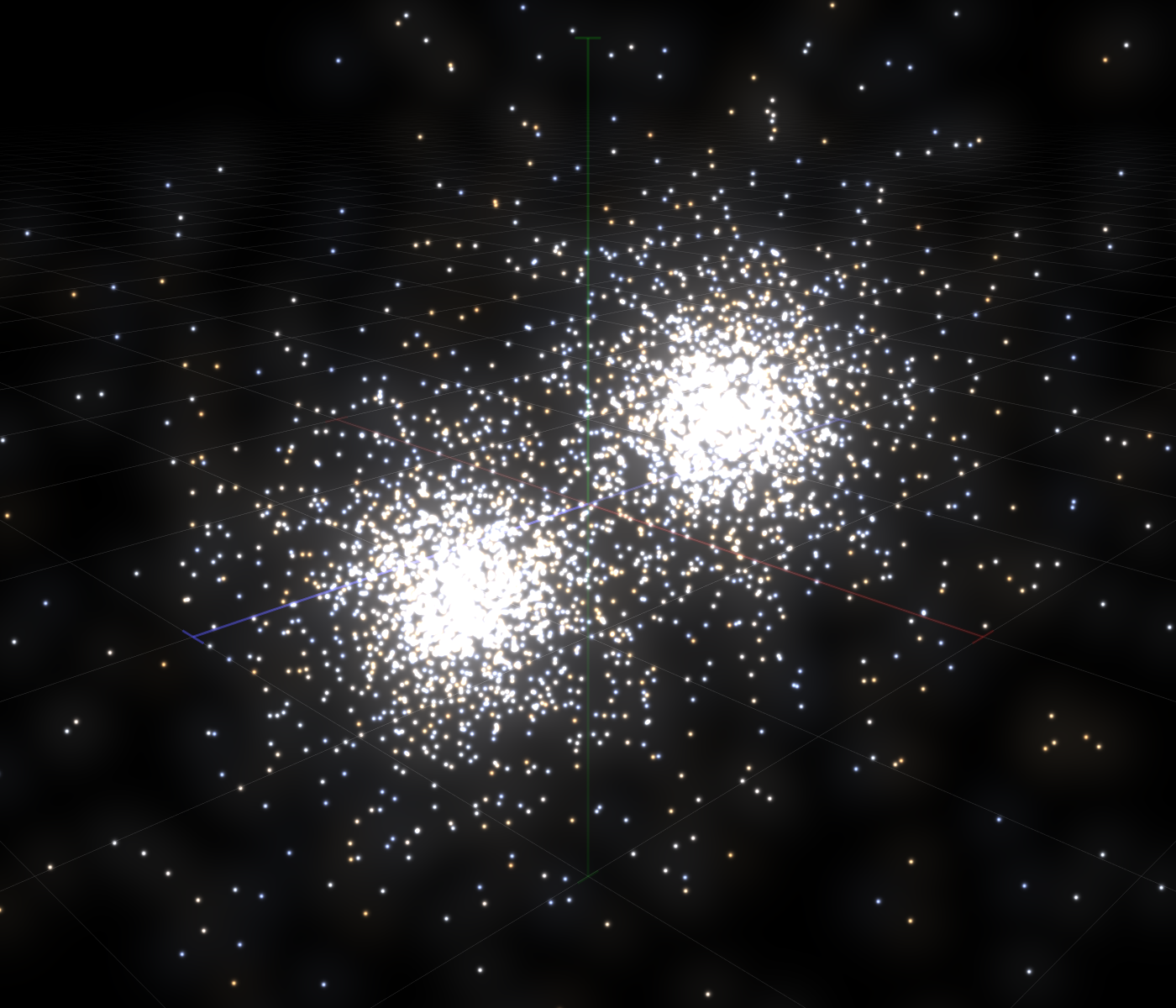}
		\subcaption{$t = 0.0$}
	\end{subfigure}
\hspace{0.95em}	
	\begin{subfigure}{0.4\textwidth}
		\centering
	\includegraphics[width=\textwidth]{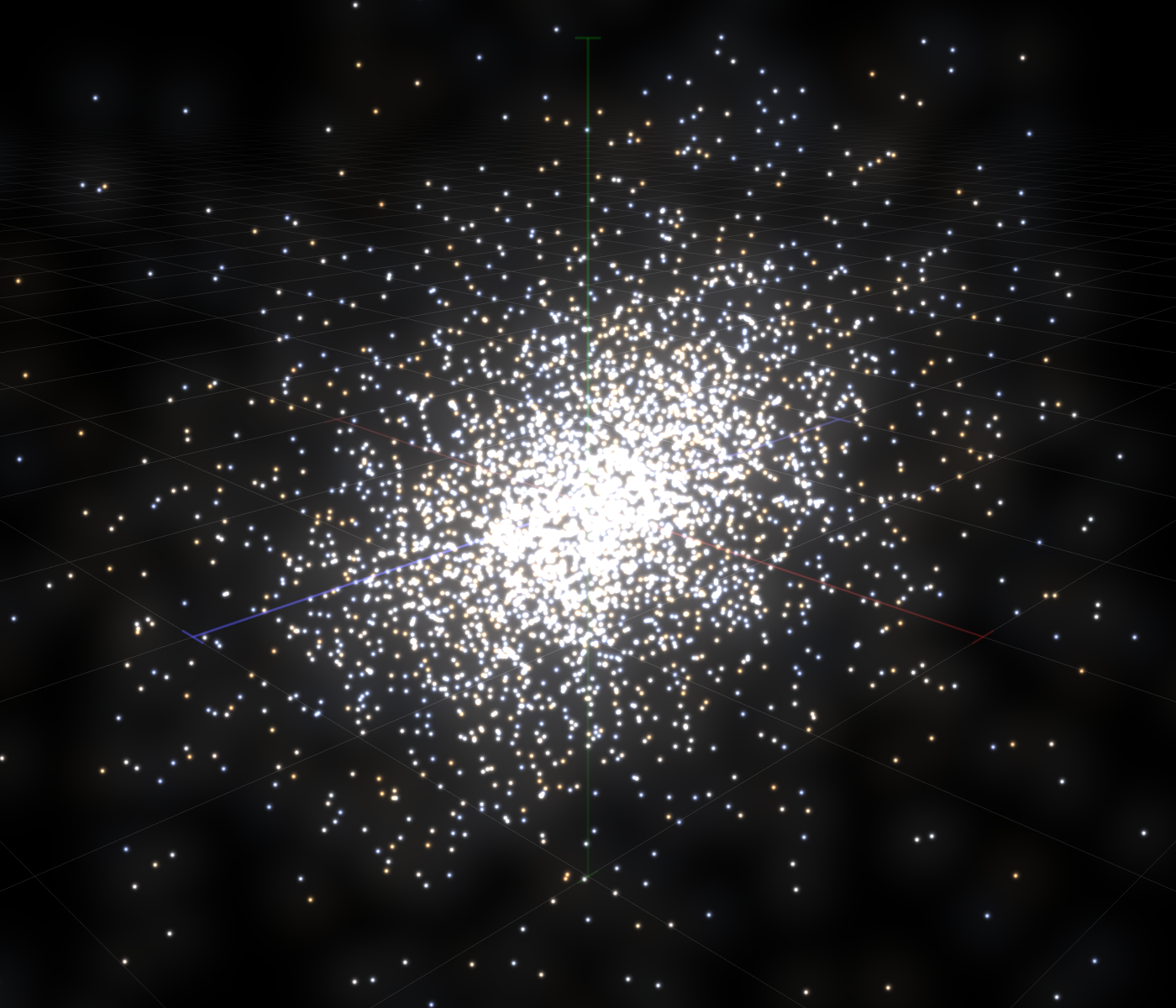}
\subcaption{$t = 2.0$}
\end{subfigure}

\vspace{0.5em}
	\begin{subfigure}{0.4\textwidth}
		\centering
	\includegraphics[width=\textwidth]{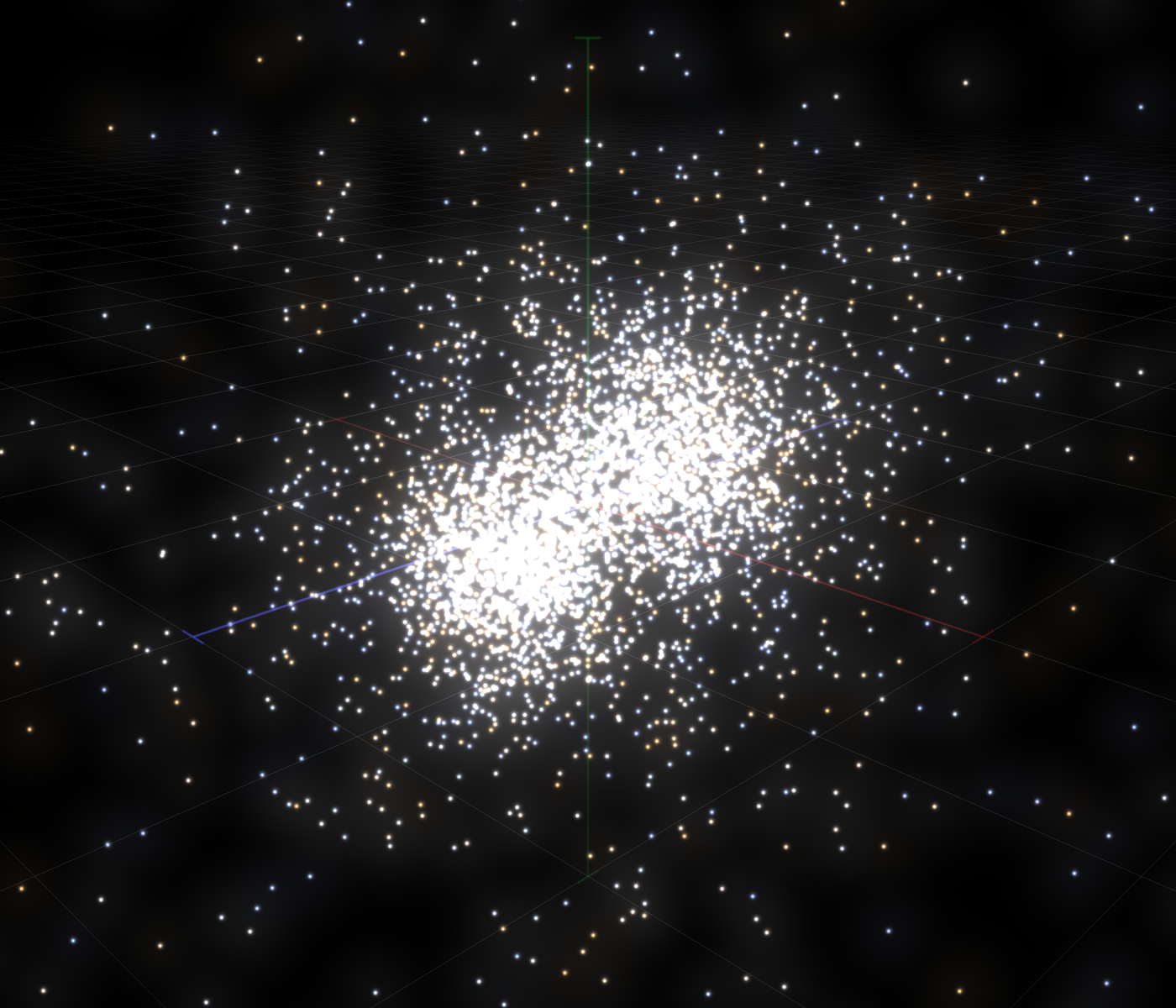}
\subcaption{$t = 4.0$}
\end{subfigure}
\hspace{0.95em}
	\begin{subfigure}{0.4\textwidth}
		\centering
	\includegraphics[width=\textwidth]{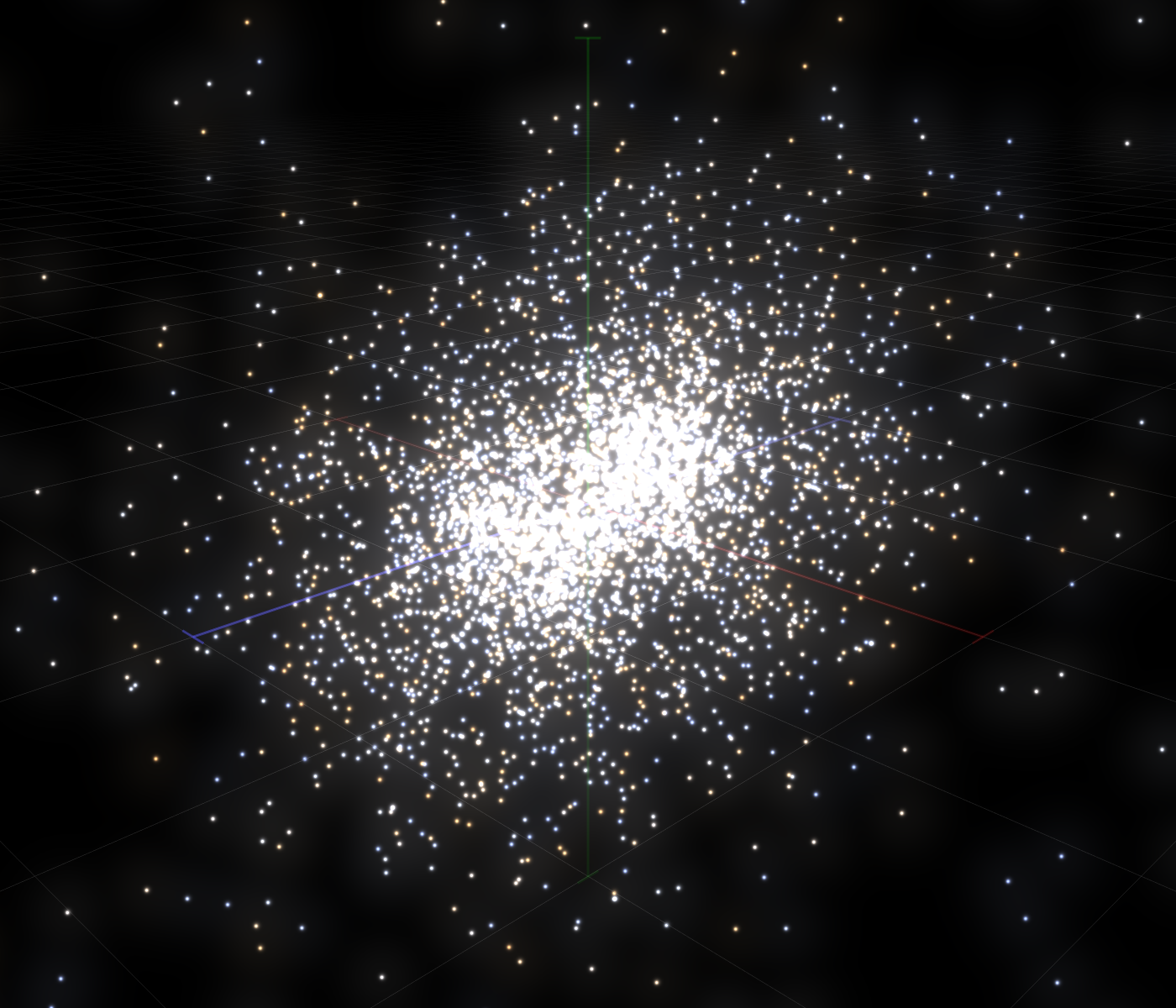}
\subcaption{$t = 6.0$}
\end{subfigure}
	\caption{The interaction and evolution of two Plummer model clusters, each with $2000$ bodies.}\label{fig:simevolution}
\end{figure}

These eight algorithms have been implemented
in C and C++ using the OpenMPI v2.1.1 implementation \cite{gabriel04:_open_mpi}.
We also implement a OpenGL 3.3 implementation of 
a 3D visualization tool for the simulations. Screenshots of which are shown
in Fig.~\ref{fig:ogltwoclusters} and Fig.~\ref{fig:simevolution}.
Our source code is freely available at 
\textcolor{blue}{\url{https://github.com/alexgbrandt/Parallel-NBody/}}.

To test these algorithms, we consider a standard test
\cite{barnes1986hierachical,singh1993parallel,DBLP:journals/tocs/SingHG95,DBLP:journals/jpdc/SinghHTGH95}
involving the interaction and eventual collision of two globular clusters.
Each spherical cluster contains equal-mass stars and follows a \textit{Plummer model} \cite{hut2007ACSvol11}.
In this model, the mass density of a core of some \textit{Plummer radius} is highly dense
and then falls off as $r^{\frac{-5}{2}}$ outside the core.
These clusters are initially separated from each other in each of the spatial dimensions
and then brought together under their mutual gravity.
In a simulation with $N$ bodies, each cluster is given $\nicefrac{N}{2}$.
Fig.~\ref{fig:simevolution} shows how this simulation evolves over time for $N = 4000$.

All experiments in this section 
simulate this interaction of two Plummer clusters, 
varying only in $N$, the number of bodies in the system,
and $p$, the number of parallel processes executing the problem.
All trials use the same random seed when constructing the clusters. 
All simulations are run in the standard units (see Section~\ref{sec:energy})
and with a multipole acceptance criterion $\theta = 0.5$, 
a time-step of 0.01, and 500 total time-steps.
Our experiments were performed on a local area network cluster of 10 nodes, 
each with two Intel Xeon X5650 processors (6 cores each; 12 cores per node),
a 12x4GB DDR3 main memory at 1.33 GHz, and running Ubuntu 18.04.4 and GCC 7.5.0.
For tests where $p \leq 10$, each process is mapped to a distinct
node of the cluster; where $p > 10$, processes are distributed evenly across
nodes and across sockets on each node. 

We begin our experimentation with validating the correctness of 
our implementation.
Recall from Section~\ref{sec:energy} that a stable
simulation will---up to numerical errors---conserve energy.
Table~\ref{tbl:energychange} shows the percent change in total energy of the system
at the end of the 500 simulation steps.
These changes represent a change in energy of less than 0.0005 in standard units. 
The original implementation of Barnes and Hut conserved
energy to about 1\% \cite{barnes1986hierachical}, thus these results are satisfactory.
Note also that for any one algorithm the change in energy was identical for
every number of processes $p$---even up to machine precision---suggesting
that the parallelization itself is correct and does not modify the operation
of the simulation itself.

\begin{table}[htb]
	\centering
	\scriptsize
	\begin{tabular}{lp{3em}rrrrr}
		\toprule
		& $N$ &    10000 &    20000 &    40000 &    60000 &    80000 \\
		\textit{Algorithm} & $p$ &          &          &          &          &          \\
		\midrule
		0 & 1   & 0.1391 & 0.1527 & 0.1538 &    -- &    -- \\
		& 100 & 0.1391 & 0.1527 & 0.1538 &  0.1415 &  0.1520 \\
		1 & 1   & 0.1324 & 0.1497 & 0.1483 & 0.1415 & 0.1520 \\
		& 100 & 0.1324 & 0.1497 & 0.1483 & 0.1415 & 0.1520 \\
		2 & 1   & 0.1324 & 0.1497 & 0.1483 & 0.1415 & 0.1520 \\
		& 100 & 0.1324 & 0.1497 & 0.1483 & 0.1415 & 0.1520 \\
		3 & 1   & 0.1324 & 0.1497 & 0.1483 & 0.1415 & 0.1520 \\
		& 100 & 0.1324 & 0.1497 & 0.1483 & 0.1415 & 0.1520 \\
		4 & 1   & 0.1324 & 0.1497 & 0.1483 & 0.1415 & 0.1520 \\
		& 100 & 0.1324 & 0.1497 & 0.1483 & 0.1415 & 0.1520 \\
		5 & 1   & 0.1324 & 0.1497 & 0.1483 & 0.1415 & 0.1520 \\
		& 100 & 0.1324 & 0.1497 & 0.1483 & 0.1415 & 0.1520 \\
		6 & 1   & 0.1324 & 0.1497 & 0.1483 & 0.1415 & 0.1520 \\
		& 100 & 0.1324 & 0.1497 & 0.1483 & 0.1415 & 0.1520 \\
		7 & 1   & 0.1324 & 0.1497 & 0.1483 & 0.1415 & 0.1520 \\
		& 100 & 0.1324 & 0.1497 & 0.1483 & 0.1415 & 0.1520 \\
		\bottomrule
	\end{tabular}
	\caption{Absolute value of percent change in total energy for various experimental trials. Change in total energy was identical for all number of processes $p$ for any one \textit{Algorithm}.
		Change in energy for the direct method (\textit{Algorithm 0}) can be attributed to the truncation error in integration and floating point error. Change in energy for all others also includes the added error of force approximation. }\label{tbl:energychange}
\end{table}

\begin{figure}[htb]
	\includegraphics[width=\textwidth]{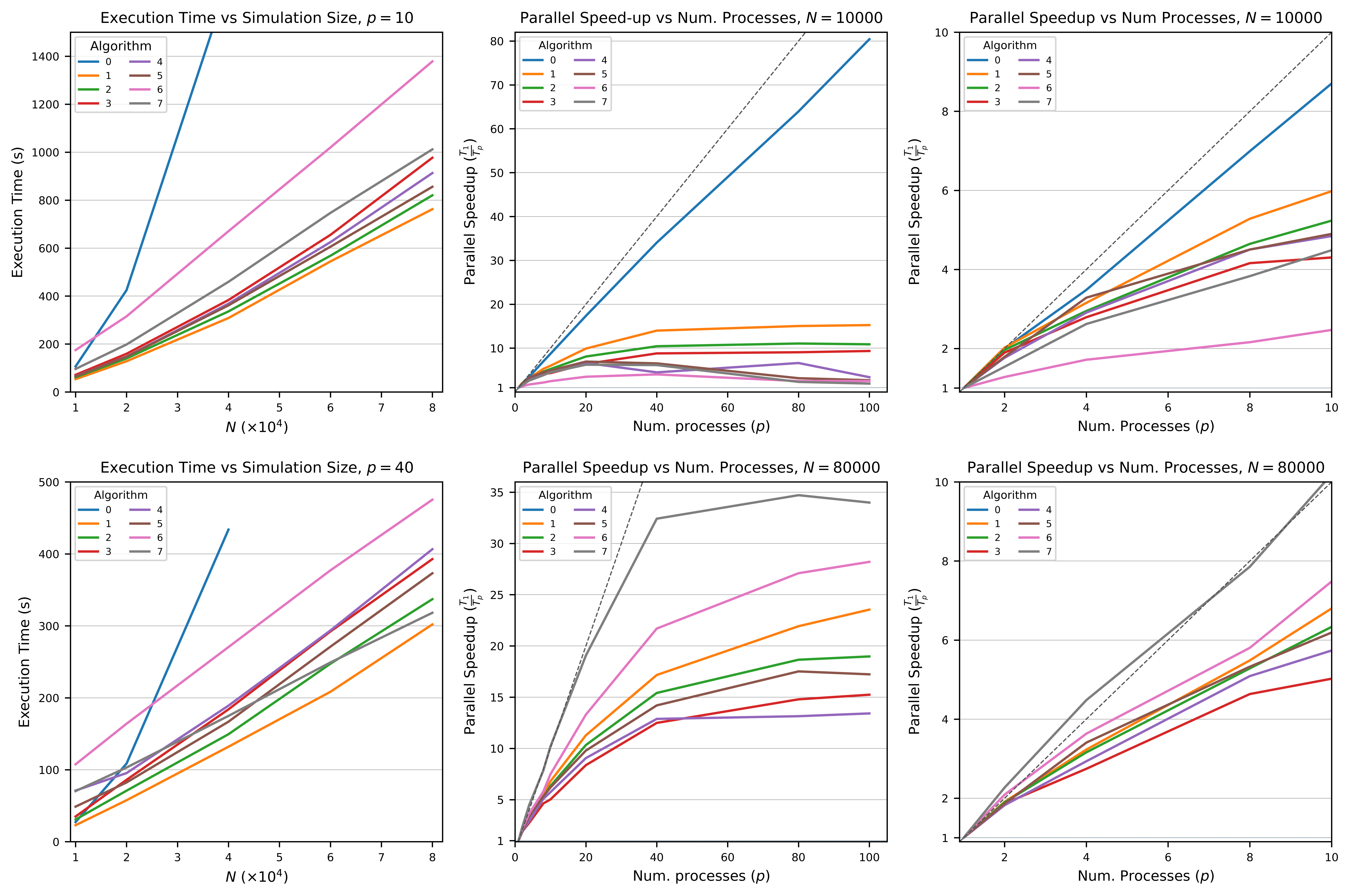}
	\caption{First column: execution time vs simulation size (number of bodies $N$) for $p=10$ and $p=40$, respectively. Second column: parallel speedup vs number of processes, for $N=10000$ and $N=80000$, respectively. No data for \textit{Algorithm 0} for $N=80000$. Third column: the same as the second column, but focusing on the range $p\in[0,10]$. Parallel speedup plots include the ideal speedup as a dashed line. \textit{Algorithm 0} achieves near linear speedup but an exorbitant running time. The more simple algorithms achieve better execution time, but worse parallel speedup, indicating that larger values of $N$ are needed for other algorithms to perform favourably.}\label{fig:plots1}
\end{figure}

Next, we consider a large test suite of our eight algorithms. 
For each algorithm we run the previously described
two cluster scenario for $N \in \{10000,20000,40000,60000,80000\}$
and $p \in \{1,2,4,8,10,20,40,80,100\}$. 
Fig.~\ref{fig:plots1} summarizes these results. 
As expected, \textit{Algorithm 0} runs much slower than the other algorithms
with its $\mathcal{O}(N^2)$ complexity. However, its 
optimal parallelization and minimal communication overhead 
lead to near linear parallel speedup.
For $N=10000$, algorithms achieve moderate speedup
for small $p$ but then communication overheads quickly dominate.
This can be seen as the up then down shape of parallel speedup curves for \textit{Algorithms 4--7}
and the flattening of the curve for the other algorithms.
For $N=80000$, the achieved speedup is much better for the more complex algorithms, 
with \textit{Algorithm 7} achieving the highest speedup, as hoped for. 
Nonetheless, speedup still quickly flattens after $p=40$. This 
again indicates communications are dominating and a larger $N$ is needed 
to take full advantage of the scalability of the algorithms. The algorithms exhibit \textit{weak scaling}
but not \textit{strong scaling}.

\clearpage

Throughout Section~\ref{sec:paralleltrees}
we presented incremental changes as an attempt
to continually improve the parallelization 
of the Barnes-Hut algorithms. Despite this,
the data shows some surprising results, listed below.
\begin{enumerate}[$(i)$]
	\item The most simple Barnes-Hut algorithm, \textit{Algorithm 1}, 
	achieves the best performance in terms of execution time.
	Further, even for $N=80000$, it achieves the third-best parallel speedup.
	The scheme is simple and does not involve much additional work for computing spatial decomposition or load-balancing, just a single gather of bodies and broadcast of the serially-constructed tree.
	However, the slopes of the execution times for $p=40$ suggests this would no longer hold 
	for slightly larger values of $N$.
	\item \textit{Algorithm 7} achieves superlinear speedup for a low to moderate number of processes. While the \textit{work law} tells us that $pT_p \geq T_1$ (the time to execute a program in parallel times the number of processes should always be more than the serial execution time), notice that \textit{Algorithm 7} in fact
	can perform less overall work when more processes are added. This is due to each process 
	computing only a local tree and the global tree never fully constructed. 
	\item \textit{Algorithm 4} achieves very poor performance. The change from \textit{Algorithm 3}
	to \textit{Algorithm 4} is the simple addition of costzones load-balancing. 
	Computing costzones is a simple $\mathcal{O}(N)$ operation which requires only a single
	iteration over the list of work estimates. This alone cannot explain the difference. 
	Rather, this experimentation suggests that the performance of a \textit{regular} gather and
	scatter---where each process sends and receives the exact same size of data---is able to achieve
	much better performance than a \textit{irregular} gather and scatter---where each process sends
	and receives a different amount of data. The costzones load-balancing requires the latter. 
	In implementation this causes the \texttt{MPI\_Gather} and \texttt{MPI\_Scatter} calls 
	to be replaced with \texttt{MPI\_Gatherv} and \texttt{MPI\_Scatterv}, respectively. Some slowdowns are expected when moving to the more generalized irregular routine \cite{DBLP:journals/tpds/TraffGT10}, 
	but this experimentation suggests the performance difference in OpenMPI is quite drastic. It could perhaps be
	mitigated with a newer version of OpenMPI. Regardless, moving to a distributed sort and costzones in \textit{Algorithm 5} removes these irregular gather and scatters and again improves performance.
\end{enumerate}

\begin{figure}
\includegraphics[width=\textwidth]{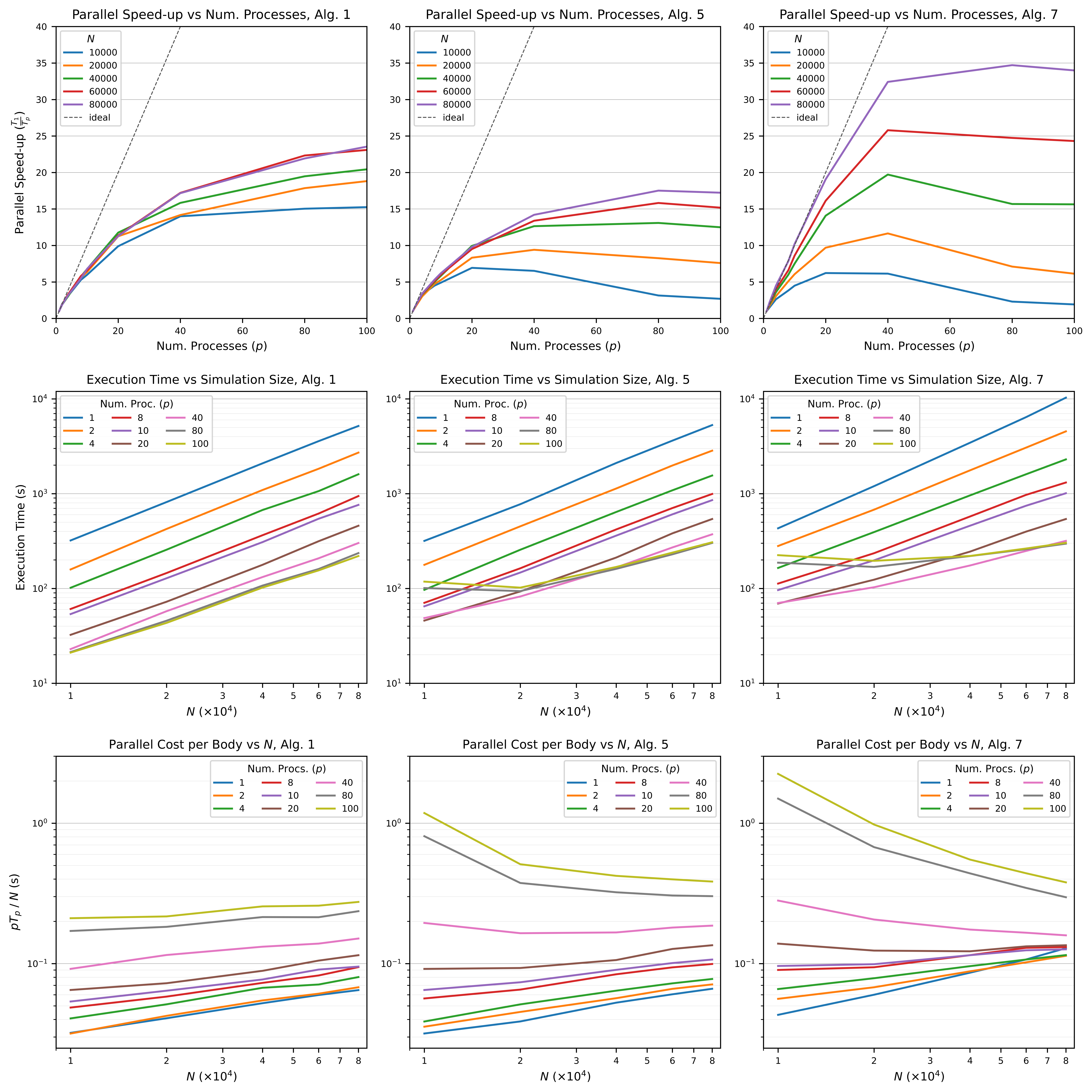}
\caption{
Top row: the parellel speedup vs number $p$ of processes for \textit{Algorithms 1, 5, 7}.
Middle row: the execution time vs number $N$ of bodies in the system for \textit{Algorithms 1, 5, 7}. Note log-log scaling.
Bottom row: the parallel cost (number $p$ of processes times execution time) per body vs number $N$ of bodies in the system for Barnes-Hut versions 1, 5, and 7. Note log-log scaling.}\label{fig:plots2}
\end{figure}

To examine more closely the performance characteristics of the different
parallelization algorithms, Fig.~\ref{fig:plots2} presents
data for \textit{Algorithm 1}, \textit{Algorithm 5}, and \textit{Algorithm 7}.
Recall that \textit{Algorithm 5} is the final variation based on costzones
and \textit{Algorithm 7} is the final variation based on hashed octree.
Fig.~\ref{fig:plots2} presents 3 plots for each algorithm: 
($i$) the parallel speedup achieved for various sizes of $N$;
($ii$) the execution time for various numbers of processes $p$
(as a log-log plot); and ($iii$) the \textit{parallel cost}---$p$ times 
the execution time in parallel $T_p$---scaled by $N$ and for 
various number of processes (as a log-log plot).

First, the parallel speedups for \textit{Algorithm 1} show that 
increasing speedup has reached a limit as $N$ increases.
Parallel speedup does not continue to increase with a 
per-process increase in problem size, thus \textit{Algorithm 1} 
has poor scalability.
\textit{Algorithm 5} improves in its speedup
with increased work per process, but overall speedup is still lacking.  
Communication costs begin to dominate. 
\textit{Algorithm 7} achieves great and continual improvements in speedup
for increasing problem size, indicating its superior scalability. 

\clearpage

Second, the execution time for each algorithm for
various process sizes is plotted against the value of $N$. 
Where these log-log plots show nearly-parallel lines
indicates that adding more processes scales down work proportionately. 
The separation between lines shows continual improvement in
speedup with adding more processes; for \textit{Algorithm 1}
the lines for $p=40,80,100$ are not well-separated again indicating
its poor scalability. For \textit{Algorithm 5} and \textit{Algorithm 7}
the curvature of the lines for high values of $p$ and low values of $N$
indicate how communication and parallelization overheads dominate the running time. 
Where the lines have flattened for larger values of $N$ indicates that the algorithm 
has reaches a sufficient problem size where the simulation work rather dominates. 
Importantly, we see that for $p \geq 80$, neither algorithm has really reached its 
full potential with these smaller data sets. 

Third, we present the value $\nicefrac{pT_p}{N}$ vs $N$ for various values of $p$.
This \textit{parallel cost} indicates how the additional work for parallelization 
and communication overheads changes with adding processes and more work per process.
The distance between the $p=1$ line and any other line indicates the
parallel overhead. For larger values of $N$ these lines should
converge but remain parallel as communication becomes
a fixed proportion of the overall running time.
For \textit{Algorithm 1} all lines have reached
the parallel stage for low to moderate values of $N$, 
indicating they have reached there limit in scalability. 
For \textit{Algorithm 5}, trends have begun to flatten for high values of $N$, 
meanwhile the vertical distance between, say, $p=100$ and $p=1$ is still quite large. 
This indicates that a large proportion of \textit{Algorithm 5}
is spend on parallel overheads. 
For large values of $N$ we see that the lines for \textit{Algorithm 7} converge, 
and even intersect. Where any line is below that of $p=1$ indicates superlinear speedup. 
For $p \geq 40$ we see that the lines are still trending downward, 
indicting that \textit{Algorithm 7} has not reached its ``large-$N$ limit'', 
and further speedup can be achieved on any fixed number of processes 
by increasing the simulation size.

\section{Conclusion}
\label{sec:conclusion}

Throughout this report we have examined 
the necessary mathematical background and 
the classical formulations of hierarchical
methods for gravitational $N$-body simulations. 
We have collected and presented the work of many papers: \cite{barnes1986hierachical}, \cite{warren1993parallel},
\cite{DBLP:journals/ijhpca/SalmonW94},  \cite{warren20142hot}, \cite{singh1993parallel}, \cite{DBLP:journals/jpdc/SinghHTGH95}, and \cite{DBLP:journals/tocs/SingHG95};
applying their algorithmic techniques to progressively
improve a parallel and distributed Barnes-Hut algorithm.
The implementation of these algorithms in C/C++ using MPI
was tested on a small compute cluster with great results. 

Overall, the experimental data suggests two things. 
First, that the costzones method and octree merge \cite{singh1993parallel,DBLP:journals/jpdc/SinghHTGH95,DBLP:journals/jpdc/SinghHTGH95} used 
in \textit{Algorithm 3}, \textit{Algorith 4}, and \textit{Algorithm 5} 
incur a high degree of parallel overhead. 
This method was first developed for a shared-memory system
where communication and inter-process synchronization is significantly cheaper. 
On distributed systems this method is less practical, although much more
simple and easy to implement.
Second, the scalability of \textit{Algorithm 7} is highly encouraging. 
While, for raw execution time, the simplest parallelization scheme of \textit{Algorithm 1}
won out, this trend would not continue for larger simulation sizes
or for a larger number of processes. The hashed octree method \cite{warren1993parallel,DBLP:journals/ijhpca/SalmonW94}
can provide superlinear speedup for large simulations and
provides excellent scaling and speedup for increasing amounts of work per process. 
Further refinement and experimentation is needed to test
the limits of this most complex but most promising algorithm.